\documentclass{article}%
\usepackage{amssymb}
\usepackage{amsmath}
\usepackage{makeidx}
\usepackage{cite}%
\usepackage{amsfonts}%
\usepackage{graphicx}
\usepackage{epsfig}%
\include{psfig.tex}

\begin{document}

\author{Hartmut Wachter\thanks{E-Mail: Hartmut.Wachter@gmx.de}\\An der Schafscheuer 56\\D-91781 Wei\ss enburg, Federal Republic of Germany}
\title{Klein-Gordon equation in $q$-de\-formed Euclidean space}
\maketitle
\date{}

\begin{abstract}
We introduce $q$-ver\-sions of the Klein-Gordon equation in the
three-di\-\-men\-\-sion\-\-al $q$-de\-formed Euclidean space. We determine
plane wave solutions to our $q$-de\-formed Klein-Gordon equations. We show
that these plane wave solutions form a complete orthogonal system. We discuss
the propagators of our $q$-de\-formed Klein-Gordon equations. We derive
continuity equations for the charge density, the energy density, and the
momentum density of a $q$-de\-formed spin-zero particle.

\end{abstract}

\section{Introduction}

It could be that space-time shows a discrete structure at small distances
\cite{Mead:1966zz,Garay:1995}, much like a solid is composed of atoms at the
microscopic level. A discrete space-time structure could mathematically be
described by a noncommutative coordinate algebra such as the $q$-de\-formed
Euclidean space \cite{Faddeev:1987ih,Fichtmueller1996,Lorek:1997eh}. We aim at
discussing quantum mechanical wave equations in $q$-de\-formed Euclidean
space. This discussion could provide clues as to whether space-time is
discrete at small distances, indeed.

In our former work, we have already considered $q$-ver\-sions of the
Schr\"{o}dinger equation for a nonrelativistic particle
\cite{Wachter:2020B,Wachter:2021A}. However, space-time should reveal its
discrete structure only at high particle energies. For this reason, we should
deal with $q$-ver\-sions of relativistic wave equations
\cite{Pillin:1993,Meyer:1994wi,Podles:1996,Blohmann:2001ph,Bachmaier:2003}. It
is possible to write down a Klein-Gordon equation or a Dirac equation in the
so-called $q$-de\-formed Minkowski space \cite{CarowWatamura:1990nk}. The
$q$-de\-formed Minkowski space, however, is such that calculations are
challenging. So, I am going to handle the problem differently: we introduce
and discuss Klein-Gordon equations in $q$-de\-formed Euclidean space.

As shown in\ Ref.~\cite{Wachter:2020A}, we can extend the algebra of
$q$-de\-formed Euclidean space by a time element. This time element is a
commutative parameter. Thus, we obtain Klein-Gordon equations in
$q$-de\-formed Euclidean space from the well-known Klein-Gordon equation if we
only replace the usual Laplace operator with its $q$-ana\-log in
$q$-de\-formed Euclidean space. These $q$-ver\-sions of the Klein-Gordon
equation do not fit with the $q$-de\-formed Poincar\'{e} symmetry
\cite{ogievetsky1992}. Nevertheless, their discussion can give us an idea of
what we might get if we describe space-time by a $q$-de\-formed quantum space.

We know from Ref.~\cite{Bauer:2003} that $q$-de\-formed partial derivatives
can act on wave functions in different ways. Thus, there is more than one
$q$-de\-formed Klein-Gordon equation in the $q$-de\-formed Euclidean space, as
shown in Chap.~\ref{LoeKleGorGleKap}. In Chap.~\ref{KapPlaWavSol}, we derive
plane wave solutions for our $q$-de\-formed Klein-Gordon equations. We also
show that these solutions form a complete system of orthogonal functions. This
fact enables us to write down $q$-ver\-sions of the propagator for a spin-zero
particle in Chap.~\ref{KapProKleGorFel}. In the subsequent chapters, we derive
$q$-de\-formed continuity equations for the probability density, the energy
density, and the momentum density of a spin-zero particle. Our reasonings also
include a $q$-de\-formed spin-zero particle interacting with an
electromagnetic field. The appendix contains some information about
mathematical tools of analysis on $q$-de\-formed Euclidean space. It will help
the reader who is not familiar with our formalism.

\section{$q$-De\-formed Klein-Gor\-don equations\label{LoeKleGorGleKap}}

A $q$-ana\-log of the Klein-Gor\-don equation should be invariant under
actions of the Hopf algebra $\mathcal{U}_{q}(\operatorname*{su}\nolimits_{2})$
(cf. App.~\ref{KapHofStr}). We assume that for $q$-de\-formed particles with
zero spin and rest mass $m$, the following $q$-ver\-sion of the Klein-Gor\-don
equation holds:%
\begin{equation}
c^{-2}\partial_{t}^{\hspace{0.01in}2}\triangleright\varphi_{R}-\hspace
{-0.01in}\nabla_{q}^{2}\triangleright\varphi_{R}+(m\hspace{0.01in}%
c)^{2}\hspace{0.01in}\varphi_{R}=0.\label{KleGorGleLin}%
\end{equation}
The $q$-de\-formed Laplace operator $\nabla_{q}^{2}$ depends on the metric of
the three-di\-men\-sion\-al $q$-de\-formed Euclidean space [also see
Eq.~(\ref{MetDreiDim}) of App.~\ref{KapQuaZeiEle}]:%
\begin{equation}
\nabla_{q}^{2}=\partial^{A}\partial_{A}=g^{AB}\partial_{B}\hspace
{0.01in}\partial_{A}.
\end{equation}

By conjugating Eq.~(\ref{KleGorGleLin}), we obtain another $q$-ver\-sion of
the Klein-Gor\-don equation [see Eq.~(\ref{RegConAbl})
in\ App.~\ref{KapParDer}]:%
\begin{equation}
\varphi_{L}\,\bar{\triangleleft}\,\hspace{0.01in}\partial_{t}^{\hspace
{0.01in}2}c^{-2}\hspace{-0.01in}-\varphi_{L}\,\bar{\triangleleft}%
\,\hspace{0.01in}\nabla_{q}^{2}+\varphi_{L}\hspace{0.01in}(m\hspace
{0.01in}c)^{2}\hspace{-0.01in}=0.\label{KleGorGleRec}%
\end{equation}
Accordingly, the wave function $\varphi_{R}$ transforms into $\varphi_{L}$ by
conjugation [see Eq.~(\ref{KonPotReiKom}) in\ App.~\ref{KapQuaZeiEle}]:%
\begin{equation}
\overline{\varphi_{R}}=\varphi_{L}.\label{KonGKWel1}%
\end{equation}

There are two types of left-ac\-tions and two types of right-ac\-tions for $q
$-de\-formed partial derivatives [see Eq.~(\ref{UnkOpeDarAbl}) and
Eq.~(\ref{KonOpeDarAbl}) in\ App.~\ref{KapParDer}]. Thus, we get further
$q$-ver\-sions of the Klein-Gor\-don equation by applying the following
substitutions in Eq.~(\ref{KleGorGleLin}) or Eq.~(\ref{KleGorGleRec}):%
\begin{equation}
\triangleright\,\leftrightarrow\,\bar{\triangleright},\qquad\bar
{\triangleleft}\,\leftrightarrow\,\triangleleft,\qquad\varphi_{R}%
\leftrightarrow\varphi_{R}^{\ast},\qquad\varphi_{L}\leftrightarrow\varphi
_{L}^{\ast}.\label{ErsKleGor}%
\end{equation}
This way, we have%
\begin{align}
c^{-2}\partial_{t}^{\hspace{0.01in}2}\,\bar{\triangleright}\,\varphi_{R}%
^{\ast}-\nabla_{q}^{2}\,\bar{\triangleright}\,\varphi_{R}^{\ast}%
+(m\hspace{0.01in}c)^{2}\hspace{0.01in}\varphi_{R}^{\ast} &
=0,\nonumber\\[0.03in]
\varphi_{L}^{\ast}\hspace{-0.01in}\triangleleft\partial_{t}^{\hspace{0.01in}%
2}c^{-2}\hspace{-0.01in}-\varphi_{L}^{\ast}\hspace{-0.01in}\triangleleft
\nabla_{q}^{2}+\varphi_{L}^{\ast}\hspace{0.01in}(m\hspace{0.01in}c)^{2} &
=0,\label{KleGorGleLin2}%
\end{align}
where $\varphi_{R}^{\ast}$ transforms into $\varphi_{L}^{\ast}$ by
conjugation:%
\begin{equation}
\overline{\varphi_{R}^{\ast}}=\varphi_{L}^{\ast}.\label{KonGKWel2}%
\end{equation}

If we want to deal with a charged particle moving in the presence of an
electromagnetic field, we have to apply the following substitutions to the $q
$-de\-formed Klein-Gor\-don equations:%
\begin{equation}
\partial_{t}\rightarrow D^{0}=\partial_{t}+\text{i\hspace{0.01in}}%
e\hspace{0.01in}A^{0},\qquad\partial^{C}\rightarrow D^{C}=\partial^{C}%
\hspace{-0.01in}-\text{i\hspace{0.01in}}e\hspace{0.01in}c^{-1}\hspace
{-0.01in}A^{C}.\label{KovAblGes}%
\end{equation}
Thus, the $q$-de\-formed Klein-Gor\-don equations for a charged particle read%
\begin{align}
c^{-2}D^{0}D^{0}\triangleright\varphi_{R} &  =D^{C}D_{C}\triangleright
\varphi_{R}-(m\hspace{0.01in}c)^{2}\hspace{0.01in}\varphi_{R},\nonumber\\
\varphi_{L}\,\bar{\triangleleft}\,D^{0}D^{0}c^{-2} &  =\varphi_{L}%
\,\bar{\triangleleft}\,D^{C}D_{C}-\varphi_{L}\hspace{0.01in}(m\hspace
{0.01in}c)^{2},\label{KleGorWec1}%
\end{align}
and%
\begin{align}
c^{-2}D^{0}D^{0}\,\bar{\triangleright}\,\varphi_{R}^{\ast} &  =D^{C}%
D_{C}\,\bar{\triangleright}\,\varphi_{R}^{\ast}-(m\hspace{0.01in}c)^{2}%
\hspace{0.01in}\varphi_{R}^{\ast},\nonumber\\
\varphi_{L}^{\ast}\hspace{-0.01in}\triangleleft D^{0}D^{0}c^{-2} &
=\varphi_{L}^{\ast}\hspace{-0.01in}\triangleleft D^{C}D_{C}-\varphi_{L}^{\ast
}\hspace{0.01in}(m\hspace{0.01in}c)^{2},\label{KleGorWec2}%
\end{align}
where the `action' of the potentials on the wave functions is defined by the star-prod\-uct [see App.~\ref{KapQuaZeiEle}]:%
\begin{align}
A^{0}\triangleright\varphi_{R} &  =A^{0}\hspace{-0.01in}\circledast\varphi
_{R}, & A^{C}\triangleright\varphi_{R} &  =A^{C}\hspace{-0.01in}%
\circledast\varphi,\nonumber\\
\varphi_{L}^{\ast}\triangleleft A^{0}\hspace{-0.01in} &  =\varphi_{L}^{\ast
}\circledast A^{0}, & \varphi_{L}^{\ast}\triangleleft A^{C}\hspace{-0.01in} &
=\varphi_{L}^{\ast}\circledast A^{C}.
\end{align}
The $q$-de\-formed Klein-Gor\-don equation for $\varphi_{R}$ again transforms
into the $q$-de\-formed Klein-Gor\-don equation for $\varphi_{L}$ if the
potentials $A^{0}$ and $A^{C}$ behave as follows under conjugation:%
\begin{equation}
\overline{A^{0}}=A_{0},\qquad\overline{A^{C}}=A_{C}.
\end{equation}
The same holds for the $q$-de\-formed Klein-Gor\-don equations in
Eq.~(\ref{KleGorWec2}).

In the following, we show that the $q$-de\-formed Klein-Gor\-don equations in
Eqs.~(\ref{KleGorWec1}) and (\ref{KleGorWec2}) are invariant under the gauge
transformations%
\begin{align}
e\hspace{0.01in}\tilde{A}^{C}\hspace{-0.01in} &  =e\hspace{0.01in}A^{C}%
\hspace{-0.01in}+\partial^{C}\hspace{-0.01in}\triangleright e\hspace
{0.01in}\chi=e\hspace{0.01in}A^{C}\hspace{-0.01in}-e\hspace{0.01in}%
\chi\triangleleft\partial^{C},\nonumber\\
e\hspace{0.01in}\tilde{A}^{0}\hspace{-0.01in} &  =e\hspace{0.01in}A^{0}%
\hspace{-0.01in}-\partial_{t}\triangleright e\hspace{0.01in}\chi
=e\hspace{0.01in}A^{0}\hspace{-0.01in}+e\hspace{0.01in}\chi\triangleleft
\partial_{t},\label{EicTraVekSkaPot}%
\end{align}
and%
\begin{align}
\tilde{\varphi}_{R}(\mathbf{x},t) &  =\exp(\text{i\hspace{0.01in}}%
e\hspace{0.01in}c^{-1}\chi)\circledast\varphi_{R}(\mathbf{x},t),\nonumber\\
\tilde{\varphi}_{L}^{\ast}(\mathbf{x},t) &  =\varphi_{L}^{\ast}(\mathbf{x}%
,t)\circledast\exp(-\text{\hspace{0.01in}i\hspace{0.01in}}e\hspace
{0.01in}c^{-1}\chi).
\end{align}
Note that $\chi$ has to be a central element of the algebra of position space.
Additionally, $\chi$ shows trivial braiding [cf. Eq.~(\ref{trivBrai}) in
App.~\ref{KapHofStr}].

That the $q$-de\-formed Klein-Gor\-don equation for $\varphi_{R}$ [cf.
Eq.~(\ref{KleGorWec1})] is invariant under the above gauge transformations is
a direct consequence of the following identities:%
\begin{align}
\tilde{D}^{C}\triangleright\tilde{\varphi}_{R} &  =\exp(\text{i\hspace
{0.01in}}e\hspace{0.01in}c^{-1}\chi)\circledast D^{C}\hspace{-0.01in}%
\triangleright\varphi_{R},\nonumber\\
\tilde{D}^{0}\triangleright\tilde{\varphi}_{R} &  =\exp(\text{i\hspace
{0.01in}}e\hspace{0.01in}c^{-1}\chi)\circledast D^{0}\hspace{-0.01in}%
\triangleright\varphi_{R}.\label{ChaIdeKovAbl}%
\end{align}
To prove the first identity, we do the following calculation
\cite{Wachter:2021A}:%
\begin{align}
\tilde{D}^{C}\triangleright\tilde{\varphi}_{R}= &  \hspace{0.04in}\partial
^{C}\triangleright\big (\exp(\text{i\hspace{0.01in}}e\hspace{0.01in}c^{-1}%
\chi)\circledast\varphi_{R}\big )\nonumber\\
&  -\big (\hspace{0.01in}\text{i\hspace{0.01in}}e\hspace{0.01in}c^{-1}%
\hspace{-0.01in}A^{C}\hspace{-0.01in}+\text{i}\hspace{0.01in}\partial
^{C}\hspace{-0.01in}\triangleright e\hspace{0.01in}c^{-1}\chi\big )\circledast
\exp(\text{i}\hspace{0.01in}e\hspace{0.01in}c^{-1}\chi)\circledast\varphi
_{R}\nonumber\\
= &  \hspace{0.04in}\partial^{C}\hspace{-0.01in}\triangleright\exp
(\text{i\hspace{0.01in}}e\hspace{0.01in}c^{-1}\chi)\circledast\varphi_{R}%
+\exp(\text{i\hspace{0.01in}}e\hspace{0.01in}c^{-1}\chi)\circledast
\partial^{C}\hspace{-0.01in}\triangleright\varphi_{R}\nonumber\\
&  -\big (\hspace{0.01in}\text{i\hspace{0.01in}}e\hspace{0.01in}A^{C}%
\hspace{-0.01in}+\text{i\hspace{0.01in}}\partial^{C}\hspace{-0.01in}%
\triangleright e\hspace{0.01in}c^{-1}\chi\big )\circledast\exp(\text{i\hspace
{0.01in}}e\hspace{0.01in}c^{-1}\chi)\circledast\varphi_{R}%
.\label{WirKanImpEichInv}%
\end{align}
The last step follows from the Leibniz rules of the $q$-de\-formed partial
derivatives and the trivial braiding of $\chi$. For the same reasons, we have%
\begin{align}
\partial^{C}\triangleright(e\hspace{0.01in}c^{-1}\chi)^{n} &  =\sum
_{j\hspace{0.01in}=\hspace{0.01in}0}^{n\hspace{0.01in}-1}\hspace
{0.01in}(e\hspace{0.01in}c^{-1}\chi)^{j}\circledast(\partial^{C}%
\hspace{-0.01in}\triangleright e\hspace{0.01in}c^{-1}\chi)\circledast
(e\hspace{0.01in}c^{-1}\chi)^{n\hspace{0.01in}-1-j}\nonumber\\
&  =n\hspace{0.02in}(\partial^{C}\hspace{-0.01in}\triangleright e\hspace
{0.01in}c^{-1}\chi)\circledast(e\hspace{0.01in}c^{-1}\chi)^{n\hspace
{0.01in}-1}\label{ParAblSkaPotN}%
\end{align}
with%
\begin{equation}
(e\hspace{0.01in}c^{-1}\chi)^{n}=\hspace{0.01in}\underset{n\text{-times}%
}{\underbrace{e\hspace{0.01in}c^{-1}\chi\circledast\ldots\circledast
e\hspace{0.01in}c^{-1}\chi}}.
\end{equation}
From Eq.~(\ref{ParAblSkaPotN}) follows:%
\begin{align}
\partial^{C}\hspace{-0.01in}\triangleright\exp(\text{i\hspace{0.01in}}%
e\hspace{0.01in}c^{-1}\chi) &  =\sum_{n\hspace{0.01in}=\hspace{0.01in}%
0}^{\infty}\frac{\text{i}^{n}}{n!}\,\partial^{C}\hspace{-0.01in}%
\triangleright(e\hspace{0.01in}c^{-1}\chi)^{n}\nonumber\\
&  =\sum_{n\hspace{0.01in}=1}^{\infty}\frac{\text{i}^{n}}{(n-1)!}%
\,(\partial^{C}\hspace{-0.01in}\triangleright e\hspace{0.01in}c^{-1}%
\chi)\circledast(e\hspace{0.01in}c^{-1}\chi)^{n\hspace{0.01in}-1}\nonumber\\
&  =\text{i}\hspace{0.01in}\partial^{C}\hspace{-0.01in}\triangleright
e\hspace{0.01in}c^{-1}\chi\circledast\sum_{n\hspace{0.01in}=\hspace{0.01in}%
0}^{\infty}\frac{\text{i}^{n}}{n!}\,(e\hspace{0.01in}c^{-1}\chi)^{n}%
\nonumber\\
&  =\text{i\hspace{0.01in}}\partial^{C}\hspace{-0.01in}\triangleright
e\hspace{0.01in}c^{-1}\chi\circledast\exp(\text{i\hspace{0.01in}}%
e\hspace{0.01in}c^{-1}\chi).\label{AblPhaFak}%
\end{align}
If we insert this result into Eq.~(\ref{WirKanImpEichInv}), we finally get:%
\begin{align}
\tilde{D}^{C}\hspace{-0.01in}\triangleright\tilde{\varphi}_{R}= &
\hspace{0.03in}\partial^{C}\hspace{-0.01in}\triangleright\exp(\text{i\hspace
{0.01in}}e\hspace{0.01in}c^{-1}\chi)\circledast\varphi_{R}+\exp(\text{i\hspace
{0.01in}}e\hspace{0.01in}c^{-1}\chi)\circledast\partial^{C}\hspace
{-0.01in}\triangleright\varphi_{R}\nonumber\\
&  -\big (\text{\hspace{0.01in}i\hspace{0.01in}}e\hspace{0.01in}c^{-1}%
\hspace{-0.01in}A^{C}\hspace{-0.01in}+\text{i\hspace{0.01in}}\partial
^{C}\hspace{-0.01in}\triangleright e\hspace{0.01in}c^{-1}\hspace{-0.01in}%
\chi\big )\circledast\exp(\text{i\hspace{0.01in}}e\hspace{0.01in}c^{-1}%
\chi)\circledast\varphi_{R}\nonumber\\
= &  \hspace{0.03in}\exp(\text{i\hspace{0.01in}}e\hspace{0.01in}c^{-1}%
\chi)\circledast\partial^{C}\hspace{-0.01in}\triangleright\varphi_{R}%
-\exp(\text{i\hspace{0.01in}}e\hspace{0.01in}c^{-1}\chi)\circledast
\text{i\hspace{0.01in}}e\hspace{0.01in}c^{-1}\hspace{-0.01in}A^{C}%
\hspace{-0.01in}\circledast\varphi_{R}\nonumber\\
= &  \hspace{0.03in}\exp(\text{i\hspace{0.01in}}e\hspace{0.01in}c^{-1}%
\chi)\circledast D^{C}\hspace{-0.01in}\triangleright\varphi_{R}%
.\label{KanImpLinWirEic}%
\end{align}
We can prove the second identity in Eq.~(\ref{ChaIdeKovAbl}) with similar
reasonings:%
\begin{align}
\tilde{D}^{0}\triangleright\tilde{\varphi}_{R}= &  \hspace{0.04in}\partial
_{t}\triangleright\big (\exp(\text{i\hspace{0.01in}}e\hspace{0.01in}c^{-1}%
\chi)\circledast\varphi_{R}\big )+\text{i\hspace{0.01in}}e\hspace{0.01in}%
A^{0}\hspace{-0.01in}\circledast\exp(\text{i}e\hspace{0.01in}c^{-1}%
\chi)\circledast\varphi_{R}\nonumber\\
&  -\text{i\hspace{0.01in}}\partial_{t}\triangleright e\hspace{0.01in}%
c^{-1}\chi\circledast\exp(\text{i\hspace{0.01in}}e\hspace{0.01in}c^{-1}%
\chi)\circledast\varphi_{R}\nonumber\\
= &  \hspace{0.04in}\partial_{t}\triangleright\exp(\text{i\hspace{0.01in}%
}e\hspace{0.01in}c^{-1}\chi)\circledast\varphi_{R}+\exp(\text{i\hspace
{0.01in}}e\hspace{0.01in}c^{-1}\chi)\circledast\partial_{t}\triangleright
\varphi_{R}\nonumber\\
&  +\text{i\hspace{0.01in}}e\hspace{0.01in}A^{0}\hspace{-0.01in}%
\circledast\exp(\text{i\hspace{0.01in}}e\hspace{0.01in}c^{-1}\chi
)\circledast\varphi_{R}-\text{i\hspace{0.01in}}\partial_{t}\hspace
{-0.01in}\triangleright e\hspace{0.01in}c^{-1}\chi\circledast\exp
(\text{i\hspace{0.01in}}e\hspace{0.01in}c^{-1}\chi)\circledast\varphi
_{R}\nonumber\\
= &  \hspace{0.04in}\exp(\text{i\hspace{0.01in}}e\hspace{0.01in}c^{-1}%
\chi)\circledast D^{0}\hspace{-0.01in}\triangleright\varphi_{R}.
\end{align}
In the last step of the calculation above, we have taken the definition of
$D^{0}$ and the following identity into account:%
\begin{equation}
\partial_{t}\triangleright\exp(\text{i\hspace{0.01in}}e\hspace{0.01in}%
c^{-1}\chi)=\text{i}\hspace{0.01in}\partial_{t}\hspace{-0.01in}\triangleright
e\hspace{0.01in}c^{-1}\chi\circledast\exp(\text{i\hspace{0.01in}}%
e\hspace{0.01in}c^{-1}\chi),
\end{equation}
In the same manner, we can prove the following identities:%
\begin{align}
\tilde{\varphi}_{L}^{\ast}\triangleleft\hspace{0.01in}\tilde{D}^{0}%
\hspace{-0.01in} &  =\varphi_{L}^{\ast}\triangleleft\hspace{0.01in}%
D^{0}\hspace{-0.01in}\circledast\exp(-\text{i\hspace{0.01in}}e\hspace
{0.01in}c^{-1}\chi),\nonumber\\
\tilde{\varphi}_{L}^{\ast}\triangleleft\hspace{0.01in}\tilde{D}^{C}%
\hspace{-0.01in} &  =\varphi_{L}^{\ast}\triangleleft\hspace{0.01in}%
D^{C}\hspace{-0.01in}\circledast\exp(-\text{i\hspace{0.01in}}e\hspace
{0.01in}c^{-1}\chi).
\end{align}
These identities imply that the $q$-de\-formed Klein-Gor\-don equation for
$\varphi_{L}^{\ast}$ [cf. Eq.~(\ref{KleGorWec2})] is also invariant under
gauge transformations.

Finally, we will show that our $q$-de\-formed Klein-Gor\-don equations become
$q$-de\-formed Schr\"{o}\-din\-ger equations if the kinetic energy is small
compared to the rest energy. Our reasonings are in complete analogy to the
undeformed case.\ Accordingly, we separate a phase factor depending on the
rest energy $m\hspace{0.01in}c^{2}$:%
\begin{align}
\varphi_{R} &  =\psi_{R}\exp(-\text{i}\hspace{0.01in}m\hspace{0.01in}%
c^{2}\hspace{0.01in}t), & \varphi_{L} &  =\psi_{L}\exp(\text{i}\hspace
{0.01in}m\hspace{0.01in}c^{2}\hspace{0.01in}t),\nonumber\\
\varphi_{R}^{\ast} &  =\psi_{R}^{\ast}\exp(-\text{i}\hspace{0.01in}%
m\hspace{0.01in}c^{2}\hspace{0.01in}t), & \varphi_{L}^{\ast} &  =\psi
_{L}^{\ast}\exp(\text{i}\hspace{0.01in}m\hspace{0.01in}c^{2}\hspace{0.01in}t).
\end{align}
In the nonrelativistic limit, the total energy $E$ of a particle is slightly
different from its rest energy. Thus, we have%
\begin{align}
|\text{i\hspace{0.01in}}\partial_{t}\triangleright\psi_{R}| &  \approx
|\psi_{R}\,(E-m\hspace{0.01in}c^{2})|\ll|\psi_{R}\,m\hspace{0.01in}%
c^{2}|,\nonumber\\
|\psi_{L}^{\ast}\triangleleft\partial_{t}\text{i}| &  \approx|(E-m\hspace
{0.01in}c^{2})\,\psi_{L}^{\ast}|\ll|m\hspace{0.01in}c^{2}\,\psi_{L}^{\ast
}|,\label{AbsRelGre}%
\end{align}
and%
\begin{align}
|e\hspace{0.01in}A^{0}\hspace{-0.01in}\circledast\psi_{R}| &  \ll
|m\hspace{0.01in}c^{2}\,\psi_{R}|,\nonumber\\
|\psi_{L}^{\ast}\circledast e\hspace{0.01in}A^{0}| &  \ll|\psi_{L}^{\ast
}\hspace{0.01in}m\hspace{0.01in}c^{2}\,|.\label{AbsRelGre2}%
\end{align}
Similar approximations hold for $\psi_{R}^{\ast}$ and $\psi_{L}$, so we can
restrict ourselves to the wave functions $\psi_{R}$ and $\psi_{L}^{\ast}$. Due
to the conditions in Eqs.~(\ref{AbsRelGre}) and (\ref{AbsRelGre2}), we can
make the following approximation:%
\begin{align}
&  (\partial_{t}+\text{i\hspace{0.01in}}e\hspace{0.01in}A^{0})(\partial
_{t}+\text{i\hspace{0.01in}}e\hspace{0.01in}A^{0})\triangleright\varphi
_{R}=\nonumber\\
&  \qquad=\big[\hspace{0.02in}\underline{\partial_{t}\partial_{t}%
\triangleright\psi_{R}}-2\hspace{0.01in}\text{i\hspace{0.01in}}\partial
_{t}\triangleright\psi_{R}\hspace{0.01in}m\hspace{0.01in}c^{2}-\psi_{R}%
\hspace{0.01in}(m\hspace{0.01in}c^{2})^{2}+\text{i\hspace{0.01in}}\partial
_{t}\triangleright e\hspace{0.01in}A^{0}\hspace{-0.01in}\circledast\psi
_{R}\nonumber\\
&  \qquad\hspace{0.17in}+\underline{2\text{\hspace{0.01in}i\hspace{0.01in}%
}e\hspace{0.01in}A^{0}\hspace{-0.01in}\circledast\partial_{t}\triangleright
\psi_{R}}+2\text{\hspace{0.01in}}A^{0}\hspace{-0.01in}\circledast\psi
_{R}\hspace{0.01in}m\hspace{0.01in}c^{2}-\underline{e\hspace{0.01in}%
A^{0}\hspace{-0.01in}\circledast e\hspace{0.01in}A^{0}\hspace{-0.01in}%
\circledast\psi_{R}}\hspace{0.02in}\big]\exp(-\text{\hspace{0.01in}i}%
\hspace{0.01in}m\hspace{0.01in}c^{2}\hspace{0.01in}t)\nonumber\\
&  \qquad\approx\big[-2\text{\hspace{0.01in}i\hspace{0.01in}}\partial
_{t}\triangleright\psi_{R}\hspace{0.01in}m\hspace{0.01in}c^{2}-\psi_{R}%
\hspace{0.01in}(m\hspace{0.01in}c^{2})^{2}+\text{i\hspace{0.01in}}\partial
_{t}\triangleright e\hspace{0.01in}A^{0}\hspace{-0.01in}\circledast\psi
_{R}\nonumber\\
&  \qquad\hspace{0.17in}+2\text{\hspace{0.01in}}e\hspace{0.01in}A^{0}%
\hspace{-0.01in}\circledast\psi_{R}\hspace{0.01in}m\hspace{0.01in}c^{2}%
\hspace{0.01in}\big]\exp(-\text{i}\hspace{0.01in}m\hspace{0.01in}c^{2}%
\hspace{0.01in}t).
\end{align}
Note that we have ignored the underlined terms in the last step. Using the
approximation above, we get a $q$-de\-formed Schr\"{o}\-din\-ger equation for
$\psi_{R}$ from the $q$-de\-formed Klein-Gor\-don equation for $\varphi_{R}$
[cf. Eq.~(\ref{KleGorGleLin})]:%
\begin{align}
\text{i\hspace{0.01in}}\partial_{t}\triangleright\psi_{R}  = &  -(2\hspace
{0.01in}m)^{-1}\hspace{0.01in}(\partial^{C}-\text{i\hspace{0.01in}}%
e\hspace{0.01in}c^{-1}\hspace{-0.01in}A^{C})(\partial_{C}-\text{i\hspace
{0.01in}}e\hspace{0.01in}c^{-1}\hspace{-0.01in}A_{C})\triangleright\psi
_{R}\nonumber\\
&  +e\hspace{0.01in}A^{0}\hspace{-0.01in}\circledast\psi_{R}+(2\hspace
{0.01in}m\hspace{0.01in}c^{2})^{-1}(\text{i\hspace{0.01in}}\partial
_{t}\triangleright e\hspace{0.01in}A^{0})\circledast\psi_{R}.\label{SchEqPsiR}%
\end{align}
Similar reasonings lead us to a $q$-de\-formed Schr\"{o}\-din\-ger equation
for the wave function $\psi_{L}^{\ast}$:%
\begin{align}
\psi_{L}^{\ast}\triangleleft\partial_{t}\text{i}  =  & -(2\hspace{0.01in}%
m)^{-1}\hspace{0.01in}\psi_{L}^{\ast}\triangleleft(\partial^{C}-\text{i\hspace
{0.01in}}e\hspace{0.01in}c^{-1}\hspace{-0.01in}A^{C})(\partial_{C}%
-\text{i\hspace{0.01in}}e\hspace{0.01in}c^{-1}\hspace{-0.01in}A_{C}%
)\nonumber\\
&  +\psi_{L}^{\ast}\circledast e\hspace{0.01in}A^{0}\hspace{-0.01in}%
+(2\hspace{0.01in}mc^{2})^{-1}\psi_{L}^{\ast}\circledast(e\hspace{0.01in}%
A^{0}\hspace{-0.01in}\triangleleft\partial_{t}\text{i}).\label{SchEqPsiSteL}%
\end{align}
These $q$-de\-formed Schr\"{o}\-din\-ger equations are of the same form as
those given in Ref.~\cite{Wachter:2020B}.

\section{Plane wave solutions\label{KapPlaWavSol}}

In Ref.~\cite{Wachter:2019A}, we have introduced $q$-de\-formed momentum
eigenfunctions (also see App.~\ref{KapExp}):%
\begin{align}
u_{\hspace{0.01in}\mathbf{p}}(\mathbf{x}) &  =\operatorname*{vol}%
\nolimits^{-1/2}\exp_{q}(\mathbf{x}|\text{i}\mathbf{p}), & u^{\mathbf{p}%
}(\mathbf{x}) &  =\operatorname*{vol}\nolimits^{-1/2}\exp_{q}(\text{i}%
^{-1}\mathbf{p}|\hspace{0.01in}\mathbf{x}),\nonumber\\
\bar{u}_{\hspace{0.01in}\mathbf{p}}(\mathbf{x}) &  =\operatorname*{vol}%
\nolimits^{-1/2}\overline{\exp}_{q}(\mathbf{x}|\text{i}\mathbf{p}), & \bar
{u}^{\mathbf{p}}(\mathbf{x}) &  =\operatorname*{vol}\nolimits^{-1/2}%
\overline{\exp}_{q}(\text{i}^{-1}\mathbf{p}|\hspace{0.01in}\mathbf{x}%
).\label{ImpEigFktqDef}%
\end{align}
The volume element is defined by [also see Eq.~(\ref{DefIntSpa}) in
App.~\ref{KapParDer}]%
\begin{equation}
\operatorname*{vol}=\hspace{-0.02in}\int\text{d}_{q}^{3}\hspace{0.01in}%
p\int\text{d}_{q}^{3}x\hspace{0.01in}\exp_{q}(\text{i}^{-1}\mathbf{p}%
|\mathbf{x}).\label{VolEleDef}%
\end{equation}
The $q$-de\-formed momentum eigenfunctions are subject to the following
eigenvalue equations [cf. Eq.~(\ref{EigGl1N}) of App.~\ref{KapExp}]:%
\begin{align}
\text{i}^{-1}\partial^{A}\triangleright u_{\hspace{0.01in}\mathbf{p}%
}(\mathbf{x}) &  =u_{\hspace{0.01in}\mathbf{p}}(\mathbf{x})\circledast p^{A},
& u^{\mathbf{p}}(\mathbf{x})\,\bar{\triangleleft}\,\partial^{A}\hspace
{0.01in}\text{i}^{-1} &  =p^{A}\circledast u^{\mathbf{p}}(\mathbf{x}%
),\nonumber\\
\text{i}^{-1}\hat{\partial}^{A}\,\bar{\triangleright}\,\bar{u}_{\hspace
{0.01in}\mathbf{p}}(\mathbf{x}) &  =\bar{u}_{\hspace{0.01in}\mathbf{p}%
}(\mathbf{x})\circledast p^{A}, & \bar{u}^{\mathbf{p}}(\mathbf{x}%
)\triangleleft\hat{\partial}^{A}\hspace{0.01in}\text{i}^{-1} &  =p^{A}%
\circledast\bar{u}^{\mathbf{p}}(\mathbf{x}).\label{EigGleImpOpeImpEigFkt0}%
\end{align}
We can also introduce `dual'\ momentum eigenfunctions [also see
Eq.~(\ref{DuaExp2}) of App.~\ref{KapExp}]:%
\begin{align}
(u^{\ast})_{\mathbf{p}}(\mathbf{x}) &  =\operatorname*{vol}\nolimits^{-1/2}%
\exp_{q}^{\ast}(\text{i}\mathbf{p}|\hspace{0.01in}\mathbf{x}), & (u^{\ast
})^{\mathbf{p}}(\mathbf{x}) &  =\operatorname*{vol}\nolimits^{-1/2}\exp
_{q}^{\ast}(\mathbf{x}|\text{i}^{-1}\mathbf{p}),\nonumber\\
(\bar{u}^{\ast})_{\mathbf{p}}(\mathbf{x}) &  =\operatorname*{vol}%
\nolimits^{-1/2}\overline{\exp}_{q}^{\ast}(\text{i}\mathbf{p}|\hspace
{0.01in}\mathbf{x}), & (\bar{u}^{\ast})^{\mathbf{p}}(\mathbf{x}) &
=\operatorname*{vol}\nolimits^{-1/2}\overline{\exp}_{q}^{\ast}(\mathbf{x}%
|\text{i}^{-1}\mathbf{p}).\label{DefDuaImpEigFktWdh}%
\end{align}
The corresponding eigenvalue equations are given by [cf.
Eq.~(\ref{EigGleExpQueAbl}) of App.~\ref{KapExp}]%
\begin{align}
(u^{\ast})_{\mathbf{p}}(\mathbf{x})\triangleleft\partial^{A}\hspace
{0.01in}\text{i}^{-1}\hspace{-0.01in} &  =p^{A}\circledast(u^{\ast
})_{\mathbf{p}}(\mathbf{x}),\nonumber\\
\text{i}^{-1}\partial^{A}\,\bar{\triangleright}\,(u^{\ast})^{\mathbf{p}%
}(\mathbf{x}) &  =(u^{\ast})^{\mathbf{p}}(\mathbf{x})\circledast
p^{A},\label{ImpEigFktqDef2}%
\end{align}
or%
\begin{align}
(\bar{u}^{\ast})_{\mathbf{p}}(\mathbf{x})\,\bar{\triangleleft}\,\hat{\partial
}^{A}\hspace{0.01in}\text{i}^{-1}\hspace{-0.01in} &  =p^{A}\circledast(\bar
{u}^{\ast})_{\mathbf{p}}(\mathbf{x}),\nonumber\\
\text{i}^{-1}\hat{\partial}^{A}\triangleright(\bar{u}^{\ast})^{\mathbf{p}%
}(\mathbf{x}) &  =(\bar{u}^{\ast})^{\mathbf{p}}(\mathbf{x})\circledast p^{A}.
\end{align}
To obtain the various expressions and identities for the momentum
eigenfunction with a bar, we apply the following substitutions:%
\begin{gather}
\triangleright\,\leftrightarrow\,\bar{\triangleright},\qquad\triangleleft
\,\leftrightarrow\,\bar{\triangleleft},\qquad\partial^{A}\,\leftrightarrow
\,\hat{\partial}^{A},\qquad u\,\leftrightarrow\,\bar{u},\nonumber\\
+\,\leftrightarrow\,-,\qquad q\,\leftrightarrow\,q^{-1}.
\end{gather}
For this reason, we need not consider the momentum eigenfunctions $\bar
{u}_{\hspace{0.01in}\mathbf{p}}$ and $(\bar{u}^{\ast})_{\hspace{0.01in}%
\mathbf{p}}$ or $\bar{u}^{\mathbf{p}}$ and $(\bar{u}^{\ast})^{\mathbf{p}}$ in
the following.

We first examine whether the $q$-de\-formed Klein-Gor\-don equations given in
Eqs.~(\ref{KleGorGleLin}) and (\ref{KleGorGleRec}) of the previous chapter
have plane wave solutions. To this end, we consider the functions%
\begin{align}
\varphi_{\mathbf{p}}(\mathbf{x},t) &  =\frac{c}{\sqrt{2}}\,u_{\hspace
{0.01in}\mathbf{p}}(\mathbf{x})\circledast\exp(-\text{i\hspace{0.01in}%
}tE_{\mathbf{p}})\circledast E_{\mathbf{p}}^{\hspace{0.01in}-1/2},\nonumber\\
\varphi^{\hspace{0.01in}\mathbf{p}}(\mathbf{x},t) &  =\frac{c}{\sqrt{2}%
}\,E_{\mathbf{p}}^{\hspace{0.01in}-1/2}\hspace{-0.01in}\circledast
\exp(\text{i\hspace{0.01in}}tE_{\mathbf{p}})\circledast u^{\mathbf{p}%
}(\mathbf{x}),\label{EbeWelKGF}%
\end{align}
with the time-de\-pen\-dent phase factors%
\begin{equation}
\exp(\pm\hspace{0.01in}\text{i\hspace{0.01in}}tE_{\mathbf{p}})=\sum
_{n\hspace{0.01in}=\hspace{0.01in}0}^{\infty}\frac{(\pm\hspace{0.01in}%
\text{i\hspace{0.01in}}tE_{\mathbf{p}})^{n}}{n!}.
\end{equation}
Note that $\varphi_{\mathbf{p}}$ and $\varphi^{\hspace{0.01in}\mathbf{p}}$ are
subject to the following identities:%
\begin{align}
\text{i}^{-1}\partial^{A}\triangleright\varphi_{\mathbf{p}}(\mathbf{x},t) &
=\varphi_{\mathbf{p}}(\mathbf{x},t)\circledast p^{A}, & \varphi^{\mathbf{p}%
}(\mathbf{x},t)\,\bar{\triangleleft}\,\partial^{A}\hspace{0.01in}\text{i}^{-1}
&  =p^{A}\circledast\varphi^{\mathbf{p}}(\mathbf{x},t),\nonumber\\
\text{i}\partial_{t}\triangleright\varphi_{\mathbf{p}}(\mathbf{x},t) &
=\varphi_{\mathbf{p}}(\mathbf{x},t)\circledast E_{\mathbf{p}}, &
\varphi^{\mathbf{p}}(\mathbf{x},t)\,\bar{\triangleleft}\,\partial_{t}%
\hspace{0.01in}\text{i} &  =E_{\mathbf{p}}\circledast\varphi^{\mathbf{p}%
}(\mathbf{x},t).\label{DerPlaWav}%
\end{align}
Inserting the functions $\varphi_{\mathbf{p}}$ or $\varphi^{\hspace
{0.01in}\mathbf{p}}$ into our $q$-de\-formed Klein-Gor\-don equations, we
obtain%
\begin{align}
0 &  =c^{-2}\partial_{t}^{\hspace{0.01in}2}\triangleright\varphi_{\mathbf{p}%
}-\nabla_{q}^{2}\triangleright\varphi_{\mathbf{p}}+(m\hspace{0.01in}%
c)^{2}\varphi_{\mathbf{p}}\nonumber\\
&  =\varphi_{\mathbf{p}}\circledast(\hspace{0.01in}p^{B}\hspace{-0.01in}%
\circledast p_{B}-c^{-2}E_{\mathbf{p}}\circledast E_{\mathbf{p}}%
+(m\hspace{0.01in}c)^{2})\label{qKleGorGleImp1}%
\end{align}
or%
\begin{align}
0 &  =\varphi^{\hspace{0.01in}\mathbf{p}}\,\bar{\triangleleft}\,\hspace
{0.01in}\partial_{t}^{\hspace{0.01in}2}c^{-2}\hspace{-0.01in}-\varphi
^{\hspace{0.01in}\mathbf{p}}\,\bar{\triangleleft}\,\hspace{0.01in}\nabla
_{q}^{2}\hspace{0.01in}+\varphi^{\hspace{0.01in}\mathbf{p}}(m\hspace
{0.01in}c)^{2}\nonumber\\
&  =(\hspace{0.01in}p^{B}\hspace{-0.01in}\circledast p_{B}-c^{-2}%
E_{\mathbf{p}}\circledast E_{\mathbf{p}}+(m\hspace{0.01in}c)^{2}%
)\circledast\varphi^{\hspace{0.01in}\mathbf{p}}.\label{qKleGorGleImp2}%
\end{align}
The plane waves in Eq.~(\ref{EbeWelKGF}) are solutions to the $q$-de\-formed
Klein-Gor\-don equations if the following \textit{en\-er\-gy-mo\-men\-tum
relation} holds:%
\begin{equation}
c^{-2}E_{\mathbf{p}}\circledast E_{\mathbf{p}}=p^{B}\hspace{-0.01in}%
\circledast p_{B}+(m\hspace{0.01in}c)^{2}.\label{EneMomRelKleGor}%
\end{equation}
Since square mass $m^{2}$ commutes with square momentum $\mathbf{p}^{2}%
(=p^{B}\hspace{-0.01in}\circledast p_{B})$, we can formally solve the
en\-er\-gy-mo\-men\-tum relation for $E_{\mathbf{p}}$. This way, we get an
infinite series:\footnote{We do not discuss conditions for the convergence of
this series.}%
\begin{equation}
E_{\mathbf{p}}=c\,(\hspace{0.01in}\mathbf{p}^{2}+(m\hspace{0.01in}%
c)^{2})^{1/2}=c\sum_{k\hspace{0.01in}=\hspace{0.01in}0}^{\infty}\binom{1/2}%
{k}\,\mathbf{p}^{2k}(m\hspace{0.01in}c)^{1-\hspace{0.01in}2k}.
\end{equation}
This result can be generalized as follows ($\alpha\in\mathbb{Q}$):%
\begin{equation}
E_{\mathbf{p}}^{\hspace{0.01in}2\alpha}=c^{2\alpha}\sum_{k\hspace
{0.01in}=\hspace{0.01in}0}^{\infty}\binom{\alpha}{k}\,\mathbf{p}^{2k}%
(m\hspace{0.01in}c)^{2(\alpha-k)}.
\end{equation}
In the formula above, we have to substitute powers of $\mathbf{p}^{2}$ by
their nor\-mal-or\-dered expressions \cite{Wachter:2020B}:%
\begin{equation}
\mathbf{p}^{2k}=\hspace{0.01in}\overset{k-\text{times}}{\overbrace
{\mathbf{p}^{2}\circledast\ldots\circledast\mathbf{p}^{2}}}=\sum
_{l\hspace{0.01in}=\hspace{0.01in}0}^{k}\hspace{0.01in}q^{-2l}(-q-q^{-1}%
)^{k-l}%
\genfrac{[}{]}{0pt}{}{k}{l}%
_{q^{4}}\,(\hspace{0.01in}p_{-})^{k-l}(\hspace{0.01in}p_{3})^{2l}%
(\hspace{0.01in}p_{+})^{k-l}.\label{EntPotP}%
\end{equation}
Note that the $q$-de\-formed binomial coefficients are defined in complete
analogy to the undeformed case [also see Eq.~(\ref{qFakDef}) in
App.~\ref{KapQuaZeiEle}]:%
\begin{equation}%
\genfrac{[}{]}{0pt}{}{n}{k}%
_{q}=\frac{[[\hspace{0.01in}n\hspace{0.01in}]]_{q}!}{[[\hspace{0.01in}%
n-k\hspace{0.01in}]]_{q}!\hspace{0.01in}[[\hspace{0.01in}k\hspace
{0.01in}]]_{q}!}.\label{qBinKoeBas}%
\end{equation}

The $q$-de\-formed momentum eigenfunctions form a complete orthogonal system
of functions \cite{Kempf:1994yd,Wachter:2019A}. In the following, we will show
that the same applies to the plane wave solutions of our $q$-de\-formed
Klein-Gordon equations.

We recall that the $q$-de\-formed momentum eigenfunctions in
Eqs.~(\ref{ImpEigFktqDef}) and (\ref{DefDuaImpEigFktWdh}) fulfill the
orthogonality relation \cite{Wachter:2019A}%
\begin{align}
\int\text{d}_{q}^{3}x\,(u^{\ast})_{\mathbf{p}}(\mathbf{x})\circledast
u_{\hspace{0.01in}\mathbf{p}^{\prime}}(\mathbf{x}) &  =\operatorname*{vol}%
\nolimits^{-1}\hspace{-0.02in}\int\text{d}_{q}^{3}x\hspace{0.01in}\exp
_{q}^{\ast}(\text{i}\mathbf{p}|\mathbf{x})\circledast\exp_{q}(\mathbf{x}%
|\text{i}\mathbf{p}^{\prime})\nonumber\\
&  =\operatorname*{vol}\nolimits^{-1}\hspace{-0.01in}\delta_{q}^{\hspace
{0.01in}3}((\ominus\hspace{0.01in}\kappa^{-1}\mathbf{p})\oplus\mathbf{p}%
^{\prime})\label{SkaProEbeDreExpWie0}%
\end{align}
or%
\begin{align}
\int\text{d}_{q}^{3}x\,u^{\mathbf{p}}(\mathbf{x})\circledast(u^{\ast
})^{\mathbf{p}^{\prime}}(\mathbf{x}) &  =\operatorname*{vol}\nolimits^{-1}%
\hspace{-0.02in}\int\text{d}_{q}^{3}x\hspace{0.01in}\exp_{q}(\text{i}%
^{-1}\mathbf{p}|\mathbf{x})\circledast\exp_{q}^{\ast}(\mathbf{x|}\text{i}%
^{-1}\mathbf{p}^{\prime})\nonumber\\
&  =\operatorname*{vol}\nolimits^{-1}\hspace{-0.01in}\delta_{q}^{\hspace
{0.01in}3}(\hspace{0.01in}\mathbf{p}\oplus(\ominus\hspace{0.01in}\kappa
^{-1}\mathbf{p}^{\prime})).\label{SkaProEbeDreExpWie1}%
\end{align}
Here $\delta_{q}^{\hspace{0.01in}3}(\hspace{0.01in}\mathbf{p})$ denotes a $q
$-de\-formed version of the three-di\-men\-sion\-al delta function:%
\begin{equation}
\delta_{q}^{\hspace{0.01in}3}(\hspace{0.01in}\mathbf{p})=\hspace{-0.01in}%
\int\text{d}_{q}^{3}x\hspace{0.01in}\exp_{q}(\text{i}^{-1}\mathbf{p}%
|\mathbf{x})=\hspace{-0.01in}\int\text{d}_{q}^{3}x\hspace{0.01in}\exp
_{q}^{\ast}(\mathbf{x|}\text{i}^{-1}\mathbf{p}).
\end{equation}
In analogy to their undeformed counterpart, the $q$-de\-formed delta function
is subject to the following identities:\footnote{The occurrence of
$\kappa^{-1}=q^{-6}$ indicates that the value of each spatial coordinate is
multiplied by that constant.}%
\begin{align}
f(\hspace{0.01in}\mathbf{y}) &  =\operatorname*{vol}\nolimits^{-1}%
\int\nolimits_{-\infty}^{+\infty}\text{d}_{q}^{3}x\,\delta_{q}^{\hspace
{0.01in}3}(\hspace{0.01in}\mathbf{y}\oplus(\ominus\hspace{0.01in}\kappa
^{-1}\mathbf{x}))\circledast f(\mathbf{x})\nonumber\\
&  =\operatorname*{vol}\nolimits^{-1}\int\nolimits_{-\infty}^{+\infty}%
\text{d}_{q}^{3}x\,\delta_{q}^{\hspace{0.01in}3}((\ominus\hspace{0.01in}%
\kappa^{-1}\mathbf{y})\oplus\mathbf{x})\circledast f(\mathbf{x})\nonumber\\
&  =\operatorname*{vol}\nolimits^{-1}\int\nolimits_{-\infty}^{+\infty}%
\text{d}_{q}^{3}x\,f(\mathbf{x})\circledast\delta_{q}^{\hspace{0.01in}%
3}((\ominus\hspace{0.01in}\kappa^{-1}\mathbf{x})\oplus\mathbf{y}))\nonumber\\
&  =\operatorname*{vol}\nolimits^{-1}\int\nolimits_{-\infty}^{+\infty}%
\text{d}_{q}^{3}x\,f(\mathbf{x})\circledast\delta_{q}^{\hspace{0.01in}%
3}(\mathbf{x}\oplus(\ominus\hspace{0.01in}\kappa^{-1}\mathbf{y}%
)).\label{AlgChaIdeqDelFkt}%
\end{align}

We can write down orthogonality relations for the plane wave solutions in
Eq.~(\ref{EbeWelKGF}) if we introduce the following functions:%
\begin{align}
(\varphi^{\ast})_{\mathbf{p}}(\mathbf{x},t) &  =\frac{c}{\sqrt{2}%
}\,E_{\mathbf{p}}^{\hspace{0.01in}-1/2}\hspace{-0.01in}\circledast
\exp(\text{i}\hspace{0.01in}tE_{\mathbf{p}})\circledast(u^{\ast})_{\mathbf{p}%
}(\mathbf{x}),\nonumber\\
(\varphi^{\ast})^{\mathbf{p}}(\mathbf{x},t) &  =\frac{c}{\sqrt{2}}%
\hspace{0.01in}(u^{\ast})^{\mathbf{p}}(\mathbf{x})\circledast\exp
(-\text{i\hspace{0.01in}}tE_{\mathbf{p}})\circledast E_{\mathbf{p}}%
^{\hspace{0.01in}-1/2}.\label{EbeWelKGFSte}%
\end{align}
The\ functions $(\varphi^{\ast})_{\mathbf{p}}$ or $(\varphi^{\ast
})^{\mathbf{p}}$ are subject to the identities%
\begin{align}
(\varphi^{\ast})_{\mathbf{p}}(\mathbf{x},t)\triangleleft\partial^{A}%
\text{i}^{-1} &  =p^{A}\circledast(\varphi^{\ast})_{\mathbf{p}}(\mathbf{x}%
,t),\nonumber\\
(\varphi^{\ast})_{\mathbf{p}}(\mathbf{x},t)\triangleleft\partial_{t}%
\hspace{0.01in}\text{i} &  =E_{\mathbf{p}}\circledast(\varphi^{\ast
})_{\mathbf{p}}(\mathbf{x},t),\label{DerPlaWavSte}%
\end{align}
or%
\begin{align}
\text{i}^{-1}\partial^{A}\,\bar{\triangleright}\,(\varphi^{\ast})^{\mathbf{p}%
}(\mathbf{x},t) &  =(\varphi^{\ast})^{\mathbf{p}}(\mathbf{x},t)\circledast
p^{A},\nonumber\\
\text{i\hspace{0.01in}}\partial_{t}\,\bar{\triangleright}\,(\varphi^{\ast
})^{\mathbf{p}}(\mathbf{x},t) &  =(\varphi^{\ast})^{\mathbf{p}}(\mathbf{x}%
,t)\circledast E_{\mathbf{p}}.
\end{align}
Moreover, $(\varphi^{\ast})_{\mathbf{p}}$ or $(\varphi^{\ast})^{\mathbf{p}}$
are plane wave solutions to the following $q$-ver\-sions of the Klein-Gor\-don
equations:%
\begin{align}
(\varphi^{\ast})_{\mathbf{p}}\hspace{-0.01in}\triangleleft\partial
_{t}^{\hspace{0.01in}2}c^{-2}-(\varphi^{\ast})_{\mathbf{p}}\triangleleft
\nabla_{q}^{2}+(\varphi^{\ast})_{\mathbf{p}}\hspace{0.01in}(m\hspace
{0.01in}c)^{2}\hspace{-0.01in} &  =0,\nonumber\\[0.03in]
c^{-2}\partial_{t}^{\hspace{0.01in}2}\triangleright\hspace{-0.01in}%
(\varphi^{\ast})^{\mathbf{p}}-\nabla_{q}^{2}\,\hspace{0.01in}\bar
{\triangleright}\,(\varphi^{\ast})^{\mathbf{p}}+(m\hspace{0.01in}c)^{2}%
\hspace{0.01in}(\varphi^{\ast})^{\mathbf{p}}\hspace{-0.01in} &
=0.\label{qKleGorGleImp3}%
\end{align}
You can see this by inserting the expressions for $(\varphi^{\ast
})_{\mathbf{p}}$ or $(\varphi^{\ast})^{\mathbf{p}}$ into the above
$q$-de\-formed Klein-Gor\-don equations. Doing so, we regain the
en\-er\-gy-mo\-men\-tum relation in Eq.~(\ref{EneMomRelKleGor}).

Taking the conjugation properties of $q$-de\-formed exponentials into account
[cf.\ Eq.~(\ref{KonEigExpQua}) in App.~\ref{KapExp}], it follows from
Eqs.~(\ref{EbeWelKGF}) and (\ref{EbeWelKGFSte}) that our plane wave solutions
behave under conjugation as follows:%
\begin{equation}
\overline{\varphi_{\mathbf{p}}}=\varphi^{\hspace{0.01in}\mathbf{p}}%
,\qquad\overline{(\varphi^{\ast})_{\mathbf{p}}}=(\varphi^{\ast})^{\mathbf{p}%
}.\label{ConPropPlaWav}%
\end{equation}

The results so far enable us to write down \textit{orthogonality relations}
for the plane wave solutions of our $q$-de\-formed Klein-Gordon equations,
i.~e.%
\begin{align}
&  \text{i\hspace{0.01in}}c^{-2}\hspace{-0.02in}\int\text{d}_{q}%
^{3}x\,(\varphi^{\ast})_{\mathbf{p}}(\mathbf{x},\pm\hspace{0.01in}%
t)\triangleleft\partial_{t}\circledast\varphi_{\mathbf{p}^{\prime}}%
(\mathbf{x},\pm\hspace{0.01in}t)\nonumber\\
&  +\text{i\hspace{0.01in}}c^{-2}\hspace{-0.02in}\int\text{d}_{q}%
^{3}x\,(\varphi^{\ast})_{\mathbf{p}}(\mathbf{x},\pm\hspace{0.01in}%
t)\circledast\partial_{t}\triangleright\varphi_{\mathbf{p}^{\prime}%
}(\mathbf{x},\pm\hspace{0.01in}t)\nonumber\\
&  \qquad\qquad=\pm\operatorname*{vol}\nolimits^{-1}\hspace{-0.01in}\delta
_{q}^{3}((\ominus\hspace{0.01in}\kappa^{-1}\mathbf{p})\oplus\mathbf{p}%
^{\prime})\label{OrtRelKleGorPhiSte}%
\end{align}
and%
\begin{align}
&  \text{i\hspace{0.01in}}c^{-2}\hspace{-0.02in}\int\text{d}_{q}^{3}%
x\,\varphi^{\hspace{0.01in}\mathbf{p}}(\mathbf{x},\pm\hspace{0.01in}%
t)\,\bar{\triangleleft}\,\partial_{t}\circledast(\varphi^{\ast})^{\mathbf{p}%
^{\prime}}(\mathbf{x},\pm\hspace{0.01in}t)\nonumber\\
&  +\text{i\hspace{0.01in}}c^{-2}\hspace{-0.02in}\int\text{d}_{q}%
^{3}x\,\varphi^{\hspace{0.01in}\mathbf{p}}(\mathbf{x},\pm\hspace
{0.01in}t)\circledast\partial_{t}\,\bar{\triangleright}\,(\varphi^{\ast
})^{\mathbf{p}^{\prime}}(\mathbf{x},\pm\hspace{0.01in}t)\nonumber\\
&  \qquad\qquad=\pm\operatorname*{vol}\nolimits^{-1}\hspace{-0.01in}\delta
_{q}^{3}(\hspace{0.01in}\mathbf{p}\oplus(\ominus\hspace{0.01in}\kappa
^{-1}\mathbf{p}^{\prime})).
\end{align}
If the signs of the time coordinates in the above expressions are different,
the integrals will vanish, i.$~$e.%
\begin{align}
&  \text{i\hspace{0.01in}}c^{-2}\hspace{-0.02in}\int\text{d}_{q}%
^{3}x\,(\varphi^{\ast})_{\mathbf{p}}(\mathbf{x},\pm\hspace{0.01in}%
t)\triangleleft\partial_{t}\circledast\varphi_{\mathbf{p}^{\prime}}%
(\mathbf{x},\mp\hspace{0.01in}t)\nonumber\\
&  +\text{i\hspace{0.01in}}c^{-2}\hspace{-0.02in}\int\text{d}_{q}%
^{3}x\,(\varphi^{\ast})_{\mathbf{p}}(\mathbf{x},\pm\hspace{0.01in}%
t)\circledast\partial_{t}\triangleright\varphi_{\mathbf{p}^{\prime}%
}(\mathbf{x},\mp\hspace{0.01in}t)=0
\end{align}
and%
\begin{align}
&  \text{i\hspace{0.01in}}c^{-2}\hspace{-0.02in}\int\text{d}_{q}^{3}%
x\,\varphi^{\hspace{0.01in}\mathbf{p}}(\mathbf{x},\pm\hspace{0.01in}%
t)\,\bar{\triangleleft}\,\partial_{t}\circledast(\varphi^{\ast})^{\mathbf{p}%
^{\prime}}(\mathbf{x},\mp\hspace{0.01in}t)\nonumber\\
&  +\text{i\hspace{0.01in}}c^{-2}\hspace{-0.02in}\int\text{d}_{q}%
^{3}x\,\varphi^{\hspace{0.01in}\mathbf{p}}(\mathbf{x},\pm\hspace
{0.01in}t)\circledast\partial_{t}\,\bar{\triangleright}\,(\varphi^{\ast
})^{\mathbf{p}^{\prime}}(\mathbf{x},\mp\hspace{0.01in}%
t)=0.\label{OrtRelKleGorPhiSteEnd}%
\end{align}

To prove the above orthogonality relation, you only need to evaluate the
corresponding integrals. We show this by the following calculation:%
\begin{align}
&  \text{i\hspace{0.01in}}c^{-2}\hspace{-0.02in}\int\text{d}_{q}%
^{3}x\,(\varphi^{\ast})_{\mathbf{p}}(\mathbf{x},\pm\hspace{0.01in}%
t)\triangleleft\partial_{t}\circledast\varphi_{\mathbf{p}^{\prime}}%
(\mathbf{x},\pm\hspace{0.01in}t)=\nonumber\\
&  \qquad=\text{i}\hspace{0.01in}2^{-1}E_{\mathbf{p}}^{\hspace{0.01in}%
-1/2}\hspace{-0.01in}\circledast\exp(\pm\hspace{0.01in}\text{i}\hspace
{0.01in}tE_{\mathbf{p}})\triangleleft\partial_{t}\nonumber\\
&  \qquad\hspace{0.17in}\circledast\int\text{d}_{q}^{3}x\,(u^{\ast
})_{\mathbf{p}}(\mathbf{x})\circledast u_{\hspace{0.01in}\mathbf{p}^{\prime}%
}(\mathbf{x})\circledast\exp(\mp\text{i\hspace{0.01in}}tE_{\mathbf{p}^{\prime
}})\circledast E_{\mathbf{p}^{\prime}}^{\hspace{0.01in}-1/2}\nonumber\\
&  \qquad=\pm\hspace{0.01in}2^{-1}E_{\mathbf{p}}^{\hspace{0.01in}-1/2}%
\hspace{-0.01in}\circledast E_{\mathbf{p}}\circledast\exp(\pm\hspace
{0.01in}\text{i}\hspace{0.01in}tE_{\mathbf{p}})\nonumber\\
&  \qquad\hspace{0.17in}\circledast\operatorname*{vol}\nolimits^{-1}\delta
_{q}^{3}((\ominus\hspace{0.01in}\kappa^{-1}\mathbf{p})\oplus\mathbf{p}%
^{\prime})\circledast\exp(\mp\text{i\hspace{0.01in}}tE_{\mathbf{p}^{\prime}%
})\circledast E_{\mathbf{p}^{\prime}}^{\hspace{0.01in}-1/2}\nonumber\\
&  \qquad=\pm\hspace{0.01in}2^{-1}E_{\mathbf{p}}^{\hspace{0.01in}1/2}%
\hspace{-0.01in}\circledast\exp(\pm\hspace{0.01in}\text{i}\hspace
{0.01in}tE_{\mathbf{p}})\circledast\exp(\mp\text{i\hspace{0.01in}%
}tE_{\mathbf{p}})\circledast E_{\mathbf{p}}^{\hspace{0.01in}-1/2}\nonumber\\
&  \qquad\hspace{0.17in}\circledast\operatorname*{vol}\nolimits^{-1}\delta
_{q}^{3}((\ominus\hspace{0.01in}\kappa^{-1}\mathbf{p})\oplus\mathbf{p}%
^{\prime})\nonumber\\
&  \qquad=\pm\hspace{0.01in}2^{-1}\operatorname*{vol}\nolimits^{-1}%
\hspace{-0.01in}\delta_{q}^{\hspace{0.01in}3}((\ominus\hspace{0.01in}%
\kappa^{-1}\mathbf{p})\oplus\mathbf{p}^{\prime}).
\end{align}
First, we have inserted the expressions for $(\varphi^{\ast})_{\mathbf{p}}$
and $\varphi_{\mathbf{p}^{\prime}}$. In the second step, we have calculated
the time derivative and applied the orthogonality relation given in
Eq.~(\ref{SkaProEbeDreExpWie0}). The second last step is a consequence of the
identities in Eq.~(\ref{AlgChaIdeqDelFkt}).

As was shown in Ref.~\cite{Wachter:2019A}, the $q$-de\-formed momentum
eigenfunctions are also subject to the following completeness relations:%
\begin{align}
\int\text{d}_{q}^{3}\hspace{0.01in}p\,u_{\hspace{0.01in}\mathbf{p}}%
(\mathbf{x},t)\circledast(u^{\ast})_{\mathbf{p}}(\hspace{0.01in}\mathbf{y},t)
& =\operatorname*{vol}\nolimits^{-1}\hspace{-0.01in}\delta_{q}^{3}%
(\mathbf{x}\oplus(\ominus\hspace{0.01in}\kappa^{-1}\mathbf{y})),\nonumber\\
\int\text{d}_{q}^{3}\hspace{0.01in}p\,(u^{\ast})^{\mathbf{p}}(\hspace
{0.01in}\mathbf{y},t)\circledast u^{\mathbf{p}}(\mathbf{x},t)  &
=\operatorname*{vol}\nolimits^{-1}\hspace{-0.01in}\delta_{q}^{3}%
((\ominus\hspace{0.01in}\kappa^{-1}\mathbf{y})\oplus\mathbf{x}%
).\label{VolRelZeiWelDreDim1}%
\end{align}
We can calculate \textit{completeness relations} for the plane wave
solutions of our $q$-de\-formed Klein-Gor\-don equations as well:%
\begin{align}
&  \int\text{d}_{q}^{3}\hspace{0.01in}p\,\varphi_{\mathbf{p}}(\mathbf{x}%
,t)\circledast(\varphi^{\ast})_{\mathbf{p}}(\mathbf{x}^{\hspace{0.01in}\prime
},t)=\frac{c^{2}}{2}\hspace{-0.01in}\int\text{d}_{q}^{3}\hspace{0.01in}%
p\,u_{\hspace{0.01in}\mathbf{p}}(\mathbf{x})\circledast E_{\mathbf{p}%
}^{\hspace{0.01in}-1}\circledast(u^{\ast})_{\mathbf{p}}(\mathbf{x}%
^{\hspace{0.01in}\prime})\nonumber\\
&  \qquad\qquad=\frac{c}{2}\hspace{-0.01in}\int\text{d}_{q}^{3}p\,u_{\hspace
{0.01in}\mathbf{p}}(\mathbf{x})\circledast(\hspace{0.01in}\mathbf{p}%
^{2}+(m\hspace{0.01in}c)^{2})^{-1/2}\circledast(u^{\ast})_{\mathbf{p}%
}(\mathbf{x}^{\hspace{0.01in}\prime})\nonumber\\
&  \qquad\qquad=\frac{c}{2}(-\nabla_{q}^{2}+(m\hspace{0.01in}c)^{2}%
)^{-1/2}\triangleright\hspace{-0.01in}\int\text{d}_{q}^{3}\hspace
{0.01in}p\,u_{\hspace{0.01in}\mathbf{p}}(\mathbf{x})\circledast(u^{\ast
})_{\mathbf{p}}(\mathbf{x}^{\hspace{0.01in}\prime})\nonumber\\
&  \qquad\qquad=\frac{c}{2\operatorname*{vol}}(-\nabla_{q}^{2}+(m\hspace
{0.01in}c)^{2})^{-1/2}\triangleright\delta_{q}^{3}(\mathbf{x}\oplus
(\ominus\hspace{0.01in}\kappa^{-1}\mathbf{x}^{\hspace{0.01in}\prime
}))\nonumber\\
&  \qquad\qquad=\frac{c}{2\operatorname*{vol}}\,\mathcal{C}_{-1/2}%
(\mathbf{x}\oplus(\ominus\hspace{0.01in}\kappa^{-1}\mathbf{x}^{\hspace
{0.01in}\prime})).
\end{align}
The second step in the above calculation is due to the en\-er\-gy-mo\-men\-tum
relation in Eq.~(\ref{EneMomRelKleGor}), and the third step is a consequence
of the eigenvalue equations in Eq.~(\ref{EigGleImpOpeImpEigFkt0}). In the
second last step, we applied the completeness relations for our $q$-de\-formed
momentum eigenfunctions [cf. Eq.~(\ref{VolRelZeiWelDreDim1})]. Finally, we
introduced the following distribution:%
\begin{align}
\mathcal{C}_{-1/2}(\mathbf{x}\oplus(\ominus\hspace{0.01in}\mathbf{x}%
^{\hspace{0.01in}\prime})) &  =(-\nabla_{q}^{2}+(m\hspace{0.01in}%
c)^{2})^{-1/2}\triangleright\delta_{q}^{3}(\mathbf{x}\oplus(\ominus
\hspace{0.01in}\mathbf{x}^{\hspace{0.01in}\prime}))\nonumber\\
&  =\sum_{k\hspace{0.01in}=\hspace{0.01in}0}^{\infty}\binom{-1/2}%
{k}\,(m\hspace{0.01in}c)^{-1-\hspace{0.01in}2k}\hspace{0.01in}(-\nabla_{q}%
^{2})^{k}\triangleright\delta_{q}^{3}(\mathbf{x}\oplus(\ominus\hspace
{0.01in}\mathbf{x}^{\hspace{0.01in}\prime})).
\end{align}
By similar reasoning, we can show the identity%
\begin{equation}
\int\text{d}_{q}^{3}\hspace{0.01in}p\,(\varphi^{\ast})^{\mathbf{p}}%
(\mathbf{x}^{\hspace{0.01in}\prime},t)\circledast\varphi^{\hspace
{0.01in}\mathbf{p}}(\mathbf{x},t)=\frac{c}{2\operatorname*{vol}}%
\,\mathcal{C}_{-1/2}^{\ast}((\ominus\hspace{0.01in}\kappa^{-1}\mathbf{x}%
^{\hspace{0.01in}\prime})\oplus\mathbf{x}),
\end{equation}
with%
\begin{align}
\mathcal{C}_{-1/2}^{\ast}((\ominus\hspace{0.01in}\mathbf{x}^{\hspace
{0.01in}\prime})\oplus\mathbf{x}) &  =\delta_{q}^{3}((\ominus\hspace
{0.01in}\mathbf{x}^{\hspace{0.01in}\prime})\oplus\mathbf{x})\,\bar
{\triangleleft}\,(-\nabla_{q}^{2}+(m\hspace{0.01in}c)^{2})^{-1/2}\nonumber\\
&  =\sum_{k=0}^{\infty}\binom{-1/2}{k}\,\delta_{q}^{3}((\ominus\hspace
{0.01in}\mathbf{x}^{\hspace{0.01in}\prime})\oplus\mathbf{x})\,\bar
{\triangleleft}\,(-\nabla_{q}^{2})^{k}\hspace{0.01in}(m\hspace{0.01in}%
c)^{-1-\hspace{0.01in}2k}.
\end{align}

Next, we consider the general solutions to the $q$-de\-formed Klein-Gor\-don
equations in Eqs.~(\ref{KleGorGleLin}), (\ref{KleGorGleRec}), and
(\ref{KleGorGleLin2}) of the previous chapter. We can write these general
solutions as expansions in terms of plane wave solutions:%
\begin{align}
\varphi_{R}(\mathbf{x},t) &  =\hspace{-0.01in}\int\text{d}_{q}^{3}%
\hspace{0.01in}p\,\big (\varphi_{\mathbf{p}}(\mathbf{x},t)\circledast
\hspace{-0.01in}f_{\mathbf{p}}^{[+]}+\varphi_{\mathbf{p}}(-\mathbf{x}%
,-t)\circledast\hspace{-0.01in}f_{\mathbf{p}}^{[-]}\big ),\nonumber\\[0.03in]
\varphi_{L}(\mathbf{x},t) &  =\hspace{-0.01in}\int\text{d}_{q}^{3}%
\hspace{0.01in}p\,\big (f_{[+]}^{\mathbf{p}}\circledast\varphi^{\hspace
{0.01in}\mathbf{p}}(\mathbf{x},t)+\hspace{-0.01in}f_{[-]}^{\mathbf{p}%
}\circledast\varphi^{\hspace{0.01in}\mathbf{p}}(-\mathbf{x}%
,-t)\big ).\label{EntWicKGFR}%
\end{align}
Likewise, we have%
\begin{align}
\varphi_{L}^{\ast}(\mathbf{x},t) &  =\sum_{\varepsilon\hspace{0.01in}%
=\hspace{0.01in}\pm}\int\text{d}_{q}^{3}\hspace{0.01in}p\,h_{\hspace
{0.01in}\mathbf{p}}^{[\varepsilon]}\circledast(\varphi^{\ast})_{\mathbf{p}%
}(\varepsilon\hspace{0.01in}\mathbf{x},\varepsilon\hspace{0.01in}%
t),\nonumber\\[0.03in]
\varphi_{R}^{\ast}(\mathbf{x},t) &  =\sum_{\varepsilon\hspace{0.01in}%
=\hspace{0.01in}\pm}\int\text{d}_{q}^{3}\hspace{0.01in}p\,(\varphi^{\ast
})^{\mathbf{p}}(\varepsilon\hspace{0.01in}\mathbf{x},\varepsilon
\hspace{0.01in}t)\circledast h_{[\varepsilon]}^{\mathbf{p}},\label{EntKGFLSte}%
\end{align}
where%
\begin{align}
\varphi_{L}^{\ast}\hspace{-0.01in}\triangleleft\partial_{t}^{\hspace{0.01in}%
2}c^{-2}\hspace{-0.01in}-\varphi_{L}^{\ast}\hspace{-0.01in}\triangleleft
\nabla_{q}^{2}+\varphi_{L}^{\ast}\hspace{0.01in}(m\hspace{0.01in}c)^{2} &
=0,\nonumber\\[0.03in]
c^{-2}\partial_{t}^{\hspace{0.01in}2}\,\bar{\triangleright}\,\varphi_{R}%
^{\ast}-\nabla_{q}^{2}\,\bar{\triangleright}\,\varphi_{R}^{\ast}%
+(m\hspace{0.01in}c)^{2}\hspace{0.01in}\varphi_{R}^{\ast} &
=0.\label{KGGleEntPhiSte2}%
\end{align}
We can calculate the coefficients in the above series expansions by the
formulas 
\begin{align}
\varepsilon h_{\hspace{0.01in}\mathbf{p}}^{[\varepsilon]}= &  \hspace
{0.03in}\text{i\hspace{0.01in}}c^{-2}\hspace{-0.02in}\int\text{d}_{q}%
^{3}x\,\varphi_{L}^{\ast}(\mathbf{x},t)\triangleleft\partial_{t}%
\circledast\varphi_{\mathbf{p}}(\varepsilon\hspace{0.01in}\mathbf{x}%
,\varepsilon\hspace{0.01in}t),\nonumber\\
&  +\text{i\hspace{0.01in}}c^{-2}\hspace{-0.02in}\int\text{d}_{q}%
^{3}x\,\varphi_{L}^{\ast}(\mathbf{x},t)\circledast\partial_{t}\triangleright
\varphi_{\mathbf{p}}(\varepsilon\hspace{0.01in}\mathbf{x},\varepsilon
\hspace{0.01in}t),\\[0.04in]
\varepsilon h_{[\varepsilon]}^{\mathbf{p}}= &  \hspace{0.03in}\text{i\hspace
{0.01in}}c^{-2}\hspace{-0.02in}\int\text{d}_{q}^{3}x\,\varphi^{\hspace
{0.01in}\mathbf{p}}(\varepsilon\hspace{0.01in}\mathbf{x},\varepsilon
\hspace{0.01in}t)\,\bar{\triangleleft}\,\partial_{t}\circledast\varphi
_{R}^{\ast}(\mathbf{x},t)\nonumber\\
&  +\text{i\hspace{0.01in}}c^{-2}\hspace{-0.02in}\int\text{d}_{q}%
^{3}x\,\varphi^{\hspace{0.01in}\mathbf{p}}(\varepsilon\hspace{0.01in}%
\mathbf{x},\varepsilon\hspace{0.01in}t)\circledast\partial_{t}\,\bar
{\triangleright}\,\varphi_{R}^{\ast}(\mathbf{x},t),\label{hplpGKF}%
\end{align}
and%
\begin{align}
\varepsilon f_{\mathbf{p}}^{[\varepsilon]}= &  \hspace{0.03in}\text{i\hspace
{0.01in}}c^{-2}\hspace{-0.02in}\int\text{d}_{q}^{3}x\,(\varphi^{\ast
})_{\mathbf{p}}(\varepsilon\hspace{0.01in}\mathbf{x},\varepsilon
\hspace{0.01in}t)\triangleleft\partial_{t}\circledast\varphi_{R}%
(\mathbf{x},t)\nonumber\\
&  +\text{i}c^{-2}\hspace{-0.02in}\int\text{d}_{q}^{3}x\,(\varphi^{\ast
})_{\mathbf{p}}(\varepsilon\hspace{0.01in}\mathbf{x},\varepsilon
\hspace{0.01in}t)\circledast\partial_{t}\triangleright\varphi_{R}%
(\mathbf{x},t),\\[0.04in]
\varepsilon f_{[\varepsilon]}^{\mathbf{p}}= &  \hspace{0.03in}\text{i\hspace
{0.01in}}c^{-2}\hspace{-0.02in}\int\text{d}_{q}^{3}x\,\varphi_{L}%
(\mathbf{x},t)\,\bar{\triangleleft}\,\partial_{t}\circledast(\varphi^{\ast
})^{\mathbf{p}}(\varepsilon\hspace{0.01in}\mathbf{x},\varepsilon
\hspace{0.01in}t)\nonumber\\
&  +\text{i\hspace{0.01in}}c^{-2}\hspace{-0.02in}\int\text{d}_{q}%
^{3}x\,\varphi_{L}(\mathbf{x},t)\circledast\partial_{t}\,\bar{\triangleright
}\,(\varphi^{\ast})^{\mathbf{p}}(\varepsilon\hspace{0.01in}\mathbf{x}%
,\varepsilon\hspace{0.01in}t).\label{fplpGKF}%
\end{align}
We briefly describe how to check the above formulas for the coefficients.
First, you insert the expansions in terms of plane wave solutions [cf.
Eq.~(\ref{EntWicKGFR}) and Eq.~(\ref{EntKGFLSte})]. This way, you can use the
orthogonality relations given in Eqs.~(\ref{OrtRelKleGorPhiSte}%
)-(\ref{OrtRelKleGorPhiSteEnd}). Taking into account the identities in
Eq.~(\ref{AlgChaIdeqDelFkt})\ will finish the proof.

There is a bilinear form for solutions to our $q$-de\-formed Klein-Gor\-don
equations. Concretely, we have%
\begin{align}
&  \text{i\hspace{0.01in}}c^{-2}\hspace{-0.02in}\int\text{d}_{q}^{3}%
x\,\varphi_{L}^{\ast}(\mathbf{x},t)\triangleleft\partial_{t}\circledast
\varphi_{R}(\mathbf{x},t)\nonumber\\
&  +\text{i\hspace{0.01in}}c^{-2}\hspace{-0.02in}\int\text{d}_{q}%
^{3}x\,\varphi_{L}^{\ast}(\mathbf{x},t)\circledast\partial_{t}\triangleright
\varphi_{R}(\mathbf{x},t)=\nonumber\\
&  \qquad\qquad=\int\text{d}_{q}^{3}p\,\big (h_{\hspace{0.01in}\mathbf{p}%
}^{[+]}\circledast f_{\mathbf{p}}^{[+]}-h_{\hspace{0.01in}\mathbf{p}}%
^{[-]}\circledast f_{\mathbf{p}}^{[-]}\big )\label{SesKGG1}%
\end{align}
and%
\begin{align}
&  \text{i}c^{-2}\hspace{-0.02in}\int\text{d}_{q}^{3}x\,\varphi_{L}%
(\mathbf{x},t)\,\bar{\triangleleft}\,\partial_{t}\circledast\varphi_{R}^{\ast
}(\mathbf{x},t)\nonumber\\
&  +\text{i}c^{-2}\hspace{-0.02in}\int\text{d}_{q}^{3}x\,\varphi
_{L}(\mathbf{x},t)\circledast\partial_{t}\,\bar{\triangleright}\,\varphi
_{R}^{\ast}(\mathbf{x},t)=\nonumber\\
&  \qquad\qquad=\int\text{d}_{q}^{3}p\,\big (f_{[+]}^{\mathbf{p}}\circledast
h_{[+]}^{\mathbf{p}}-f_{[-]}^{\mathbf{p}}\circledast h_{[-]}^{\mathbf{p}%
}\big ).\label{SesKGG2}%
\end{align}
We briefly explain how we get the expressions in momentum space from those in
position space. First, we write each wave function in position space as
expansion in terms of plane wave solutions. This way, we can apply the
orthogonality relations given in Eqs.~(\ref{OrtRelKleGorPhiSte}%
)-(\ref{OrtRelKleGorPhiSteEnd}). By using the identities in
Eq.~(\ref{AlgChaIdeqDelFkt}), we finally get the expression in momentum space.
Note that the bilinear form in Eq.~(\ref{SesKGG1}) or Eq.~(\ref{SesKGG2})
vanishes if we require%
\begin{equation}
h_{\hspace{0.01in}\mathbf{p}}^{[+]}=h_{\hspace{0.01in}\mathbf{p}}^{[-]},\quad
f_{\mathbf{p}}^{[+]}=f_{\mathbf{p}}^{[-]}\qquad\text{or}\qquad h_{[+]}%
^{\mathbf{p}}=h_{[-]}^{\mathbf{p}},\quad f_{[+]}^{\mathbf{p}}=f_{[-]}%
^{\mathbf{p}}.\label{BedNeuTei}%
\end{equation}

\section{Free propagators\label{KapProKleGorFel}}

The series expansions in Eqs.~(\ref{EntWicKGFR}) and (\ref{EntKGFLSte}) of the
previous chapter show us that the solutions to our $q$-de\-formed
Klein-Gor\-don equations consist of two parts with different signs of energy,
i.~e.%
\begin{align}
\varphi_{R} &  =(\varphi_{R})^{[+]}+(\varphi_{R})^{[-]},\nonumber\\
\varphi_{L} &  =(\varphi_{L})^{[+]}+(\varphi_{L})^{[-]},
\end{align}
and%
\begin{align}
\varphi_{R}^{\ast} &  =(\varphi_{R}^{\ast})^{[+]}+(\varphi_{R}^{\ast}%
)^{[-]},\nonumber\\
\varphi_{L}^{\ast} &  =(\varphi_{L}^{\ast})^{[+]}+(\varphi_{L}^{\ast})^{[-]}.
\end{align}
The propagators for the Klein-Gor\-don equations should take a form so that
the positive energy solution runs forward in time while the negative energy
solution runs backward in time. Hence, we seek propagators $\Delta_{R}$ and
$\Delta_{L}$ that satisfy the identities\footnote{$\theta(t)$ stands for the
Heaviside step function.}%
\begin{align}
&  \text{i\hspace{0.01in}}c^{-2}\hspace{-0.02in}\int\text{d}_{q}^{3}%
x\,\Delta_{R}(\mathbf{x}^{\hspace{0.01in}\prime}\hspace{-0.01in}%
,t^{\hspace{0.01in}\prime}\hspace{-0.01in};\mathbf{x},t)\triangleleft
\partial_{t}\circledast\varphi_{R}(\mathbf{x},t)\nonumber\\
&  +\text{i\hspace{0.01in}}c^{-2}\hspace{-0.02in}\int\text{d}_{q}^{3}%
x\,\Delta_{R}(\mathbf{x}^{\hspace{0.01in}\prime}\hspace{-0.01in}%
,t^{\hspace{0.01in}\prime}\hspace{-0.01in};\mathbf{x},t)\circledast
\partial_{t}\triangleright\varphi_{R}(\mathbf{x},t)=\nonumber\\
&  \qquad\qquad=\theta(t^{\hspace{0.01in}\prime}\hspace{-0.02in}%
-t)\,(\varphi_{R})^{[+]}(\mathbf{x}^{\hspace{0.01in}\prime}\hspace
{-0.01in},t^{\hspace{0.01in}\prime})-\theta(t-t^{\hspace{0.01in}\prime
})\,(\varphi_{R})^{[-]}(\mathbf{x}^{\hspace{0.01in}\prime}\hspace
{-0.01in},t^{\hspace{0.01in}\prime})\label{ChaIdeKGPro1}%
\end{align}
or%
\begin{align}
&  \text{i\hspace{0.01in}}c^{-2}\hspace{-0.02in}\int\text{d}_{q}^{3}%
x\,\varphi_{L}(\mathbf{x},t)\,\bar{\triangleleft}\,\partial_{t}\circledast
\Delta_{L}(\mathbf{x},t;\mathbf{x}^{\hspace{0.01in}\prime}\hspace
{-0.01in},t^{\hspace{0.01in}\prime})\nonumber\\
&  +\text{i\hspace{0.01in}}c^{-2}\hspace{-0.02in}\int\text{d}_{q}%
^{3}x\,\varphi_{L}(\mathbf{x},t)\circledast\partial_{t}\,\bar{\triangleright
}\,\Delta_{L}(\mathbf{x},t;\mathbf{x}^{\hspace{0.01in}\prime}\hspace
{-0.01in},t^{\hspace{0.01in}\prime})=\nonumber\\
&  \qquad\qquad=\theta(t^{\hspace{0.01in}\prime}\hspace{-0.02in}%
-t)\,(\varphi_{L})^{[+]}(\mathbf{x}^{\hspace{0.01in}\prime}\hspace
{-0.01in},t^{\hspace{0.01in}\prime})-\theta(t-t^{\hspace{0.01in}\prime
})\,(\varphi_{L})^{[-]}(\mathbf{x}^{\hspace{0.01in}\prime}\hspace
{-0.01in},t^{\hspace{0.01in}\prime}).\label{ChaIdeKGPro2}%
\end{align}
Propagators with these properties are given by%
\begin{align}
\Delta_{R}(\mathbf{x}^{\hspace{0.01in}\prime}\hspace{-0.01in},t^{\hspace
{0.01in}\prime}\hspace{-0.01in};\mathbf{x},t) &  =\theta(t^{\hspace
{0.01in}\prime}\hspace{-0.02in}-t)\int\text{d}_{q}^{3}\hspace{0.01in}%
p\,\varphi_{\mathbf{p}}(\mathbf{x}^{\hspace{0.01in}\prime}\hspace
{-0.01in},t^{\hspace{0.01in}\prime})\circledast(\varphi^{\ast})_{\mathbf{p}%
}(\mathbf{x},t)\nonumber\\
&  \hspace{0.15in}+\theta(t-t^{\hspace{0.01in}\prime})\int\text{d}_{q}%
^{3}\hspace{0.01in}p\,\varphi_{\mathbf{p}}(\mathbf{x}^{\hspace{0.01in}\prime
}\hspace{-0.01in},-\hspace{0.01in}t^{\hspace{0.01in}\prime})\circledast
(\varphi^{\ast})_{\mathbf{p}}(\mathbf{x},-\hspace{0.01in}%
t)\label{EntFreProKleGorFelDelL}%
\end{align}
and%
\begin{align}
\Delta_{L}(\mathbf{x},t;\mathbf{x}^{\hspace{0.01in}\prime}\hspace
{-0.01in},t^{\hspace{0.01in}\prime}) &  =\theta(t^{\hspace{0.01in}\prime
}\hspace{-0.02in}-t)\int\text{d}_{q}^{3}\hspace{0.01in}p\,(\varphi^{\ast
})^{\mathbf{p}}(\mathbf{x},t)\circledast\varphi^{\hspace{0.01in}\mathbf{p}%
}(\mathbf{x}^{\hspace{0.01in}\prime}\hspace{-0.01in},t^{\hspace{0.01in}\prime
})\nonumber\\
&  \hspace{0.15in}+\theta(t-t^{\hspace{0.01in}\prime})\int\text{d}_{q}%
^{3}\hspace{0.01in}p\,(\varphi^{\ast})^{\mathbf{p}}(\mathbf{x},-\hspace
{0.01in}t)\circledast\varphi^{\hspace{0.01in}\mathbf{p}}(\mathbf{x}%
^{\hspace{0.01in}\prime}\hspace{-0.01in},-\hspace{0.01in}t^{\hspace
{0.01in}\prime}).\label{EntFreProKleGorFelDelR}%
\end{align}
To show that the expression in Eq.~(\ref{EntFreProKleGorFelDelL}) or
Eq.~(\ref{EntFreProKleGorFelDelR}) fulfills the identity in
Eq.~(\ref{ChaIdeKGPro1}) or Eq.~(\ref{ChaIdeKGPro2}), we proceed as follows.
First, we replace the propagator in Eq.~(\ref{ChaIdeKGPro1}) or
Eq.~(\ref{ChaIdeKGPro2}) by its expression in
Eq.~(\ref{EntFreProKleGorFelDelL}) or Eq.~(\ref{EntFreProKleGorFelDelR}). In
the same way, we write the wave functions $\varphi_{R}$ and $\varphi_{L}$ as
expansions in terms of plane wave solutions [cf. Eq.~(\ref{EntWicKGFR}) of the
previous chapter]. Then, we can apply the orthogonality relations given in
Eqs.~(\ref{OrtRelKleGorPhiSte})-(\ref{OrtRelKleGorPhiSteEnd}) of
Chap.~\ref{KapPlaWavSol}. Using the identities in Eq.~(\ref{AlgChaIdeqDelFkt})
of Chap.~\ref{KapPlaWavSol}, we regain the plane wave expansions. We have the
additional factor $\theta(t^{\hspace{0.01in}\prime}\hspace{-0.02in}-t)$ for
the plane waves with positive energy. For the plane waves with negative
energy, we have the factor $-\theta(t-t^{\hspace{0.01in}\prime})$.

We can also introduce so-called `dual' propagators%
\begin{align}
\Delta_{R}^{\ast}(\mathbf{x}^{\hspace{0.01in}\prime}\hspace{-0.01in}%
,t^{\hspace{0.01in}\prime}\hspace{-0.01in};\mathbf{x},t)= &  \hspace
{0.03in}\theta(t^{\hspace{0.01in}\prime}\hspace{-0.02in}-t)\int\text{d}%
_{q}^{3}\hspace{0.01in}p\,(\varphi^{\ast})^{\mathbf{p}}(\mathbf{x}%
^{\hspace{0.01in}\prime}\hspace{-0.01in},t^{\hspace{0.01in}\prime}%
)\circledast\varphi^{\hspace{0.01in}\mathbf{p}}(\mathbf{x},t)\nonumber\\
&  +\theta(t-t^{\hspace{0.01in}\prime})\int\text{d}_{q}^{3}\hspace
{0.01in}p\,(\varphi^{\ast})^{\mathbf{p}}(\mathbf{x}^{\hspace{0.01in}\prime
}\hspace{-0.01in},-\hspace{0.01in}t^{\hspace{0.01in}\prime})\circledast
\varphi^{\hspace{0.01in}\mathbf{p}}(\mathbf{x},-\hspace{0.01in}t)
\end{align}
and%
\begin{align}
\Delta_{L}^{\ast}(\mathbf{x},t;\mathbf{x}^{\hspace{0.01in}\prime}%
\hspace{-0.01in},t^{\hspace{0.01in}\prime})= &  \hspace{0.03in}\theta
(t^{\hspace{0.01in}\prime}\hspace{-0.02in}-t)\int\text{d}_{q}^{3}%
\hspace{0.01in}p\,\varphi_{\mathbf{p}}(\mathbf{x},t)\circledast(\varphi^{\ast
})_{\mathbf{p}}(\mathbf{x}^{\hspace{0.01in}\prime}\hspace{-0.01in}%
,t^{\hspace{0.01in}\prime})\nonumber\\
&  +\theta(t-t^{\hspace{0.01in}\prime})\int\text{d}_{q}^{3}\hspace
{0.01in}p\,\varphi_{\mathbf{p}}(\mathbf{x},-\hspace{0.01in}t)\circledast
(\varphi^{\ast})_{\mathbf{p}}(\mathbf{x}^{\hspace{0.01in}\prime}%
\hspace{-0.01in},-\hspace{0.01in}t^{\hspace{0.01in}\prime}%
).\label{EntFreProKleGorFelDelRSte}%
\end{align}
These new propagators are subject to%
\begin{align}
&  \text{i\hspace{0.01in}}c^{-2}\hspace{-0.02in}\int\text{d}_{q}^{3}%
x\,\varphi_{L}^{\ast}(\mathbf{x},t)\triangleleft\partial_{t}\circledast
\Delta_{L}^{\ast}(\mathbf{x},t;\mathbf{x}^{\hspace{0.01in}\prime}%
\hspace{-0.01in},t^{\hspace{0.01in}\prime})\nonumber\\
&  +\text{i\hspace{0.01in}}c^{-2}\hspace{-0.02in}\int\text{d}_{q}%
^{3}x\,\varphi_{L}^{\ast}(\mathbf{x},t)\circledast\partial_{t}\triangleright
\Delta_{L}^{\ast}(\mathbf{x},t;\mathbf{x}^{\hspace{0.01in}\prime}%
\hspace{-0.01in},t^{\hspace{0.01in}\prime})=\nonumber\\
&  \qquad\qquad=\theta(t^{\hspace{0.01in}\prime}\hspace{-0.02in}%
-t)\,(\varphi_{L}^{\ast})^{[+]}(\mathbf{x}^{\hspace{0.01in}\prime}%
\hspace{-0.01in},t^{\hspace{0.01in}\prime})-\theta(t-t^{\hspace{0.01in}\prime
})\,(\varphi_{L}^{\ast})^{[-]}(\mathbf{x}^{\hspace{0.01in}\prime}%
\hspace{-0.01in},t^{\hspace{0.01in}\prime})\label{ChaGleKleGorProSteR}%
\end{align}
or%
\begin{align}
&  \text{i\hspace{0.01in}}c^{-2}\hspace{-0.02in}\int\text{d}_{q}^{3}%
x\,\Delta_{R}^{\ast}(\mathbf{x}^{\hspace{0.01in}\prime}\hspace{-0.01in}%
,t^{\hspace{0.01in}\prime}\hspace{-0.01in};\mathbf{x},t)\,\bar{\triangleleft
}\,\partial_{t}\circledast\varphi_{R}^{\ast}(\mathbf{x},t)\nonumber\\
&  +\text{i\hspace{0.01in}}c^{-2}\hspace{-0.02in}\int\text{d}_{q}^{3}%
x\,\Delta_{R}^{\ast}(\mathbf{x}^{\hspace{0.01in}\prime}\hspace{-0.01in}%
,t^{\hspace{0.01in}\prime}\hspace{-0.01in};\mathbf{x},t)\circledast
\partial_{t}\,\bar{\triangleright}\,\varphi_{R}^{\ast}(\mathbf{x}%
,t)=\nonumber\\
&  \qquad\qquad=\theta(t^{\hspace{0.01in}\prime}\hspace{-0.02in}%
-t)\,(\varphi_{R}^{\ast})^{[+]}(\mathbf{x}^{\hspace{0.01in}\prime}%
\hspace{-0.01in},t^{\hspace{0.01in}\prime})-\theta(t-t^{\hspace{0.01in}\prime
})\,(\varphi_{R}^{\ast})^{[-]}(\mathbf{x}^{\hspace{0.01in}\prime}%
\hspace{-0.01in},t^{\hspace{0.01in}\prime}).
\end{align}

Our $q$-de\-formed propagators transform into each other by conjugation:%
\begin{align}
\overline{\Delta_{R}(\mathbf{x}^{\hspace{0.01in}\prime}\hspace{-0.01in}%
,t^{\hspace{0.01in}\prime}\hspace{-0.01in};\mathbf{x},t)} &  =\Delta
_{L}(\mathbf{x},t;\mathbf{x}^{\hspace{0.01in}\prime}\hspace{-0.01in}%
,t^{\hspace{0.01in}\prime}),\nonumber\\
\overline{\Delta_{L}^{\ast}(\mathbf{x}^{\hspace{0.01in}\prime}\hspace
{-0.01in},t^{\hspace{0.01in}\prime}\hspace{-0.01in};\mathbf{x},t)} &
=\Delta_{R}^{\ast}(\mathbf{x},t;\mathbf{x}^{\hspace{0.01in}\prime}%
\hspace{-0.01in},t^{\hspace{0.01in}\prime}).
\end{align}
These identities follow from Eqs.~(\ref{EntFreProKleGorFelDelL}%
)-(\ref{EntFreProKleGorFelDelRSte}) if we consider the conjugation properties
of $q$-de\-formed volume integrals and $q$-de\-formed plane wave solutions
[also see Eq.~(\ref{KonEigSteProFkt}) in App.~\ref{KapQuaZeiEle}%
,\ Eq.~(\ref{KonEigVolInt}) in App.~\ref{KapParDer}, and
Eq.~(\ref{ConPropPlaWav}) in Chap.~\ref{KapPlaWavSol}].

Our $q$-de\-formed propagators are also subject to the equations%
\begin{align}
&  ((m\hspace{0.01in}c)^{2}+c^{-2}\partial_{t}^{\hspace{0.01in}2}-\nabla
_{q}^{2})\triangleright\Delta_{R}(\mathbf{x},t;\mathbf{x}^{\hspace
{0.01in}\prime}\hspace{-0.01in},t^{\hspace{0.01in}\prime})=\nonumber\\
&  \qquad\qquad=-\text{\hspace{0.01in}i\hspace{0.01in}}\operatorname*{vol}%
\nolimits^{-1}\hspace{-0.01in}\delta(t-t^{\hspace{0.01in}\prime})\,\delta
_{q}^{3}(\mathbf{x}\oplus(\ominus\hspace{0.01in}\kappa^{-1}\mathbf{x}%
^{\hspace{0.01in}\prime})),\label{DifGleKGFLR0}\\[0.07in]
&  \Delta_{L}(\mathbf{x}^{\hspace{0.01in}\prime}\hspace{-0.01in}%
,t^{\hspace{0.01in}\prime}\hspace{-0.01in};\mathbf{x},t)\,\bar{\triangleleft
}\,(\partial_{t}^{\hspace{0.01in}2}c^{-2}\hspace{-0.01in}-\nabla_{q}%
^{2}+(m\hspace{0.01in}c)^{2})=\nonumber\\
&  \qquad\qquad=\text{i\hspace{0.01in}}\operatorname*{vol}\nolimits^{-1}%
\hspace{-0.01in}\delta_{q}^{3}((\ominus\hspace{0.01in}\kappa^{-1}%
\mathbf{x}^{\hspace{0.01in}\prime})\oplus\mathbf{x})\,\delta(t^{\hspace
{0.01in}\prime}-t),\label{DifGleKGFLR}%
\end{align}
or%
\begin{align}
&  \Delta_{L}^{\ast}(\mathbf{x}^{\hspace{0.01in}\prime}\hspace{-0.01in}%
,t^{\hspace{0.01in}\prime}\hspace{-0.01in};\mathbf{x},t)\,\bar{\triangleleft
}\,(\partial_{t}^{\hspace{0.01in}2}c^{-2}\hspace{-0.01in}-\nabla_{q}%
^{2}+(m\hspace{0.01in}c)^{2})=\nonumber\\
&  \qquad\qquad=\text{i\hspace{0.01in}}\operatorname*{vol}\nolimits^{-1}%
\hspace{-0.01in}\delta_{q}^{3}((\ominus\hspace{0.01in}\kappa^{-1}%
\mathbf{x}^{\hspace{0.01in}\prime})\oplus\mathbf{x})\,\delta(t^{\hspace
{0.01in}\prime}-t),\\[0.07in]
&  ((m\hspace{0.01in}c)^{2}+c^{-2}\partial_{t}^{\hspace{0.01in}2}-\nabla
_{q}^{2})\triangleright\Delta_{R}^{\ast}(\mathbf{x},t;\mathbf{x}%
^{\hspace{0.01in}\prime}\hspace{-0.01in},t^{\hspace{0.01in}\prime
})=\nonumber\\
&  \qquad\qquad=-\text{\hspace{0.01in}i\hspace{0.01in}}\operatorname*{vol}%
\nolimits^{-1}\hspace{-0.01in}\delta(t-t^{\hspace{0.01in}\prime})\,\delta
_{q}^{3}(\mathbf{x}\oplus(\ominus\hspace{0.01in}\kappa^{-1}\mathbf{x}%
^{\hspace{0.01in}\prime})).
\end{align}
We show how the above identities are derived using Eq.~(\ref{DifGleKGFLR0}) as
an example. To this end, we write the propagator $\Delta_{R}$ as%
\begin{equation}
\Delta_{R}(\mathbf{x},t;\mathbf{x}^{\hspace{0.01in}\prime}\hspace
{-0.01in},t^{\hspace{0.01in}\prime})=\theta(t-t^{\hspace{0.01in}\prime
})\hspace{0.01in}D_{R}(\mathbf{x},t;\mathbf{x}^{\hspace{0.01in}\prime}%
\hspace{-0.01in},t^{\hspace{0.01in}\prime})+\theta(t^{\hspace{0.01in}\prime
}\hspace{-0.02in}-t)\hspace{0.01in}D_{R}(\mathbf{x},-t;\mathbf{x}%
^{\hspace{0.01in}\prime}\hspace{-0.01in},-t^{\hspace{0.01in}\prime
})\label{KGProAnt}%
\end{equation}
with%
\begin{equation}
D_{R}(\mathbf{x},t;\mathbf{x}^{\hspace{0.01in}\prime}\hspace{-0.01in}%
,t^{\hspace{0.01in}\prime})=\int\text{d}_{q}^{3}\hspace{0.01in}p\,\varphi
_{\mathbf{p}}(\mathbf{x},t)\circledast(\varphi^{\ast})_{\mathbf{p}}%
(\mathbf{x}^{\hspace{0.01in}\prime}\hspace{-0.01in},t^{\hspace{0.01in}\prime
}).\label{AusKLDE}%
\end{equation}
Using the identity%
\begin{equation}
\partial_{t}\triangleright\theta(t-t^{\hspace{0.01in}\prime})=\delta
(t-t^{\hspace{0.01in}\prime})=\delta(t^{\hspace{0.01in}\prime}\hspace
{-0.02in}-t),\label{AblDelFktZei}%
\end{equation}
we can do the following calculation:%
\begin{align}
&  \partial_{t}^{\hspace{0.01in}2}\triangleright\Delta_{R}(\mathbf{x}%
,t;\mathbf{x}^{\prime\hspace{0.01in}}\hspace{-0.01in},t^{\hspace{0.01in}%
\prime})=\nonumber\\
&  \qquad=[\hspace{0.01in}\partial_{t}\triangleright\delta(t-t^{\hspace
{0.01in}\prime})]\hspace{0.01in}[D_{R}(\mathbf{x},t;\mathbf{x}^{\hspace
{0.01in}\prime}\hspace{-0.01in},t^{\hspace{0.01in}\prime})-D_{R}%
(\mathbf{x},-\hspace{0.01in}t;\mathbf{x}^{\hspace{0.01in}\prime}%
\hspace{-0.01in},-\hspace{0.01in}t^{\hspace{0.01in}\prime})]\nonumber\\
&  \qquad\hspace{0.15in}+2\hspace{0.01in}\delta(t-t^{\hspace{0.01in}\prime
})\,\partial_{t}\triangleright\hspace{-0.01in}[D_{R}(\mathbf{x},t;\mathbf{x}%
^{\hspace{0.01in}\prime}\hspace{-0.01in},t^{\hspace{0.01in}\prime}%
)-D_{R}(\mathbf{x},-\hspace{0.01in}t;\mathbf{x}^{\hspace{0.01in}\prime}%
\hspace{-0.01in},-\hspace{0.01in}t^{\hspace{0.01in}\prime})]\nonumber\\
&  \qquad\hspace{0.15in}+\theta(t-t^{\hspace{0.01in}\prime})\,\partial
_{t}^{\hspace{0.01in}2}\triangleright\hspace{-0.01in}D_{R}(\mathbf{x}%
,t;\mathbf{x}^{\hspace{0.01in}\prime}\hspace{-0.01in},t^{\hspace{0.01in}%
\prime})\nonumber\\
&  \qquad\hspace{0.15in}+\theta(t^{\hspace{0.01in}\prime}\hspace
{-0.01in}-t)\,\partial_{t}^{\hspace{0.01in}2}\triangleright\hspace
{-0.01in}D_{R}(\mathbf{x},-\hspace{0.01in}t;\mathbf{x}^{\hspace{0.01in}\prime
}\hspace{-0.01in},-\hspace{0.01in}t^{\hspace{0.01in}\prime}%
).\label{ZwiRecPhoProZei}%
\end{align}
With the identities%
\begin{equation}
\lbrack\partial_{t}\hspace{0.01in}\triangleright\delta(t)]\hspace
{0.01in}f(t)=f(t)\hspace{0.01in}[-\hspace{0.01in}\partial_{t}\hspace
{0.01in}\triangleright\hspace{-0.01in}f(t)]
\end{equation}
and [cf. Eq.~(\ref{qKleGorGleImp1}) of the previous chapter and
Eq.~(\ref{AusKLDE})]%
\begin{equation}
c^{-2}\partial_{t}^{\hspace{0.01in}2}\triangleright\hspace{-0.01in}%
D_{R}(\mathbf{x},t;\mathbf{x}^{\hspace{0.01in}\prime}\hspace{-0.01in}%
,t^{\hspace{0.01in}\prime})=(\Delta_{q}-(m\hspace{0.01in}c)^{2})\triangleright
\hspace{-0.01in}D_{R}(\mathbf{x},t;\mathbf{x}^{\hspace{0.01in}\prime}%
\hspace{-0.01in},t^{\hspace{0.01in}\prime}),\label{MaxGleKL}%
\end{equation}
we can write the result of Eq.~(\ref{ZwiRecPhoProZei}) as follows:%
\begin{align}
&  c^{-2}\partial_{t}^{\hspace{0.01in}2}\triangleright\hspace{-0.01in}%
\Delta_{R}(\mathbf{x},t;\mathbf{x}^{\hspace{0.01in}\prime}\hspace
{-0.01in},t^{\hspace{0.01in}\prime})=\nonumber\\
&  \qquad=c^{-2}\delta(t-t^{\hspace{0.01in}\prime})\,\partial_{t}%
\triangleright\hspace{-0.01in}[D_{R}(\mathbf{x},t;\mathbf{x}^{\hspace
{0.01in}\prime}\hspace{-0.01in},t^{\hspace{0.01in}\prime})-D_{R}%
(\mathbf{x},-\hspace{0.01in}t;\mathbf{x}^{\hspace{0.01in}\prime}%
\hspace{-0.01in},-\hspace{0.01in}t^{\hspace{0.01in}\prime})]\nonumber\\
&  \qquad\hspace{0.15in}+\theta(t-t^{\hspace{0.01in}\prime})\,(\Delta
_{q}-(m\hspace{0.01in}c)^{2})\triangleright\hspace*{-0.01in}D_{R}%
(\mathbf{x},t;\mathbf{x}^{\hspace{0.01in}\prime}\hspace{-0.01in}%
,t^{\hspace{0.01in}\prime})\nonumber\\
&  \qquad\hspace{0.15in}+\theta(t^{\hspace{0.01in}\prime}\hspace
{-0.01in}-t)\,(\Delta_{q}-(m\hspace{0.01in}c)^{2})\triangleright
\hspace*{-0.01in}D_{R}(\mathbf{x},-\hspace{0.01in}t;\mathbf{x}^{\hspace
{0.01in}\prime}\hspace{-0.01in},-\hspace{0.01in}t^{\hspace{0.01in}\prime
}).\label{ZwiRecChaGleProL}%
\end{align}
If we consider Eq.~(\ref{AusKLDE}), we can also rewrite the expressions
depending on the time derivative:%
\begin{align}
&  c^{-2}\delta(t-t^{\hspace{0.01in}\prime})\,\partial_{t}\triangleright
\hspace*{-0.01in}D_{R}(\mathbf{x},\pm\hspace{0.01in}t;\mathbf{x}%
^{\hspace{0.01in}\prime}\hspace{-0.01in},\pm\hspace{0.01in}t^{\hspace
{0.01in}\prime})=\nonumber\\
&  \qquad=c^{-2}\delta(t-t^{\hspace{0.01in}\prime})\int\text{d}_{q}^{3}%
\hspace{0.01in}p\,\varphi_{\mathbf{p}}(\mathbf{x},\pm\hspace{0.01in}%
t)\circledast(\mp\hspace{0.01in}\text{i}E_{\mathbf{p}})\circledast
(\varphi^{\ast})_{\mathbf{p}}(\mathbf{x}^{\hspace{0.01in}\prime}%
\hspace{-0.01in},\pm\hspace{0.01in}t^{\hspace{0.01in}\prime})\nonumber\\
&  \qquad=\mp\frac{\text{i}}{2}\delta(t-t^{\hspace{0.01in}\prime})\int
\text{d}_{q}^{3}\hspace{0.01in}p\,u_{\hspace*{0.01in}\mathbf{p}}%
(\mathbf{x})\circledast(u^{\ast})_{\mathbf{p}}(\mathbf{x}^{\prime})\nonumber\\
&  \qquad=\mp\frac{\text{i}}{2\operatorname*{vol}}\hspace{0.01in}%
\delta(t-t^{\hspace{0.01in}\prime})\hspace{0.01in}\delta_{q}^{3}%
(\mathbf{x}\oplus(\ominus\hspace{0.01in}\kappa^{-1}\mathbf{x}^{\prime
})).\label{UmfDelZeiAblKL}%
\end{align}
We briefly describe the above calculation. The action of the time derivative
on the wave function $\varphi_{\hspace*{0.01in}\mathbf{p}}$ gives the factor
$\mp\hspace{0.01in}$i$E_{\mathbf{p}}$. Moreover, it holds $t=t^{\hspace
{0.01in}\prime}$ due to the delta function $\delta(t-t^{\hspace{0.01in}\prime
})$. For this reason, the time-de\-pen\-dent phase factors in the wave
functions $\varphi_{\mathbf{p}}$ and$\ (\varphi^{\ast})_{\mathbf{p}}$ cancel
each other out. In the last step, we applied the completeness relations for
our $q$-de\-formed momentum eigenfunctions [cf. Eq.~(\ref{VolRelZeiWelDreDim1}%
) of the previous chapter]. Plugging the result of Eq.~(\ref{UmfDelZeiAblKL})
into Eq.~(\ref{ZwiRecChaGleProL}), we finally get:%
\begin{align}
&  c^{-2}\partial_{t}^{\hspace*{0.01in}2}\triangleright\hspace*{-0.01in}%
\Delta_{R}(\mathbf{x},t;\mathbf{x}^{\hspace{0.01in}\prime}\hspace
{-0.01in},t^{\hspace{0.01in}\prime})=\nonumber\\
&  \qquad=-\hspace{0.01in}\text{i}\operatorname*{vol}\nolimits^{-1}%
\hspace*{-0.01in}\delta(t-t^{\hspace{0.01in}\prime})\hspace{0.01in}\delta
_{q}^{3}(\mathbf{x}\oplus(\ominus\hspace{0.01in}\kappa^{-1}\mathbf{x}^{\prime
}))\nonumber\\
&  \qquad\hspace{0.15in}+\Delta_{q}\triangleright\Delta_{R}(\mathbf{x}%
,t;\mathbf{x}^{\hspace{0.01in}\prime}\hspace{-0.01in},t^{\hspace{0.01in}%
\prime})-(m\hspace{0.01in}c)^{2}\hspace{0.01in}\Delta_{R}(\mathbf{x}%
,t;\mathbf{x}^{\hspace{0.01in}\prime}\hspace{-0.01in},t^{\hspace{0.01in}%
\prime}).
\end{align}

The identities in Eqs.~(\ref{DifGleKGFLR0}) and (\ref{DifGleKGFLR}) show that
the functions%
\begin{align}
\phi_{R}(\mathbf{x},t) &  =\varphi_{R}(\mathbf{x},t)+\text{i}\int\text{d}%
^{3}x^{\prime}\text{d}t^{\hspace{0.01in}\prime}\,\Delta_{R}(\mathbf{x}%
,t;\mathbf{x}^{\hspace{0.01in}\prime}\hspace{-0.01in},t^{\hspace{0.01in}%
\prime})\circledast\mathcal{\varrho}(\mathbf{x}^{\hspace{0.01in}\prime}%
\hspace{-0.01in},t^{\hspace{0.01in}\prime}),\nonumber\\[0.12in]
\phi_{L}(\mathbf{x},t) &  =\varphi_{L}(\mathbf{x},t)-\text{i}\int\text{d}%
^{3}x^{\prime}\text{d}t^{\hspace{0.01in}\prime}\,\mathcal{\varrho}%
(\mathbf{x}^{\hspace{0.01in}\prime}\hspace{-0.01in},t^{\hspace{0.01in}\prime
})\circledast\Delta_{L}(\mathbf{x}^{\hspace{0.01in}\prime}\hspace
{-0.01in},t^{\hspace{0.01in}\prime}\hspace{-0.01in};\mathbf{x}%
,t)\label{LoeInhKleGorGle}%
\end{align}
are solutions to the following $q$-de\-formed inhomogenous Klein-Gor\-don
equations:%
\begin{align}
c^{-2}\partial_{t}^{\hspace{0.01in}2}\triangleright\phi_{R}-\nabla_{q}%
^{2}\triangleright\phi_{R}+(m\hspace{0.01in}c)^{2}\hspace{0.01in}\phi_{R} &
=\mathcal{\varrho}.\nonumber\\[0.05in]
\phi_{L}\,\bar{\triangleleft}\,\hspace{0.01in}\partial_{t}^{\hspace{0.01in}%
2}c^{-2}\hspace{-0.01in}-\phi_{L}\,\bar{\triangleleft}\,\hspace{0.01in}%
\nabla_{q}^{2}+\phi_{L}\hspace{0.01in}(m\hspace{0.01in}c)^{2} &
=\mathcal{\varrho}.
\end{align}
Note that the functions $\varphi_{R}$ and $\varphi_{L}$ in
Eq.~(\ref{LoeInhKleGorGle}) are solutions to free $q$-de\-formed
Klein-Gor\-don equations [cf. Eqs.~(\ref{KleGorGleLin}) and
(\ref{KleGorGleRec}) in Chap.~\ref{LoeKleGorGleKap}]. Similarly, the functions%
\begin{align}
\phi_{R}^{\ast}(\mathbf{x},t) &  =\varphi_{R}^{\ast}(\mathbf{x},t)+\text{i}%
\int\text{d}^{3}x^{\prime}\text{d}t^{\hspace{0.01in}\prime}\,\Delta_{R}^{\ast
}(\mathbf{x},t;\mathbf{x}^{\hspace{0.01in}\prime}\hspace{-0.01in}%
,t^{\hspace{0.01in}\prime})\circledast\mathcal{\varrho}(\mathbf{x}%
^{\hspace{0.01in}\prime}\hspace{-0.01in},t^{\hspace{0.01in}\prime
}),\nonumber\\[0.12in]
\phi_{L}^{\ast}(\mathbf{x},t) &  =\varphi_{L}^{\ast}(\mathbf{x},t)-\text{i}%
\int\text{d}^{3}x^{\prime}\text{d}t^{\hspace{0.01in}\prime}\,\mathcal{\varrho
}(\mathbf{x}^{\hspace{0.01in}\prime}\hspace{-0.01in},t^{\hspace{0.01in}\prime
})\circledast\Delta_{L}^{\ast}(\mathbf{x}^{\hspace{0.01in}\prime}%
\hspace{-0.01in},t^{\hspace{0.01in}\prime}\hspace{-0.01in};\mathbf{x}%
,t)\label{LoeInhKleGorGleLRSte}%
\end{align}
satisfy the following $q$-de\-formed versions of the inhomogenous
Klein-Gor\-don equation:%
\begin{align}
\phi_{L}^{\ast}\hspace{-0.01in}\triangleleft\partial_{t}^{\hspace{0.01in}%
2}c^{-2}\hspace{-0.01in}-\phi_{L}^{\ast}\hspace{-0.01in}\triangleleft
\Delta_{q}+\phi_{L}^{\ast}\hspace{0.01in}(m\hspace{0.01in}c)^{2} &
=\mathcal{\varrho},\nonumber\\[0.05in]
c^{-2}\partial_{t}^{\hspace{0.01in}2}\,\hspace{0.01in}\bar{\triangleright
}\,\phi_{R}^{\ast}-\Delta_{q}\,\bar{\triangleright}\,\phi_{R}^{\ast}%
+(m\hspace{0.01in}c)^{2}\hspace{0.01in}\phi_{R}^{\ast} &  =\mathcal{\varrho}.
\end{align}
Once again, $\varphi_{R}^{\ast}$ and $\varphi_{L}^{\ast}$ are solutions to
free $q$-de\-formed Klein-Gor\-don equations [cf. Eq.~(\ref{KleGorGleLin2}) in
Chap.~\ref{LoeKleGorGleKap}].

We can write down a momentum space form of each $q$-de\-formed Klein-Gordon
propagator. Concretely, we have%
\begin{align}
&  \Delta_{R}(\mathbf{x}^{\hspace{0.01in}\prime}\hspace{-0.01in}%
,t^{\hspace{0.01in}\prime}\hspace{-0.01in};\mathbf{x},t)=\int\text{d}%
E\operatorname*{e}\nolimits^{-\text{i}(t^{\prime}-\hspace{0.01in}t)E}%
\hspace{-0.01in}\int\text{d}_{q}^{3}\hspace{0.01in}p\,u_{\hspace
{0.01in}\mathbf{p}}(\mathbf{x}^{\hspace{0.01in}\prime})\circledast\Delta
_{R}(E,\mathbf{p})\circledast(u^{\ast})_{\mathbf{p}}(\mathbf{x}%
),\nonumber\\[0.16in]
&  \Delta_{L}(\mathbf{x},t;\mathbf{x}^{\hspace{0.01in}\prime}\hspace
{-0.01in},t^{\hspace{0.01in}\prime})=\int\text{d}E\operatorname*{e}%
\nolimits^{-\text{i}(t^{\prime}-\hspace{0.01in}t)E}\hspace{-0.01in}%
\int\text{d}_{q}^{3}\hspace{0.01in}p\,(u^{\ast})^{\mathbf{p}}(\mathbf{x}%
)\circledast\Delta_{L}(E,\mathbf{p})\circledast u^{\mathbf{p}}(\mathbf{x}%
^{\hspace{0.01in}\prime})\label{MomKGProR}%
\end{align}
with%
\begin{equation}
\Delta_{R}(E,\mathbf{p})=-\Delta_{L}(E,\mathbf{p})=\frac{\text{i\hspace
{0.01in}}c^{2}}{2\pi}(E^{\hspace{0.01in}2}-E_{\mathbf{p}}^{\hspace{0.01in}%
2}+\text{i\hspace{0.01in}}\varepsilon)^{-1}.\label{MomProDelR}%
\end{equation}
Note that we can write the expression in Eq.~(\ref{MomProDelR}) as a series in
normal-or\-dered monomials of momentum coordinates. To get this expansion, we
first write the fraction as a power series of $\mathbf{p}^{2}$ and then use
the formula in Eq.~(\ref{EntPotP}) of the previous chapter.

In the following, we show how to derive the momentum space form of the
$q$-de\-formed Klein-Gordon propagators. We demonstrate our considerations
using the example of the propagator $\Delta_{R}(\mathbf{x}^{\hspace
{0.01in}\prime}\hspace{-0.01in},t^{\hspace{0.01in}\prime}\hspace
{-0.01in};\mathbf{x},t)$. First, we need the following identity for the
Heaviside function:%
\begin{equation}
\theta(t)=\frac{\text{i}}{2\pi}\lim_{\varepsilon\hspace*{0.01in}%
\rightarrow\hspace{0.01in}0^{+}}\int\text{d}E\,\frac{\operatorname{e}%
^{-\text{i}tE}}{E+\text{i\hspace{0.01in}}\varepsilon}.\label{FouEntTheFkt}%
\end{equation}
With this identity, we can carry out the following calculation:%
\begin{align}
&  \theta(t^{\hspace{0.01in}\prime}\hspace{-0.02in}-t)\,E_{\mathbf{p}%
}^{\hspace{0.01in}-1}\hspace{-0.01in}\circledast\exp(-\text{i\hspace{0.01in}%
}(t^{\hspace{0.01in}\prime}\hspace{-0.02in}-t)E_{\mathbf{p}})=\nonumber\\
&  \qquad=\frac{\text{i}}{2\pi}\lim_{\varepsilon\hspace{0.01in}\rightarrow
\hspace{0.01in}0^{+}}\int\text{d}E\,\frac{1}{E+\text{i\hspace{0.01in}%
}\varepsilon}\,E_{\mathbf{p}}^{\hspace{0.01in}-1}\circledast\exp
(-\text{i\hspace{0.01in}}(t^{\hspace{0.01in}\prime}\hspace{-0.02in}%
-t)(E+\hspace{-0.01in}E_{\mathbf{p}}))\nonumber\\
&  \qquad=\frac{\text{i}}{2\pi}\lim_{\varepsilon\hspace{0.01in}\rightarrow
\hspace{0.01in}0^{+}}\int\text{d}E\,\operatorname{e}^{-\text{i}(t^{\prime
}-\hspace{0.01in}t)E}\hspace{0.01in}(E-\hspace{-0.01in}E_{\mathbf{p}%
}+\text{i\hspace{0.01in}}\varepsilon)^{-1}\circledast E_{\mathbf{p}}%
^{\hspace{0.01in}-1}.\label{UmfPhoPro1}%
\end{align}
As a final step of the above calculation, we have substituted $E$ as the
variable of integration by $E+\hspace{-0.01in}E_{\mathbf{p}}$. In the same
way, we get:%
\begin{align}
&  \theta(t-t^{\hspace{0.01in}\prime})\,E_{\mathbf{p}}^{\hspace{0.01in}%
-1}\hspace{-0.01in}\circledast\exp(\text{i\hspace{0.01in}}(t^{\hspace
{0.01in}\prime}\hspace{-0.02in}-t)E_{\mathbf{p}})=\nonumber\\
&  \qquad=\frac{-\text{\hspace{0.01in}i}}{2\pi}\lim_{\varepsilon
\hspace{0.01in}\rightarrow\hspace{0.01in}0^{+}}\int\text{d}E\,\operatorname{e}%
^{-\text{i}(t^{\prime}-\hspace{0.01in}t)E}\hspace{0.01in}(E+\hspace
{-0.01in}E_{\mathbf{p}}-\text{i\hspace{0.01in}}\varepsilon)^{-1}\circledast
E_{\mathbf{p}}^{\hspace{0.01in}-1}.\label{UmfPhoPro2}%
\end{align}
Due to the results in Eqs.~(\ref{UmfPhoPro1}) and (\ref{UmfPhoPro2}), we can
rewrite the expression for the propagator $\Delta_{R}$ in Eq.~(\ref{KGProAnt})
as follows:%
\begin{align}
&  \Delta_{R}(\mathbf{x}^{\hspace{0.01in}\prime}\hspace{-0.01in}%
,t^{\hspace{0.01in}\prime}\hspace{-0.01in};\mathbf{x},t)=\nonumber\\
&  \qquad=\theta(t^{\hspace{0.01in}\prime}\hspace{-0.01in}-t)\,D_{R}%
(\mathbf{x}^{\hspace{0.01in}\prime}\hspace{-0.01in},t^{\hspace{0.01in}\prime
}\hspace{-0.01in};\mathbf{x},t)+\theta(t-t^{\hspace{0.01in}\prime}%
)\,D_{R}(\mathbf{x}^{\hspace{0.01in}\prime}\hspace{-0.01in},-\hspace
{0.01in}t^{\hspace{0.01in}\prime};\mathbf{x},-\hspace{0.01in}t)\nonumber\\
&  \qquad=\int\text{d}_{q}^{3}\hspace{0.01in}p\,u_{\hspace{0.01in}\mathbf{p}%
}(\mathbf{x}^{\hspace{0.01in}\prime})\circledast\Delta_{R}(E,\mathbf{p}%
;t^{\hspace{0.01in}\prime}\hspace{-0.01in},t)\circledast(u^{\ast}%
)_{\mathbf{p}}(\mathbf{x})\label{ZwiRechProR}%
\end{align}
with%
\begin{align}
&  c^{-2}\hspace{0.01in}\Delta_{R}(E,\mathbf{p};t^{\hspace{0.01in}\prime
}\hspace{-0.01in},t)=\nonumber\\
&  \qquad=\frac{1}{2}\hspace{0.01in}\theta(t^{\hspace{0.01in}\prime}%
\hspace{-0.02in}-t)\,E_{\mathbf{p}}^{\hspace{0.01in}-1}\hspace{-0.01in}%
\circledast\exp(-\text{i\hspace{0.01in}}(t^{\hspace{0.01in}\prime}%
\hspace{-0.02in}-t)\hspace{0.01in}E_{\mathbf{p}})\nonumber\\
&  \qquad\hspace{0.17in}+\frac{1}{2}\hspace{0.01in}\theta(t-t^{\hspace
{0.01in}\prime})\,E_{\mathbf{p}}^{\hspace{0.01in}-1}\hspace{-0.01in}%
\circledast\exp(\text{i\hspace{0.01in}}(t^{\hspace{0.01in}\prime}%
\hspace{-0.02in}-t)\hspace{0.01in}E_{\mathbf{p}})\nonumber\\
&  \qquad=\frac{\text{i}}{4\pi}\lim_{\varepsilon\hspace{0.01in}\rightarrow
\hspace{0.01in}0^{+}}\int\text{d}E\,\operatorname{e}^{-\text{i}(t^{\prime
}-\hspace{0.01in}t)E}[(E-\hspace{-0.01in}E_{\mathbf{p}}+\text{i\hspace
{0.01in}}\varepsilon)^{-1}\hspace{-0.01in}\nonumber\\
&  \qquad\hspace{0.17in}\qquad\qquad\qquad\qquad\qquad\hspace{0.05in}%
-(E+\hspace{-0.01in}E_{\mathbf{p}}\hspace{-0.01in}-\text{i\hspace{0.01in}%
}\varepsilon)^{-1}]\circledast\hspace{-0.01in}E_{\mathbf{p}}^{\hspace
{0.01in}-1}.
\end{align}
We can combine the two fractions in square brackets:%
\begin{align}
\frac{1}{E-\hspace{-0.01in}E_{\mathbf{p}}+\text{i\hspace{0.01in}}\varepsilon
}-\frac{1}{E+\hspace{-0.01in}E_{\mathbf{p}}\hspace{-0.01in}-\text{i\hspace
{0.01in}}\varepsilon} &  =\frac{E+\hspace{-0.01in}E_{\mathbf{p}}%
\hspace{-0.01in}-\text{i\hspace{0.01in}}\varepsilon-\hspace{-0.01in}%
E+\hspace{-0.01in}E_{\mathbf{p}}\hspace{-0.01in}-\text{i\hspace{0.01in}%
}\varepsilon}{E^{\hspace{0.01in}2}-(E_{\mathbf{p}}\hspace{-0.01in}%
-\text{i\hspace{0.01in}}\varepsilon)^{2}}\nonumber\\
&  =\frac{2E_{\mathbf{p}}\hspace{-0.01in}-2\text{\hspace{0.01in}%
i\hspace{0.01in}}\varepsilon}{E^{\hspace{0.01in}2}-E_{\mathbf{p}}%
^{\hspace{0.01in}2}+2\text{\hspace{0.01in}i\hspace{0.01in}}\varepsilon
E_{\mathbf{p}}\hspace{-0.01in}-\varepsilon^{2}}.
\end{align}
This way, we end up with the following expression:%
\begin{equation}
\Delta_{R}(E,\mathbf{p};t^{\hspace{0.01in}\prime}\hspace{-0.01in}%
,t)=\frac{\text{i\hspace{0.01in}}c^{2}}{2\pi}\lim_{\varepsilon\hspace
{0.01in}\rightarrow\hspace{0.01in}0^{+}}\int\text{d}E\,(E^{\hspace{0.01in}%
2}-E_{\mathbf{p}}^{\hspace{0.01in}2}+\text{i\hspace{0.01in}}\varepsilon
)^{-1}\operatorname{e}^{-\text{i}(t^{\prime}-t)E}.
\end{equation}
If we substitute the expression above into Eq.~(\ref{ZwiRechProR}), we can
finally read off the momentum space form of the $q$-de\-formed Klein-Gor\-don
propagator $\Delta_{R}$.

\section{Charge conservation\label{KapCharKon}}

In this chapter, we derive continuity equations for our $q$-de\-formed
Klein-Gor\-don equations. To this end, we first multiply the Klein-Gor\-don
equation for $\varphi_{R}$ [cf. Eq.~(\ref{KleGorGleLin}) in
Chap.~\ref{LoeKleGorGleKap}] on the left by $\varphi_{L}^{\ast}$ and the
Klein-Gor\-don equation for $\varphi_{L}^{\ast}$ [cf. Eq.~(\ref{KleGorGleLin2}%
) in Chap.~\ref{LoeKleGorGleKap}] on the right by $\varphi_{R}$. Subtracting
the two equations multiplied by $\varphi_{L}^{\ast}$ or $\varphi_{R}$, we get
the following identity:%
\begin{equation}
\varphi_{L}^{\ast}\circledast c^{-2}\partial_{t}^{\hspace{0.01in}%
2}\triangleright\varphi_{R}-\varphi_{L}^{\ast}\hspace{-0.01in}\triangleleft
\partial_{t}^{\hspace{0.01in}2}c^{-2}\hspace{-0.01in}\circledast\varphi
_{R}-\varphi_{L}^{\ast}\circledast\nabla_{q}^{2}\triangleright\varphi
_{R}+\varphi_{L}^{\ast}\triangleleft\nabla_{q}^{2}\circledast\hspace
{0.01in}\varphi_{R}=0.\label{DifMulKGG}%
\end{equation}
We rewrite the expressions depending on time derivatives as follows:%
\begin{align}
&  \varphi_{L}^{\ast}\circledast c^{-2}\partial_{t}^{\hspace{0.01in}%
2}\triangleright\varphi_{R}-\varphi_{L}^{\ast}\hspace{-0.01in}\triangleleft
\partial_{t}^{\hspace{0.01in}2}c^{-2}\hspace{-0.01in}\circledast\varphi
_{R}=\nonumber\\
&  \qquad=\partial_{t}\triangleright\hspace{-0.01in}\big [\varphi_{L}^{\ast
}\circledast c^{-2}\partial_{t}\triangleright\varphi_{R}+\varphi_{L}^{\ast
}\hspace{-0.01in}\triangleleft\partial_{t}\hspace{0.01in}c^{-2}\hspace
{-0.01in}\circledast\varphi_{R}\big ].\label{AntTim}%
\end{align}
To rewrite the expressions in Eq.~(\ref{DifMulKGG}) depending on $\nabla
_{q}^{2}$, we need $q$-ver\-sions of Green's theorem \cite{Wachter:2021A},
i.~e.%
\begin{align}
&  \psi\triangleleft\nabla_{q}^{2}\circledast\phi-\psi\circledast\nabla
_{q}^{2}\triangleright\phi=\nonumber\\
&  \qquad=-\hspace{0.01in}\partial^{C}\triangleright\left[  q^{-2}%
\hspace{0.01in}\psi\triangleleft(\mathcal{L}_{\partial}){^{A}}_{\hspace
{-0.01in}C}\hspace{0.02in}\partial^{B}\hspace{-0.01in}\circledast\phi
+\psi\triangleleft(\mathcal{L}_{\partial}){^{A}}_{\hspace{-0.01in}%
C}\circledast\partial^{B}\triangleright\phi\right]  \hspace{0.02in}%
g_{AB}\label{QGreIde0}%
\end{align}
and%
\begin{align}
&  \psi\triangleleft\nabla_{q}^{2}\circledast\phi-\psi\circledast\nabla
_{q}^{2}\triangleright\phi=\nonumber\\
&  \qquad=g_{AB}\hspace{-0.01in}\left[  \psi\triangleleft\partial
^{A}\circledast(\mathcal{L}_{\partial}){^{B}}_{\hspace{-0.01in}C}%
\triangleright\phi+q^{2}\hspace{0.01in}\psi\circledast\partial^{A}%
(\mathcal{L}_{\partial}){^{B}}_{\hspace{-0.01in}C}\triangleright\phi\right]
\triangleleft\partial^{C}.\label{QGreIde1}%
\end{align}
Substituting $\psi$ and $\phi$ by $\varphi_{L}^{\ast}$ or $\varphi_{R}$, we
get:%
\begin{gather}
\varphi_{L}^{\ast}\circledast\nabla_{q}^{2}\triangleright\varphi_{R}%
-\varphi_{L}^{\ast}\hspace{-0.01in}\triangleleft\nabla_{q}^{2}\circledast
\varphi_{R}=\nonumber\\
=\partial^{C}\triangleright\big [q^{-2}\varphi_{L}^{\ast}\triangleleft
(\mathcal{L}_{\partial}){^{A}}_{C}\,\partial_{A}\hspace{-0.01in}%
\circledast\varphi_{R}+\varphi_{L}^{\ast}\triangleleft(\mathcal{L}_{\partial
}){^{A}}_{C}\circledast\partial_{A}\hspace{-0.01in}\triangleright\varphi
_{R}\big ].
\end{gather}
By combining this result with that of Eq.~(\ref{AntTim}), we finally get the
continuity equation%
\begin{equation}
\partial_{t}\triangleright\hspace{-0.01in}\rho(\mathbf{x},t)+\partial
^{A}\triangleright j_{A}(\mathbf{x},t)=0\label{KonGleKGG1}%
\end{equation}
with%
\begin{align}
j_{A}(\mathbf{x},t) &  =\,-\frac{\text{i\hspace{0.01in}}e}{2\hspace{0.01in}%
m}\big [q^{-2}\varphi_{L}^{\ast}\triangleleft(\mathcal{L}_{\partial}){^{C}%
}_{\hspace{-0.01in}A}\,\partial_{C}\hspace{-0.01in}\circledast\varphi
_{R}+\varphi_{L}^{\ast}\triangleleft(\mathcal{L}_{\partial}){^{C}}%
_{\hspace{-0.01in}A}\circledast\partial_{C}\hspace{-0.01in}\triangleright
\varphi_{R}\big ],\nonumber\\[0.05in]
\rho(\mathbf{x},t) &  =\hspace{0.05in}\frac{\text{i\hspace{0.01in}}e}%
{2\hspace{0.01in}m\hspace{0.01in}c^{2}}\big [\varphi_{L}^{\ast}\hspace
{-0.01in}\circledast\partial_{t}\triangleright\varphi_{R}+\varphi_{L}^{\ast
}\hspace{-0.01in}\triangleleft\partial_{t}\circledast\varphi_{R}%
\big ].\label{LadDich1}%
\end{align}

By conjugating Eq.~(\ref{KonGleKGG1}) [cf. Eq.~(\ref{KonEigSteProFkt}) in
App.~\ref{KapQuaZeiEle},\ Eq.~(\ref{RegConAbl}) in App.~\ref{KapParDer},
Eq.~(\ref{KonLMat}) in App.~\ref{KapHofStr} as well as Eqs.~(\ref{KonGKWel1})
and (\ref{KonGKWel2}) in Chap.~\ref{LoeKleGorGleKap}], we obtain the
continuity equation for the other $q$-ver\-sion of the Klein-Gordon equation,
i.~e.%
\begin{equation}
\rho^{\ast}(\mathbf{x},t)\,\bar{\triangleleft}\,\partial_{t}+(\hspace
{0.01in}j^{\ast})^{A}(\mathbf{x},t)\,\bar{\triangleleft}\,\partial
_{A}=0\label{KonGleKGG2}%
\end{equation}
with%
\begin{align}
\overline{\rho(\mathbf{x},t)}=\rho^{\ast}(\mathbf{x},t)= &  \hspace
{0.05in}\frac{\text{i\hspace{0.01in}}e}{2\hspace{0.01in}m\hspace{0.01in}c^{2}%
}\big [\varphi_{L}\hspace{-0.01in}\circledast\partial_{t}\,\bar{\triangleright
}\,\varphi_{R}^{\ast}+\varphi_{L}\,\bar{\triangleleft}\,\hspace{0.01in}%
\partial_{t}\circledast\varphi_{R}^{\ast}\big ],\label{LadDich2}\\[0.05in]
\overline{j_{A}(\mathbf{x},t)}=(\hspace{0.01in}j^{\ast})^{A}(\mathbf{x},t)= &
\,-\frac{\text{i\hspace{0.01in}}e}{2\hspace{0.01in}m}\hspace{0.01in}%
q^{-2}g_{CD}\hspace{0.02in}\varphi_{L}\circledast\partial^{C}\hspace
{0.01in}(\mathcal{\bar{L}}_{\partial}){^{D}}_{\hspace{-0.01in}E}%
\,\bar{\triangleright}\,\varphi_{R}^{\ast}\hspace{0.02in}g^{EA}\nonumber\\
&  -\frac{\text{i\hspace{0.01in}}e}{2\hspace{0.01in}m}\hspace{0.01in}%
g_{CD}\hspace{0.02in}\varphi_{L}\,\bar{\triangleleft}\,\partial^{C}%
\circledast(\mathcal{\bar{L}}_{\partial}){^{D}}_{\hspace{-0.01in}E}%
\,\bar{\triangleright}\,\varphi_{R}^{\ast}\hspace{0.02in}g^{EA}%
.\label{StromDich2}%
\end{align}

We get further continuity equations for our $q$-de\-formed Klein-Gor\-don
equations by applying the following substitutions to the expressions in
Eqs.~(\ref{KonGleKGG1})-(\ref{LadDich2}):%
\begin{gather}
\triangleright\,\leftrightarrow\,\bar{\triangleright},\qquad\triangleleft
\,\leftrightarrow\,\bar{\triangleleft},\qquad q\,\leftrightarrow
\,q^{-1},\nonumber\\
\mathcal{\bar{L}}_{\partial}\,\leftrightarrow\,\mathcal{L}_{\partial}%
,\qquad\varphi_{R}\leftrightarrow\varphi_{R}^{\ast},\qquad\varphi
_{L}\leftrightarrow\varphi_{L}^{\ast}.\label{ErsRegQKleGor}%
\end{gather}

In analogy to the undeformed case, $\rho(\mathbf{x},t)$ or $\rho^{\ast
}(\mathbf{x},t)$ is the charge density of a $q$-de\-formed spin-zero particle,
and $j_{A}(\mathbf{x},t)$ or $j_{A}^{\ast}(\mathbf{x},t)$ is the corresponding
current density. By integrating $\rho(\mathbf{x},t)$ or $\rho^{\ast
}(\mathbf{x},t)$ over all space, we regain the bilinear forms in
Eq.~(\ref{SesKGG1}) or Eq.~(\ref{SesKGG2}) of Chap.~\ref{KapPlaWavSol}. In
this respect, the condition in Eq.~(\ref{BedNeuTei}) of
Chap.~\ref{KapPlaWavSol} describes a $q$-de\-formed spin-zero particle without
any electric charge. Using the continuity equation in Eq.~(\ref{KonGleKGG1})
or Eq.~(\ref{KonGleKGG2}), we can show that the overall charge of a
$q$-de\-formed spin-zero particle is constant over time [cf. Eq.~(\ref{StoThe}%
) in App.~\ref{KapParDer}]:%
\begin{align}
&  \frac{\text{i\hspace{0.01in}}e}{2\hspace{0.01in}m\hspace{0.01in}c^{2}%
}\hspace{0.01in}\partial_{t}\hspace{0.01in}\triangleright\hspace{-0.01in}%
\int\text{d}_{q}^{3}x\,\big [\varphi_{L}^{\ast}\hspace{-0.01in}\circledast
\partial_{t}\triangleright\hspace{-0.01in}\varphi_{R}+\varphi_{L}^{\ast
}\hspace{-0.01in}\triangleleft\partial_{t}\circledast\varphi_{R}%
\big ]=\nonumber\\
&  \qquad=\partial_{t}\hspace{0.01in}\triangleright\hspace{-0.01in}%
\int\text{d}_{q}^{3}x\,\rho(\mathbf{x},t)=\hspace{-0.01in}\int\text{d}_{q}%
^{3}x\,\partial_{t}\triangleright\hspace{-0.01in}\rho(\mathbf{x},t)\nonumber\\
&  \qquad=-\hspace{-0.01in}\int\text{d}_{q}^{3}x\,\partial^{A}\triangleright
j_{A}(\mathbf{x},t)=0.
\end{align}

We can modify the above reasonings such that they apply to a charged spin-zero
particle moving in the presence of an electromagnetic field. To this end, we
substitute the time derivative $\partial_{t}$ and each partial derivative
$\partial^{C}$ in Eq.~(\ref{DifMulKGG}) by the corresponding covariant
derivative $D^{0}$ or $D^{C}$ [cf. Eq.~(\ref{KovAblGes}) in
Chap.~\ref{LoeKleGorGleKap}]:%
\begin{gather}
\varphi_{L}^{\ast}\circledast c^{-2}D^{0}D^{0}\triangleright\varphi
_{R}-\varphi_{L}^{\ast}\hspace{-0.01in}\triangleleft D^{0}D^{0}c^{-2}%
\hspace{-0.01in}\circledast\varphi_{R}\nonumber\\
-\varphi_{L}^{\ast}\circledast D^{C}D_{C}\triangleright\varphi_{R}+\varphi
_{L}^{\ast}\triangleleft D^{C}D_{C}\circledast\hspace{0.01in}\varphi
_{R}=0.\label{AusGleCharDen}%
\end{gather}
We have shown in Ref.~\cite{Wachter:2021A} that the covariant derivatives
$D^{C}$ are subject to the following indentities:%
\begin{align}
&  \psi\triangleleft D^{C}D_{C}\circledast\phi-\psi\circledast D^{C}%
D_{C}\triangleright\phi=\nonumber\\
&  \qquad=-\hspace{0.02in}\partial^{F}\triangleright\left[  q^{-2}%
\hspace{0.01in}\psi\triangleleft(\mathcal{L}_{\partial}){^{B}}_{\hspace
{-0.01in}F}\hspace{0.01in}D^{C}\circledast\phi+\psi\triangleleft
(\mathcal{L}_{\partial}){^{B}}_{\hspace{-0.01in}F}\circledast D^{C}%
\triangleright\phi\right]  \hspace{0.02in}g_{BC}\nonumber\\
&  \qquad=g_{BC}\hspace{-0.01in}\left[  \psi\triangleleft D^{B}\circledast
(\mathcal{L}_{\partial}){^{C}}_{\hspace{-0.01in}F}\triangleright\phi
+q^{2}\hspace{0.01in}\psi\circledast D^{B}(\mathcal{L}_{\partial}){^{C}%
}_{\hspace{-0.01in}F}\triangleright\phi\right]  \triangleleft\partial
^{F}.\label{GreIdeKinImp}%
\end{align}
A direct calculation shows that a similar identity holds for the covariant
time derivative $D^{0}$:%
\begin{align}
&  \varphi_{L}^{\ast}\circledast c^{-2}D^{0}D^{0}\triangleright\varphi
_{R}-\varphi_{L}^{\ast}\hspace{-0.01in}\triangleleft D^{0}D^{0}c^{-2}%
\hspace{-0.01in}\circledast\varphi_{R}=\nonumber\\
&  \qquad=\partial_{t}\triangleright\hspace{-0.01in}\big [\varphi_{L}^{\ast
}\circledast c^{-2}D^{0}\triangleright\varphi_{R}+\varphi_{L}^{\ast}%
\hspace{-0.01in}\triangleleft D^{0}\hspace{0.01in}c^{-2}\hspace{-0.01in}%
\circledast\varphi_{R}\big ].
\end{align}
The above identities enable us to rewrite Eq.~(\ref{AusGleCharDen}) as a
continuity equation. We obtain the new charge density and the new current
density from Eq.~(\ref{LadDich1}) or Eq.~(\ref{LadDich2}) by substituting the
covariant derivatives $D^{0}$ and $D^{C}$ for the time derivative
$\partial_{t}$ or the $q$-de\-formed partial derivatives $\partial^{C}$. We
get%
\begin{equation}
\partial_{t}\triangleright\hspace{-0.01in}\rho(\mathbf{x},t)+\partial
^{A}\triangleright j_{A}(\mathbf{x},t)=0
\end{equation}
with%
\begin{align}
\rho(\mathbf{x},t)= &  \,\frac{\text{i\hspace{0.01in}}e}{2\hspace
{0.01in}m\hspace{0.01in}c^{2}}\big [\varphi_{L}^{\ast}\hspace{-0.01in}%
\circledast D^{0}\triangleright\varphi_{R}+\varphi_{L}^{\ast}\hspace
{-0.01in}\triangleleft D^{0}\circledast\varphi_{R}\big ]\nonumber\\[0.05in]
= &  \,\frac{\text{i\hspace{0.01in}}e}{2\hspace{0.01in}m\hspace{0.01in}c^{2}%
}\big [\varphi_{L}^{\ast}\hspace{-0.01in}\circledast\partial_{t}%
\triangleright\varphi_{R}+\varphi_{L}^{\ast}\hspace{-0.01in}\triangleleft
\partial_{t}\circledast\varphi_{R}\big ]\nonumber\\
&  \,-\frac{e^{2}}{m\hspace{0.01in}c^{2}}\hspace{0.01in}\varphi_{L}^{\ast
}\hspace{-0.01in}\circledast A^{0}\circledast\varphi_{R}%
\end{align}
and%
\begin{align}
j_{B}(\mathbf{x},t)= &  -\frac{\text{i\hspace{0.01in}}e}{2\hspace{0.01in}%
m}\big [q^{-2}\varphi_{L}^{\ast}\triangleleft(\mathcal{L}_{\partial}){^{C}%
}_{\hspace{-0.01in}B}\,D_{C}\hspace{-0.01in}\circledast\varphi_{R}+\varphi
_{L}^{\ast}\triangleleft(\mathcal{L}_{\partial}){^{C}}_{\hspace{-0.01in}%
B}\circledast D_{C}\hspace{-0.01in}\triangleright\varphi_{R}\big ]\nonumber\\
= &  -\frac{\text{i\hspace{0.01in}}e}{2\hspace{0.01in}m}\big [q^{-2}%
\varphi_{L}^{\ast}\triangleleft(\mathcal{L}_{\partial}){^{C}}_{\hspace
{-0.01in}B}\,\partial_{C}\hspace{-0.01in}\circledast\varphi_{R}+\varphi
_{L}^{\ast}\triangleleft(\mathcal{L}_{\partial}){^{C}}_{\hspace{-0.01in}%
B}\circledast\partial_{C}\hspace{-0.01in}\triangleright\varphi_{R}%
\big ]\nonumber\\
&  -\frac{e^{2}}{2\hspace{0.01in}m\hspace{0.01in}c}\hspace{0.01in}%
(q^{-2}+1)\hspace{0.01in}\varphi_{L}^{\ast}\triangleleft(\mathcal{L}%
_{\partial}){^{C}}_{\hspace{-0.01in}B}\circledast A_{C}\circledast\varphi_{R}.
\end{align}
Conjugating the above identities and expressions gives us%
\begin{equation}
\rho^{\ast}(\mathbf{x},t)\,\bar{\triangleleft}\,\partial_{t}+(\hspace
{0.01in}j^{\ast})^{A}(\mathbf{x},t)\,\bar{\triangleleft}\,\partial_{A}=0
\end{equation}
with%
\begin{align}
\overline{\rho(\mathbf{x},t)}=\rho^{\ast}(\mathbf{x},t)= &  \hspace
{0.05in}\frac{\text{i\hspace{0.01in}}e}{2\hspace{0.01in}m\hspace{0.01in}c^{2}%
}\big [\varphi_{L}\hspace{-0.01in}\circledast D_{0}\,\bar{\triangleright
}\,\varphi_{R}^{\ast}+\varphi_{L}\,\bar{\triangleleft}\,\hspace{0.01in}%
D_{0}\circledast\varphi_{R}^{\ast}\big ],\\[0.05in]
\overline{j_{A}(\mathbf{x},t)}=(j^{\ast})^{A}(\mathbf{x},t)= &  \,-\frac
{\text{i\hspace{0.01in}}e}{2\hspace{0.01in}m}\hspace{0.01in}q^{-2}%
g_{CD}\hspace{0.02in}\varphi_{L}\circledast D^{C}\hspace{0.01in}%
(\mathcal{\bar{L}}_{\partial}){^{D}}_{\hspace{-0.01in}E}\,\bar{\triangleright
}\,\varphi_{R}^{\ast}\hspace{0.02in}g^{EA}\nonumber\\
&  -\frac{\text{i\hspace{0.01in}}e}{2\hspace{0.01in}m}\hspace{0.01in}%
g_{CD}\hspace{0.02in}\varphi_{L}\,\bar{\triangleleft}\,D^{C}\circledast
(\mathcal{\bar{L}}_{\partial}){^{D}}_{\hspace{-0.01in}E}\,\bar{\triangleright
}\,\varphi_{R}^{\ast}\hspace{0.02in}g^{EA}.
\end{align}

\section{Conservation of energy\label{KapConEne}}

We interpret the expression [cf. Eqs.~(\ref{EntWicKGFR}) and (\ref{EntKGFLSte}%
) in\ Chap.~\ref{KapPlaWavSol}]%
\begin{align}
\langle E\rangle_{\varphi}= &  \,\frac{1}{2}\sum_{\varepsilon\hspace
{0.01in}=\hspace{0.01in}\pm}\int\text{d}_{q}^{3}\hspace{0.01in}p\,h_{\hspace
{0.01in}\mathbf{p}}^{[\varepsilon]}\circledast E_{\mathbf{p}}\circledast
f_{\mathbf{p}}^{[\varepsilon]}\nonumber\\
= &  \,-\frac{1}{2\hspace{0.01in}c^{2}}\hspace{-0.01in}\int\text{d}_{q}%
^{3}x\,\varphi_{L}^{\ast}(\mathbf{x},t)\triangleleft\partial_{t}%
\circledast\partial_{t}\triangleright\hspace{-0.01in}\varphi_{R}%
(\mathbf{x},t)\nonumber\\
&  \,-\frac{1}{2}\int\text{d}_{q}^{3}x\,\varphi_{L}^{\ast}(\mathbf{x}%
,t)\triangleleft\partial^{A}\hspace{-0.01in}\circledast\partial_{A}%
\triangleright\varphi_{R}(\mathbf{x},t)\nonumber\\
&  \,+\frac{1}{2}\int\text{d}_{q}^{3}x\,\varphi_{L}^{\ast}(\mathbf{x}%
,t)\circledast(m\hspace{0.01in}c)^{2}\hspace{0.01in}\varphi_{R}(\mathbf{x}%
,t)\label{AusFelEneKGFGes}%
\end{align}
as the expectation value for the \textit{energy of a }$q$\textit{-de\-formed
spin-zero particle} if the following identity holds ($\varepsilon=\pm$):%
\begin{equation}
\overline{h_{\hspace{0.01in}\mathbf{p}}^{[\varepsilon]}}=f_{\mathbf{p}%
}^{[\varepsilon]}.\label{ErwBedKGF}%
\end{equation}
In addition to this, we require:%
\begin{equation}
\overline{f_{\mathbf{p}}^{[\pm]}}=f_{[\mp]}^{\mathbf{p}}\text{\quad and\quad
}\overline{h_{\hspace{0.01in}\mathbf{p}}^{[\pm]}}=h_{[\mp]}^{\mathbf{p}%
}.\label{ConEntKoe}%
\end{equation}
Due to the above condition, conjugating Eq.~(\ref{AusFelEneKGFGes}) gives
another expression for the energy of a $q$-de\-formed spin-zero particle [cf.
also Eq.~(\ref{KonEigSteProFkt}) in App.~\ref{KapQuaZeiEle}%
,\ Eqs.~(\ref{RegConAbl}) and (\ref{KonEigVolInt}) in App.~\ref{KapParDer} as
well as Eqs.~(\ref{KonGKWel1}) and (\ref{KonGKWel2}) in
Chap.~\ref{LoeKleGorGleKap}]:%
\begin{align}
\langle E^{\ast}\rangle_{\varphi}= &  \,\frac{1}{2}\sum_{\varepsilon
\hspace{0.01in}=\hspace{0.01in}\pm}\int\text{d}_{q}^{3}\hspace{0.01in}%
p\,f_{[\varepsilon]}^{\mathbf{p}}\circledast E_{\mathbf{p}}\circledast
h_{[\varepsilon]}^{\mathbf{p}}\nonumber\\
= &  \,-\frac{1}{2\hspace{0.01in}c^{2}}\int\text{d}_{q}^{3}x\,\varphi
_{L}(\mathbf{x},t)\triangleleft\partial_{t}\circledast\partial_{t}%
\triangleright\hspace{-0.01in}\varphi_{R}^{\ast}(\mathbf{x},t)\nonumber\\
&  \,-\frac{1}{2}\int\text{d}_{q}^{3}x\,\varphi_{L}(\mathbf{x},t)\triangleleft
\partial^{A}\hspace{-0.01in}\circledast\partial_{A}\triangleright\varphi
_{R}^{\ast}(\mathbf{x},t)\nonumber\\
&  \,+\frac{1}{2}\int\text{d}_{q}^{3}x\,\varphi_{L}(\mathbf{x},t)\circledast
(m\hspace{0.01in}c)^{2}\hspace{0.01in}\varphi_{R}^{\ast}(\mathbf{x}%
,t).\label{EneKleGor}%
\end{align}
In analogy to Eq.~(\ref{ErwBedKGF}), it holds:%
\begin{equation}
\overline{h_{[\varepsilon]}^{\mathbf{p}}}=f_{[\varepsilon]}^{\mathbf{p}%
}.\label{BedEntAmpKGFN}%
\end{equation}

In Eqs.~(\ref{AusFelEneKGFGes}) and (\ref{EneKleGor}), we have written down
the expectation value for the energy, both in momentum space and in position
space. We briefly explain how to get the expression in momentum space from
that in position space. As a first step, we have to substitute the wave
functions in position space by their series expansions [cf.
Eqs.~(\ref{EntWicKGFR}) and (\ref{EntKGFLSte}) of Chap.~\ref{KapPlaWavSol}].
This way, we can apply the time derivatives and $q$-de\-formed partial
derivatives to our $q$-de\-formed plane wave solutions [cf
Eqs.~(\ref{DerPlaWav}) and (\ref{DerPlaWavSte}) in Chap.~\ref{KapPlaWavSol}].
With the help of the en\-er\-gy-mo\-men\-tum relation [cf.
Eq.~(\ref{EneMomRelKleGor}) in Chap.~\ref{KapPlaWavSol}], the orthogonality
relations for $q$-de\-formed plane waves [cf. Eqs.~(\ref{SkaProEbeDreExpWie0})
and (\ref{SkaProEbeDreExpWie1}) in Chap.~\ref{KapPlaWavSol}], and the
identities for $q$-de\-formed delta functions [cf. Eq.~(\ref{AlgChaIdeqDelFkt}%
) of Chap.~\ref{KapPlaWavSol}], we finally end up with the expression in
momentum space.

Next, we derive continuity equations for the \textit{energy density of a free
}$q$\textit{-de\-formed spin-zero particle}. A look at
Eq.~(\ref{AusFelEneKGFGes}) shows that the energy density takes on the
following form:%
\begin{align}
\mathcal{H}= &  -\frac{1}{2\hspace{0.01in}c^{2}}\,\varphi_{L}^{\ast
}(\mathbf{x},t)\triangleleft\partial_{t}\circledast\partial_{t}\triangleright
\hspace{-0.01in}\varphi_{R}(\mathbf{x},t)\nonumber\\
&  -\frac{1}{2}\,\varphi_{L}^{\ast}(\mathbf{x},t)\triangleleft\partial
^{C}\hspace{-0.01in}\circledast\partial_{C}\triangleright\varphi
_{R}(\mathbf{x},t)\nonumber\\
&  +\frac{1}{2}\,\varphi_{L}^{\ast}(\mathbf{x},t)\circledast(m\hspace
{0.01in}c)^{2}\hspace{0.01in}\varphi_{R}(\mathbf{x},t).\label{EneDen}%
\end{align}
We can calculate the time derivative of the energy density by applying the
usual Leibniz rule:%
\begin{align}
\partial_{t}\hspace{0.01in}\triangleright\mathcal{H}= &  \,\frac{1}%
{2\hspace{0.01in}c^{2}}\,\varphi_{L}^{\ast}\triangleleft\partial_{t}%
\hspace{0.01in}\partial_{t}\circledast\partial_{t}\triangleright
\hspace{-0.01in}\varphi_{R}-\frac{1}{2\hspace{0.01in}c^{2}}\,\varphi_{L}%
^{\ast}\triangleleft\partial_{t}\circledast\partial_{t}\hspace{0.01in}%
\partial_{t}\triangleright\hspace{-0.01in}\varphi_{R}\nonumber\\
&  \,+\frac{1}{2}\,\varphi_{L}^{\ast}\triangleleft\partial_{t}\hspace
{0.01in}\partial^{C}\hspace{-0.01in}\circledast\partial_{C}\triangleright
\varphi_{R}-\frac{1}{2}\,\varphi_{L}^{\ast}\triangleleft\partial^{C}%
\hspace{-0.01in}\circledast\partial_{C}\hspace{0.01in}\partial_{t}%
\triangleright\hspace{-0.01in}\varphi_{R}\nonumber\\
&  \,-\frac{1}{2}\,\varphi_{L}^{\ast}\triangleleft\partial_{t}\circledast
(m\hspace{0.01in}c)^{2}\hspace{0.01in}\varphi_{R}+\frac{1}{2}\,\varphi
_{L}^{\ast}\circledast(m\hspace{0.01in}c)^{2}\partial_{t}\triangleright
\hspace{-0.01in}\varphi_{R}.
\end{align}
Using the $q$-de\-formed Klein-Gor\-don equations for $\varphi_{R}$ and
$\varphi_{L}^{\ast}$ [cf. Eqs.~(\ref{KleGorGleLin})\ and (\ref{KleGorGleLin2})
in Chap.~\ref{LoeKleGorGleKap}], we can rewrite the above equation as follows:%
\begin{align}
\partial_{t}\hspace{0.01in}\triangleright\mathcal{H}= &  \,\frac{1}%
{2}\,\varphi_{L}^{\ast}\triangleleft\partial^{C}\partial_{C}\circledast
\partial_{t}\triangleright\hspace{-0.01in}\varphi_{R}-\frac{1}{2\hspace
{0.01in}}\,\varphi_{L}^{\ast}\triangleleft\partial_{t}\circledast\partial
^{C}\partial_{C}\triangleright\hspace{-0.01in}\varphi_{R}\nonumber\\
&  \,+\frac{1}{2}\,\varphi_{L}^{\ast}\triangleleft\partial_{t}\hspace
{0.01in}\partial^{C}\hspace{-0.01in}\circledast\partial_{C}\triangleright
\varphi_{R}-\frac{1}{2}\,\varphi_{L}^{\ast}\triangleleft\partial^{C}%
\hspace{-0.01in}\circledast\partial_{C}\hspace{0.01in}\partial_{t}%
\triangleright\hspace{-0.01in}\varphi_{R}.\label{ZeiAblEneDic}%
\end{align}
The Leibniz rules in Eq.~(\ref{AllVerRelParAblEle1}) of App.~\ref{KapHofStr}
imply the following identities for actions of $q$-de\-formed partial
derivatives (see Ref.~\cite{Wachter:2021A}):%
\begin{align}
\psi\circledast\partial^{C}\triangleright\phi &  =\partial_{(2)}%
^{C}\triangleright\big [\psi\triangleleft\partial_{(1)}^{C}\circledast
\phi\big ]\nonumber\\
&  =\partial^{B}\triangleright\left[  \psi\triangleleft(\mathcal{L}_{\partial
}){^{C}}_{\hspace{-0.01in}B}\circledast\phi\right]  +\psi\triangleleft
\partial^{C}\hspace{-0.01in}\circledast\phi\label{UmgLeiReg1}%
\end{align}
and%
\begin{align}
\psi\triangleleft\partial^{C}\circledast\phi &  =[\psi\circledast
\partial_{(1)}^{C}\triangleright\phi]\triangleleft\partial_{(2)}%
^{C}\nonumber\\
&  =[\psi\circledast(\mathcal{L}_{\partial}){^{C}}_{\hspace{-0.01in}%
B}\triangleright\phi]\triangleleft\partial^{B}+\psi\circledast\partial
^{C}\hspace{-0.01in}\triangleright\phi.\label{UmgLeiReg2}%
\end{align}
With the help of these identities, we can write the right-hand side of
Eq.~(\ref{ZeiAblEneDic}) as divergency:%
\begin{align}
\partial_{t}\hspace{0.01in}\triangleright\mathcal{H}= &  \,\frac{1}%
{2}\,\big [\varphi_{L}^{\ast}\triangleleft\partial_{t}\circledast
(\mathcal{L}_{\partial}){^{B}}_{\hspace{-0.01in}C}\hspace{0.01in}\partial
_{B}\triangleright\hspace{-0.01in}\varphi_{R}\big ]\triangleleft\partial
^{C}\nonumber\\
&  \,-\frac{1}{2}\,\partial^{C}\triangleright\big [\varphi_{L}^{\ast
}\triangleleft\partial^{B}(\mathcal{L}_{\partial}){^{D}}_{\hspace{-0.01in}%
C}\,g_{BD}\circledast\partial_{t}\triangleright\varphi_{R}\big ].
\end{align}
Eq.~(\ref{UmgLeiReg1}) together with Eq.~(\ref{UmgLeiReg2}) implies the
following identity:%
\begin{equation}
\lbrack\psi\circledast(\mathcal{L}_{\partial}){^{C}}_{\hspace{-0.01in}%
B}\triangleright\phi]\triangleleft\partial^{B}=-\hspace{0.01in}\partial
^{B}\triangleright\left[  \psi\triangleleft(\mathcal{L}_{\partial}){^{C}%
}_{\hspace{-0.01in}B}\circledast\phi\right]  .\label{VerLMatAbl}%
\end{equation}
Moreover, we have shown in Ref.~\cite{Wachter:2021A} that the L-ma\-tri\-ces
are subject to%
\begin{equation}
g_{BD}\hspace{0.01in}\partial^{B}(\mathcal{L}_{\partial}){^{D}}_{\hspace
{-0.01in}C}=q^{-2}(\mathcal{L}_{\partial}){^{B}}_{\hspace{-0.01in}C}%
\hspace{0.02in}\partial^{D}g_{BD}\hspace{0.01in}.\label{VerAplLMat}%
\end{equation}
With the help of Eq.~(\ref{VerLMatAbl}) and Eq.~(\ref{VerAplLMat}), we finally
get the continuity equation%
\begin{equation}
\partial_{t}\hspace{0.01in}\triangleright\mathcal{H}+\partial^{C}%
\triangleright S_{C}=0\label{KonEngDic}%
\end{equation}
with the following \textit{energy flux density}:%
\begin{align}
S_{C}= &  \,\frac{1}{2}\,\varphi_{L}^{\ast}\triangleleft\partial
_{t}(\mathcal{L}_{\partial}){^{B}}_{\hspace{-0.01in}C}\circledast\partial
_{B}\triangleright\hspace{-0.01in}\varphi_{R}\nonumber\\
&  \,+\frac{1}{2}\,q^{-2}\,\varphi_{L}^{\ast}\triangleleft(\mathcal{L}%
_{\partial}){^{B}}_{\hspace{-0.01in}C}\,\partial_{B}\circledast\partial
_{t}\triangleright\hspace{-0.01in}\varphi_{R}.\label{EneStrSich}%
\end{align}

Integrating Eq.~(\ref{KonEngDic}) over all space leads to a surface term at
spatial infinity. However, this surface term will be zero since the wave
functions vanish at spatial infinity. Accordingly, the total energy of a
$q$-de\-formed spin-zero particle is constant over time:%
\begin{equation}
\partial_{t}\triangleright\hspace{-0.02in}\int\text{d}_{q}^{3}\hspace
{0.01in}x\,\mathcal{H}(\mathbf{x},t)=-\int\hspace{-0.01in}\text{d}_{q}%
^{3}\hspace{0.01in}x\,\partial^{C}\triangleright S_{C}=0.
\end{equation}

By conjugating Eqs.~(\ref{EneDen}), (\ref{KonEngDic}), and (\ref{EneStrSich}),
we get another expression for the energy density and a corresponding
continuity equation. Concretely, we have%
\begin{equation}
\mathcal{H}^{\ast}\,\bar{\triangleleft}\,\hspace{0.01in}\partial_{t}+(S^{\ast
})^{C}\,\bar{\triangleleft}\,\hspace{0.01in}\partial_{C}=0
\end{equation}
with%
\begin{align}
\mathcal{H}^{\ast}= &  \,-\frac{1}{2\hspace{0.01in}c^{2}}\,\varphi_{L}%
\,\bar{\triangleleft}\,\partial_{t}\circledast\partial_{t}\,\bar
{\triangleright}\,\varphi_{R}^{\ast}-\frac{1}{2}\,\varphi_{L}\,\bar
{\triangleleft}\,\partial^{C}\hspace{-0.01in}\circledast\partial_{C}%
\,\bar{\triangleright}\,\varphi_{R}^{\ast}\nonumber\\
&  \,+\frac{1}{2}\,\varphi_{L}\circledast(m\hspace{0.01in}c)^{2}%
\hspace{0.01in}\varphi_{R}^{\ast}%
\end{align}
and
\begin{align}
(S^{\ast})^{C}\hspace{-0.01in}= &  \,\frac{1}{2}\,g_{BD}\hspace{0.02in}%
\varphi_{L}\,\bar{\triangleleft}\,\hspace{0.01in}\partial^{B}\circledast
(\mathcal{\bar{L}}_{\partial}){^{D}}_{\hspace{-0.01in}E}\hspace{0.02in}%
\partial_{t}\,\bar{\triangleright}\,\varphi_{R}^{\ast}\hspace{0.02in}%
g^{EC}\nonumber\\
&  \,+\frac{1}{2}\,q^{-2}\,g_{BD}\hspace{0.02in}\varphi_{L}\,\bar
{\triangleleft}\,\hspace{0.01in}\partial_{t}\,\circledast\partial^{B}%
\hspace{0.01in}(\mathcal{\bar{L}}_{\partial}){^{D}}_{\hspace{-0.01in}E}%
\,\bar{\triangleright}\,\varphi_{R}^{\ast}\hspace{0.02in}g^{EC}.
\end{align}

Once again, we can apply the rules in Eq.~(\ref{ErsRegQKleGor}) of
Chap.~\ref{KapCharKon} to the results of this chapter.\ This way, we get
further $q$-ver\-sions of the continuity equation for the energy density of a
spin-zero particle.

\section{Conservation of momentum}

The expectation value for the \textit{momentum of a }$q$\textit{-de\-formed
spin-zero particle} is given by the following expression:%
\begin{align}
\langle\hspace{0.01in}p^{A}\rangle= &  \,\frac{1}{2}\sum_{\varepsilon
\hspace{0.01in}=\hspace{0.01in}\pm}\int\text{d}_{q}^{3}\hspace{0.01in}%
p\,h_{\hspace{0.01in}\mathbf{p}}^{[\varepsilon]}\circledast p^{A}\circledast
f_{\mathbf{p}}^{[\varepsilon]}\nonumber\\
= &  \,\frac{1}{2\hspace{0.01in}c^{2}}\hspace{-0.01in}\int\text{d}_{q}%
^{3}x\,\varphi_{L}^{\ast}(\mathbf{x},t)\triangleleft\partial_{t}%
\circledast\partial^{A}\triangleright\varphi_{R}(\mathbf{x},t)\nonumber\\
&  \,+\frac{1}{2\hspace{0.01in}c^{2}}\hspace{-0.01in}\int\text{d}_{q}%
^{3}x\,\varphi_{L}^{\ast}(\mathbf{x},t)\triangleleft\partial^{A}%
\hspace{-0.01in}\circledast\partial_{t}\triangleright\hspace{-0.01in}%
\varphi_{R}(\mathbf{x},t).\label{ErwImpKGF}%
\end{align}
Conjugating Eq.~(\ref{ErwImpKGF}) leads to another expression for the
expectation value of a $q$-de\-formed spin-zero particle [cf. 
Eq.~(\ref{KonEigSteProFkt}) in App.~\ref{KapQuaZeiEle},\ Eqs.~(\ref{RegConAbl}%
) and (\ref{KonEigVolInt}) in App.~\ref{KapParDer}, Eqs.~(\ref{KonGKWel1}) and
(\ref{KonGKWel2}) in Chap.~\ref{LoeKleGorGleKap}, and Eq.~(\ref{ConEntKoe}) in
Chap.~\ref{KapConEne}]:%
\begin{align}
\langle\hspace{0.01in}p_{A}^{\ast}\rangle=\overline{\langle\hspace
{0.01in}p^{A}\rangle}= &  \,\frac{1}{2}\sum_{\varepsilon\hspace{0.01in}%
=\hspace{0.01in}\pm}\int\text{d}_{q}^{3}\hspace{0.01in}p\,f_{[\varepsilon
]}^{\mathbf{p}}\circledast p_{A}\circledast h_{[\varepsilon]}^{\mathbf{p}%
}\nonumber\\
= &  \,\frac{1}{2\hspace{0.01in}c^{2}}\hspace{-0.01in}\int\text{d}_{q}%
^{3}x\,\varphi_{L}(\mathbf{x},t)\,\bar{\triangleleft}\,\partial_{t}%
\circledast\partial_{A}\,\bar{\triangleright}\,\varphi_{R}^{\ast}%
(\mathbf{x},t)\nonumber\\
&  \,+\frac{1}{2\hspace{0.01in}c^{2}}\hspace{-0.01in}\int\text{d}_{q}%
^{3}x\,\varphi_{L}(\mathbf{x},t)\,\bar{\triangleleft}\,\partial_{A}%
\circledast\partial_{t}\,\hspace{0.01in}\bar{\triangleright}\,\varphi
_{R}^{\ast}(\mathbf{x},t).\label{ErwImpKGFKon}%
\end{align}

In Eqs.~(\ref{ErwImpKGF}) and (\ref{ErwImpKGFKon}), we have written down the
expectation value for the momentum of a $q$\textit{-}de\-formed spin-zero
particle, both in momentum space and in position space. We can derive the
expression in momentum space from that in position space by the same
reasonings we have already applied to the expectation value for the energy of
a $q$\textit{-}de\-formed spin-zero particle [cf. Eqs.~(\ref{AusFelEneKGFGes})
and (\ref{EneKleGor}) ot the previous chapter].

We introduce the following expressions for the \textit{momentum density of }a
$q$-de\-formed spin-zero particle:%
\begin{align}
i^{A} &  =\frac{1}{2\hspace{0.01in}c^{2}}\hspace{0.01in}\varphi_{L}^{\ast
}\triangleleft\partial_{t}\circledast\partial^{A}\triangleright\varphi
_{R}+\frac{1}{2\hspace{0.01in}c^{2}}\hspace{0.01in}\varphi_{L}^{\ast
}\triangleleft\partial^{A}\hspace{-0.01in}\circledast\partial_{t}%
\triangleright\hspace{-0.01in}\varphi_{R},\nonumber\\[0.06in]
i_{A}^{\ast} &  =\frac{1}{2\hspace{0.01in}c^{2}}\,\varphi_{L}\,\bar
{\triangleleft}\,\partial_{t}\circledast\partial_{A}\,\bar{\triangleright
}\,\varphi_{R}^{\ast}+\frac{1}{2\hspace{0.01in}c^{2}}\,\varphi_{L}%
\,\bar{\triangleleft}\,\partial_{A}\circledast\partial_{t}\,\hspace
{0.01in}\bar{\triangleright}\,\varphi_{R}^{\ast}.
\end{align}
Using Eqs.~(\ref{KonGKWel1}) and (\ref{KonGKWel2}) in
Chap.~\ref{LoeKleGorGleKap} together with Eq.~(\ref{RegConAbl}) in
Chap.~\ref{KapParDer}, we can show that the two expressions for the momentum
density transform into each other by conjugation:%
\begin{equation}
\overline{i^{A}}=i_{A}^{\ast}.
\end{equation}

Next, we calculate the time derivative of the momentum density $i^{A}$:%
\begin{align}
\partial_{t}\triangleright i^{A}= &  -\frac{1}{2\hspace{0.01in}c^{2}}%
\hspace{0.01in}\varphi_{L}^{\ast}\triangleleft\partial_{t}\hspace
{0.01in}\partial_{t}\circledast\partial^{A}\triangleright\varphi_{R}+\frac
{1}{2\hspace{0.01in}c^{2}}\hspace{0.01in}\varphi_{L}^{\ast}\triangleleft
\partial_{t}\circledast\partial^{A}\partial_{t}\triangleright\varphi
_{R}\nonumber\\
&  -\frac{1}{2\hspace{0.01in}c^{2}}\hspace{0.01in}\varphi_{L}^{\ast
}\triangleleft\partial_{t}\hspace{0.01in}\partial^{A}\hspace{-0.01in}%
\circledast\partial_{t}\triangleright\hspace{-0.01in}\varphi_{R}+\frac
{1}{2\hspace{0.01in}c^{2}}\hspace{0.01in}\varphi_{L}^{\ast}\triangleleft
\partial^{A}\hspace{-0.01in}\circledast\partial_{t}\hspace{0.01in}\partial
_{t}\triangleright\hspace{-0.01in}\varphi_{R}.
\end{align}
We use the $q$-de\-formed Klein-Gor\-don equations for $\varphi_{R}$ and
$\varphi_{L}^{\ast}$ [cf. Eqs.~(\ref{KleGorGleLin})\ and (\ref{KleGorGleLin2})
in Chap.~\ref{LoeKleGorGleKap}] to rewrite the first and last term on the
right-hand side of the above equation:%
\begin{align}
\partial_{t}\triangleright i^{A}= &  -\frac{1}{2}\hspace{0.01in}(\varphi
_{L}^{\ast}\triangleleft\partial^{B}\partial_{B}-\varphi_{L}^{\ast}%
(m\hspace{0.01in}c)^{2})\circledast\partial^{A}\triangleright\varphi
_{R}\nonumber\\
&  +\frac{1}{2}\hspace{0.01in}\varphi_{L}^{\ast}\triangleleft\partial
^{A}\hspace{-0.01in}\circledast(\partial^{B}\partial_{B}\triangleright
\hspace{-0.01in}\varphi_{R}-(m\hspace{0.01in}c)^{2}\varphi_{R})\nonumber\\
&  +\frac{1}{2\hspace{0.01in}c^{2}}\hspace{0.01in}\varphi_{L}^{\ast
}\triangleleft\partial_{t}\circledast\partial^{A}\partial_{t}\triangleright
\varphi_{R}-\frac{1}{2\hspace{0.01in}c^{2}}\hspace{0.01in}\varphi_{L}^{\ast
}\triangleleft\partial_{t}\hspace{0.01in}\partial^{A}\hspace{-0.01in}%
\circledast\partial_{t}\triangleright\hspace{-0.01in}\varphi_{R}%
.\label{ZeiImpDic}%
\end{align}
By applying the identities in Eqs.~(\ref{UmgLeiReg1}) and (\ref{UmgLeiReg2})
of the previous chapter, the first expression on the right-hand side takes on
the following form:%
\begin{align}
&  (\varphi_{L}^{\ast}\triangleleft\partial^{B}\partial_{B}-\varphi_{L}^{\ast
}(m\hspace{0.01in}c)^{2})\circledast\partial^{A}\triangleright\varphi
_{R}=\nonumber\\
&  \qquad=(\varphi_{L}^{\ast}\triangleleft\partial^{A}\partial^{B}\partial
_{B}-\varphi_{L}^{\ast}\triangleleft\partial^{A}(m\hspace{0.01in}%
c)^{2})\circledast\varphi_{R}\nonumber\\
&  \qquad\hspace{0.17in}+\partial^{D}\triangleright\big [\big (\varphi
_{L}^{\ast}\triangleleft\partial^{B}\partial_{B}-\varphi_{L}^{\ast}%
(m\hspace{0.01in}c)^{2}\big )(\mathcal{L}_{\partial}){^{A}}_{\hspace
{-0.01in}D}\big ]\circledast\varphi_{R}.
\end{align}
Similar reasoning leads to%
\begin{align}
\varphi_{L}^{\ast}\triangleleft\partial_{t}\circledast\partial^{A}\partial
_{t}\triangleright\varphi_{R}= &  \,\varphi_{L}^{\ast}\triangleleft
\partial^{A}\partial_{t}\circledast\partial_{t}\triangleright\varphi
_{R}\nonumber\\
&  \,+\partial^{D}\triangleright\big [\varphi_{L}^{\ast}\triangleleft
\partial_{t}\hspace{0.01in}(\mathcal{L}_{\partial}){^{A}}_{\hspace{-0.01in}%
D}\circledast\partial_{t}\triangleright\hspace{-0.01in}\varphi_{R}\big ]
\end{align}
and%
\begin{align}
\varphi_{L}^{\ast}\triangleleft\partial^{A}\hspace{-0.01in}\circledast
\partial^{B}\partial_{B}\triangleright\hspace{-0.01in}\varphi_{R}= &
\,\varphi_{L}^{\ast}\triangleleft\partial^{A}\partial^{B}\hspace
{-0.01in}\circledast\partial_{B}\triangleright\hspace{-0.01in}\varphi
_{R}\nonumber\\
&  \,+\partial^{D}\triangleright\big [\varphi_{L}^{\ast}\triangleleft
\partial^{A}(\mathcal{L}_{\partial}){^{B}}_{\hspace{-0.01in}D}\circledast
\partial_{B}\triangleright\hspace{-0.01in}\varphi_{R}\big ]\nonumber\\
= &  \,\varphi_{L}^{\ast}\triangleleft\partial^{A}\partial^{B}\partial
_{B}\hspace{-0.01in}\circledast\varphi_{R}\nonumber\\
&  \,+\partial^{D}\triangleright\big [\varphi_{L}^{\ast}\triangleleft
\partial^{A}\partial^{B}(\mathcal{L}_{\partial}){^{C}}_{\hspace{-0.01in}%
D}\hspace{0.01in}g_{BC}\circledast\varphi_{R}\big ]\nonumber\\
&  \,+\partial^{D}\triangleright\big [\varphi_{L}^{\ast}\triangleleft
\partial^{A}(\mathcal{L}_{\partial}){^{B}}_{\hspace{-0.01in}D}\circledast
\partial_{B}\triangleright\hspace{-0.01in}\varphi_{R}\big ].
\end{align}
Putting these results together and considering the identity%
\begin{equation}
(\partial^{B}\partial_{B}-(m\hspace{0.01in}c)^{2})\hspace{0.01in}%
(\mathcal{L}_{\partial}){^{A}}_{\hspace{-0.01in}D}=q^{-4}\hspace
{0.01in}(\mathcal{L}_{\partial}){^{A}}_{\hspace{-0.01in}D}\hspace
{0.01in}(\partial^{B}\partial_{B}-(m\hspace{0.01in}c)^{2}),
\end{equation}
we can write the right-hand side of Eq.~(\ref{ZeiImpDic}) as divergency. This
way, we end up with the continuity equation%
\begin{equation}
\partial_{t}\triangleright i^{A}+\partial^{D}\triangleright T_{D}%
^{A}=0,\label{ConMomDen}%
\end{equation}
where the \textit{stress-tensor} takes on the following form:%
\begin{align}
T_{D}^{A}= &  \,-\frac{1}{2}\hspace{0.01in}\varphi_{L}^{\ast}\triangleleft
\partial^{A}\partial^{B}(\mathcal{L}_{\partial}){^{C}}_{\hspace{-0.01in}%
D}\hspace{0.01in}g_{BC}\circledast\varphi_{R}\nonumber\\
&  \,+\frac{1}{2\hspace{0.01in}q^{4}}\hspace{0.01in}\varphi_{L}^{\ast
}\triangleleft(\mathcal{L}_{\partial}){^{A}}_{\hspace{-0.01in}D}%
\hspace{0.01in}\partial^{B}\partial_{B}\circledast\varphi_{R}\nonumber\\
&  \,-\frac{1}{2\hspace{0.01in}q^{4}}\hspace{0.01in}\varphi_{L}^{\ast
}\triangleleft(\mathcal{L}_{\partial}){^{A}}_{\hspace{-0.01in}D}%
\hspace{0.01in}(mc)^{2}\circledast\varphi_{R}\nonumber\\
&  \,-\frac{1}{2}\hspace{0.01in}\varphi_{L}^{\ast}\triangleleft\partial
^{A}(\mathcal{L}_{\partial}){^{B}}_{\hspace{-0.01in}D}\circledast\partial
_{B}\triangleright\varphi_{R}\nonumber\\
&  \,-\frac{1}{2\hspace{0.01in}c^{2}}\hspace{0.01in}\varphi_{L}^{\ast
}\triangleleft\partial_{t}\hspace{0.01in}(\mathcal{L}_{\partial}){^{A}%
}_{\hspace{-0.01in}D}\circledast\partial_{t}\triangleright\varphi
_{R}.\label{SpaTen}%
\end{align}
Conjugating Eqs.~(\ref{ConMomDen}) and (\ref{SpaTen}) gives us another
continuity equation, i.~e.%
\begin{equation}
(i^{\ast})_{A}\,\bar{\triangleleft}\,\partial_{t}+(T^{\ast})_{A}^{D}%
\,\bar{\triangleleft}\,\partial_{D}=0
\end{equation}
and%
\begin{align}
(T^{\ast})_{A}^{D}= &  \,-\frac{1}{2}\hspace{0.01in}\varphi_{L}\circledast
(\mathcal{\bar{L}}_{\partial}){^{B}}_{\hspace{-0.01in}C}\hspace{0.01in}%
\partial_{B}\partial_{A}\,\bar{\triangleright}\,\varphi_{R}^{\ast}%
\hspace{0.02in}g^{CD}\nonumber\\
&  \,+\frac{1}{2\hspace{0.01in}q^{4}}\hspace{0.01in}g_{AE}\hspace
{0.02in}\varphi_{L}\circledast\partial^{B}\partial_{B}\hspace{0.01in}%
(\mathcal{\bar{L}}_{\partial}){^{E}}_{\hspace{-0.01in}C}\,\bar{\triangleright
}\,\varphi_{R}^{\ast}\hspace{0.02in}g^{CD}\nonumber\\
&  \,-\frac{1}{2\hspace{0.01in}q^{4}}\hspace{0.01in}g_{AE}\hspace
{0.02in}\varphi_{L}\circledast(mc)^{2}\hspace{0.01in}(\mathcal{\bar{L}%
}_{\partial}){^{E}}_{\hspace{-0.01in}C}\,\bar{\triangleright}\,\varphi
_{R}^{\ast}\hspace{0.02in}g^{CD}\nonumber\\
&  \,-\frac{1}{2}\hspace{0.01in}g_{BE}\hspace{0.02in}\varphi_{L}%
\,\bar{\triangleleft}\,\partial^{B}\hspace{0.01in}(\mathcal{\bar{L}}%
_{\partial}){^{E}}_{\hspace{-0.01in}C}\circledast\partial_{A}\,\bar
{\triangleright}\,\varphi_{R}^{\ast}\hspace{0.02in}g^{CD}\nonumber\\
&  \,-\frac{1}{2\hspace{0.01in}c^{2}}\hspace{0.01in}g_{AE}\hspace
{0.02in}\varphi_{L}\,\bar{\triangleleft}\,\partial_{t}\circledast
(\mathcal{\bar{L}}_{\partial}){^{E}}_{\hspace{-0.01in}C}\hspace{0.01in}%
\partial_{t}\,\bar{\triangleright}\,\varphi_{R}^{\ast}\hspace{0.02in}g^{CD}.
\end{align}

We can get further $q$-de\-formed continuity equations from the above results
using the rules in Eq.~(\ref{ErsRegQKleGor}) of Chap.~\ref{KapCharKon}.

\appendix

\section{Star-products\label{KapQuaZeiEle}}

The three-di\-men\-sion\-al $q$-de\-formed Euclidean space $\mathbb{R}_{q}%
^{3}$ has the generators $X^{+}$, $X^{3}$, and $X^{-}$, subject to the
following commutation relations \cite{Lorek:1997eh}:%
\begin{align}
X^{3}X^{+} &  =q^{2}X^{+}X^{3},\nonumber\\
X^{3}X^{-} &  =q^{-2}X^{-}X^{3},\nonumber\\
X^{-}X^{+} &  =X^{+}X^{-}+(q-q^{-1})\hspace{0.01in}X^{3}X^{3}%
.\label{RelQuaEukDre}%
\end{align}
We can extend the algebra of $\mathbb{R}_{q}^{3}$ by a time element $X^{0}$,
which commutes with the generators $X^{+}$, $X^{3}$, and $X^{-}$
\cite{Wachter:2020A}:%
\begin{equation}
X^{0}X^{A}=X^{A}X^{0},\text{\qquad}A\in\{+,3,-\}.\label{ZusRelExtDreEukQUa}%
\end{equation}
In the following, $\mathbb{R}_{q}^{3,t}$ denotes the algebra generated by
$X^{i}$ with $i\in\{0,+,3,-\}$.

There is a $q$-ver\-sion of the three-di\-men\-sion\-al Euclidean metric
$g^{AB}$ with its inverse $g_{AB}$ \cite{Lorek:1997eh} (rows and columns are
arranged in the order $+,3,-$):%
\begin{equation}
g_{AB}=g^{AB}=\left(
\begin{array}
[c]{ccc}%
0 & 0 & -\hspace{0.01in}q\\
0 & 1 & 0\\
-\hspace{0.01in}q^{-1} & 0 & 0
\end{array}
\right)  .\label{MetDreiDim}%
\end{equation}
We can use the $q$-de\-formed metric to raise and lower indices:%
\begin{equation}
X_{A}=g_{AB}\hspace{0.01in}X^{B},\qquad X^{A}=g^{AB}X_{B}.\label{HebSenInd}%
\end{equation}

The algebra $\mathbb{R}_{q}^{3,t}$ has a semilinear, involutive, and
anti-mul\-ti\-plica\-tive mapping, which we call \textit{quantum space
conjugation}. If we indicate conjugate elements of a quantum space by a
bar,\footnote{A bar over a complex number indicates complex conjugation.} we
can write the properties of quantum space conjugation as follows
($\alpha,\beta\in\mathbb{C}$ and $u,v\in\mathbb{R}_{q}^{3,t}$):%
\begin{equation}
\overline{\alpha\,u+\beta\,v}=\overline{\alpha}\,\overline{u}+\overline{\beta
}\,\overline{v},\quad\overline{\overline{u}}=u,\quad\overline{u\,v}%
=\overline{v}\,\overline{u}.
\end{equation}
The commutation relations in Eq.~(\ref{RelQuaEukDre}) and
Eq.~(\ref{ZusRelExtDreEukQUa}) are invariant under conjugation if the
following applies \cite{Wachter:2020A}:%
\begin{equation}
\overline{X^{A}}=X_{A}=g_{AB}\hspace{0.01in}X^{B},\qquad\overline{X^{0}}%
=X_{0}.\label{ConSpaKoo}%
\end{equation}

We can write each element $F\in$ $R_{q}^{3,t}$ as an expansion in terms of
nor\-mal-or\-dered monomials\textbf{\ }(\textit{Poincar\'{e}-Birkhoff-Witt
property}):%
\begin{equation}
F=\sum\limits_{n_{+},\ldots,\hspace{0.01in}n_{0}}a_{\hspace{0.01in}n_{+}%
\ldots\hspace{0.01in}n_{0}}\,(X^{+})^{n_{+}}(X^{3})^{n_{3}}(X^{-})^{n_{-}%
}(X^{0})^{n_{0}},\quad\quad a_{\hspace{0.01in}n_{+}\ldots\hspace{0.01in}n_{0}%
}\in\mathbb{C}.
\end{equation}
There is a vector space isomorphism%
\begin{equation}
\mathcal{W}:\mathbb{C}[\hspace{0.01in}x^{+},x^{3},x^{-},t\hspace
{0.01in}]\rightarrow\mathbb{R}_{q}^{3,t}\label{VecRauIsoInv}%
\end{equation}
with%
\begin{equation}
\mathcal{W}\left(  (x^{+})^{n_{+}}(x^{3})^{n_{3}}(x^{-})^{n_{-}}%
t^{\hspace{0.01in}n_{0}}\right)  =(X^{+})^{n_{+}}(X^{3})^{n_{3}}(X^{-}%
)^{n_{-}}(X^{0})^{n_{0}}.\label{StePro0}%
\end{equation}
In general, we have%
\begin{equation}
\mathbb{C}[\hspace{0.01in}x^{+},x^{3},x^{-},t\hspace{0.01in}]\ni f\mapsto
F\in\mathbb{R}_{q}^{3,t},
\end{equation}
where%
\begin{align}
f &  =\sum\limits_{n_{+},\ldots,\hspace{0.01in}n_{0}}a_{\hspace{0.01in}%
n_{+}\ldots\hspace{0.01in}n_{0}}\,(x^{+})^{n_{+}}(x^{3})^{n_{3}}(x^{-}%
)^{n_{-}}t^{\hspace{0.01in}n_{0}},\nonumber\\
F &  =\sum\limits_{n_{+},\ldots,\hspace{0.01in}n_{0}}a_{\hspace{0.01in}%
n_{+}\ldots\hspace{0.01in}n_{0}}\,(X^{+})^{n_{+}}(X^{3})^{n_{3}}(X^{-}%
)^{n_{-}}(X^{0})^{n_{0}}.\label{AusFfNorOrd}%
\end{align}
The vector space isomorphism $\mathcal{W}$ is nothing else than the
\textit{Moyal-Weyl mapping} giving an operator $F$ to a com\-plex-val\-ued
function $f$
\cite{Bayen:1977ha,1997q.alg.....9040K,Madore:2000en,Moyal:1949sk}.

We can extend this vector space isomorphism to an algebra isomorphism. To this
end, we introduce the so-called \textit{star-prod\-uct},\textit{\ }subject to
the following homomorphism condition:%
\begin{equation}
\mathcal{W}\left(  f\circledast g\right)  =\mathcal{W}\left(  f\right)
\cdot\mathcal{W}\left(  \hspace{0.01in}g\right)  .\label{HomBedWeyAbb}%
\end{equation}
Since the Mo\-yal-Weyl mapping is invertible, we can write the star-prod\-uct
as follows:%
\begin{equation}
f\circledast g=\mathcal{W}^{\hspace{0.01in}-1}\big (\,\mathcal{W}\left(
f\right)  \cdot\mathcal{W}\left(  \hspace{0.01in}g\right)
\big ).\label{ForStePro}%
\end{equation}

The product of two nor\-mal-or\-dered monomials can again be written as an
expansion in terms of nor\-mal-or\-dered monomials (see
Ref.~\cite{Wachter:2002A} for details):%
\begin{equation}
(X^{+})^{n_{+}}\ldots\hspace{0.01in}(X^{0})^{n_{0}}\cdot(X^{+})^{m_{+}}%
\ldots\hspace{0.01in}(X^{0})^{m_{0}}=\sum_{\underline{k}\hspace{0.01in}%
=\hspace{0.01in}0}B_{\underline{k}}^{\hspace{0.01in}\underline{n}%
,\underline{m}}\,(X^{+})^{k_{+}}\ldots\hspace{0.01in}(X^{0})^{k_{0}%
}.\label{EntProMon}%
\end{equation}
This expansion leads to a general formula for the star-prod\-uct of two power
series in commutative space-time coordinates ($\lambda=q-q^{-1}$):\
\begin{gather}
f(\mathbf{x},t)\circledast g(\mathbf{x},t)=\nonumber\\
\sum_{k\hspace{0.01in}=\hspace{0.01in}0}^{\infty}\lambda^{k}\hspace
{0.01in}\frac{(x^{3})^{2k}}{[[k]]_{q^{4}}!}\,q^{2(\hat{n}_{3}\hspace
{0.01in}\hat{n}_{+}^{\prime}+\,\hat{n}_{-}\hat{n}_{3}^{\prime})}%
D_{q^{4},\hspace{0.01in}x^{-}}^{k}f(\mathbf{x},t)\,D_{q^{4},\hspace
{0.01in}x^{\prime+}}^{k}g(\mathbf{x}^{\prime},t)\big|_{x^{\prime}%
\rightarrow\hspace{0.01in}x}.\label{StaProForExp}%
\end{gather}
The expression\ above depends on the operators%
\begin{equation}
\hat{n}_{A}=x^{A}\frac{\partial}{\partial x^{A}}\label{NOpeDef}%
\end{equation}
and the so-called Jackson derivatives \cite{Jackson:1910yd}:%
\begin{equation}
D_{q^{k},\hspace{0.01in}x}\,f=\frac{f(q^{k}x)-f(x)}{q^{k}x-x}.
\end{equation}
Moreover, the $q$-numbers are given by%
\begin{equation}
\lbrack\lbrack a]]_{q}=\frac{1-q^{a}}{1-q},
\end{equation}
and the $q$-fac\-to\-ri\-als are defined in complete analogy to the undeformed
case:%
\begin{equation}
\lbrack\lbrack\hspace{0.01in}n]]_{q}!=[[1]]_{q}\hspace{0.01in}[[2]]_{q}%
\ldots\lbrack\lbrack\hspace{0.01in}n-1]]_{q}\hspace{0.01in}[[\hspace
{0.01in}n]]_{q},\qquad\lbrack\lbrack0]]_{q}!=1.\label{qFakDef}%
\end{equation}

We require that $\mathcal{W}^{\hspace{0.01in}-1}$ is a $\ast$-al\-ge\-bra
homomorphism. This way, we can define a new conjugation on the commutative
space-time algebra\textbf{\ }$\mathbb{C}[\hspace{0.01in}x^{+},x^{3}%
,x^{-},t\hspace{0.01in}]$:%
\begin{equation}
\mathcal{W}(\hspace{0.01in}\overline{f}\hspace{0.01in})=\overline
{\mathcal{W}(f)}\qquad\Leftrightarrow\text{\qquad}\overline{f}=\mathcal{W}%
^{-1}\big (\hspace{0.01in}\overline{\mathcal{W}(f)}\hspace{0.01in}%
\big ).\label{ConAlgIso}%
\end{equation}
It follows from Eq.~(\ref{ConSpaKoo})\ and Eq.~(\ref{ConAlgIso}) that $\bar
{f}$ takes the following form \cite{Wachter:2007A,Wachter:2020A}:%
\begin{align}
\overline{f(\mathbf{x},t)} &  =\sum\nolimits_{\underline{n}}\bar{a}%
_{n_{+},n_{3},n_{-},n_{0}}\,(-\hspace{0.01in}q\hspace{0.01in}x^{-})^{n_{+}%
}(\hspace{0.01in}x^{3})^{n_{3}}(-\hspace{0.01in}q^{-1}x^{+})^{n_{-}}%
\hspace{0.01in}t^{\hspace{0.01in}n_{0}}\nonumber\\
&  =\sum\nolimits_{\underline{n}}(-\hspace{0.01in}q)^{n_{-}-\hspace
{0.02in}n_{+}}\hspace{0.01in}\bar{a}_{n_{-},n_{3},n_{+},n_{0}}\,(\hspace
{0.01in}x^{+})^{n_{+}}(\hspace{0.01in}x^{3})^{n_{3}}(\hspace{0.01in}%
x^{-})^{n_{-}}\hspace{0.01in}t^{\hspace{0.01in}n_{0}}\nonumber\\
&  =\bar{f}(\mathbf{x},t).\label{KonPotReiKom}%
\end{align}
The coefficient $\bar{a}_{n_{+},n_{3},n_{-},n_{0}}$ is the complex conjugate
of $a_{n_{+},n_{3},n_{-},n_{0}}$. The star-pro\-duct behaves as follows under
quantum space conjugation:%
\begin{equation}
\overline{f\circledast g}=\overline{g}\circledast\overline{f}%
.\label{KonEigSteProFkt}%
\end{equation}

\section{Partial derivatives and integrals\label{KapParDer}}

There are partial derivatives for $q$-de\-formed space-time coordinates
\cite{CarowWatamura:1990zp,Wess:1990vh}. These $q$-de\-formed partial
derivatives satisfy the same commutation relations as the
covariant coordinate generators $X_{i}$:%
\begin{gather}
\partial_{0}\hspace{0.01in}\partial_{+}=\hspace{0.01in}\partial_{+}%
\hspace{0.01in}\partial_{0},\quad\partial_{0}\hspace{0.01in}\partial
_{-}=\hspace{0.01in}\partial_{-}\hspace{0.01in}\partial_{0},\quad\partial
_{0}\hspace{0.01in}\partial_{3}=\partial_{3}\hspace{0.01in}\partial
_{0},\nonumber\\
\partial_{+}\hspace{0.01in}\partial_{3}=q^{2}\partial_{3}\hspace
{0.01in}\partial_{+},\quad\partial_{3}\hspace{0.01in}\partial_{-}%
=\hspace{0.01in}q^{2}\partial_{-}\hspace{0.01in}\partial_{3},\nonumber\\
\partial_{+}\hspace{0.01in}\partial_{-}-\partial_{-}\hspace{0.01in}%
\partial_{+}=\hspace{0.01in}\lambda\hspace{0.01in}\partial_{3}\hspace
{0.01in}\partial_{3}.
\end{gather}
The commutation relations above are invariant under conjugation if the
following conjugation properties hold:\footnote{The indices of partial
derivatives are raised and lowered in the same way as those of coordinates
[see Eq.~(\ref{HebSenInd}) in Chap.~\ref{KapQuaZeiEle}].}%
\begin{equation}
\overline{\partial_{A}}=-\hspace{0.01in}\partial^{A}=-g^{AB}\partial
_{B},\qquad\overline{\partial_{0}}=-\hspace{0.01in}\partial^{0}=-\hspace
{0.01in}\partial_{0}.\label{KonAbl}%
\end{equation}

There are two ways of commuting $q$-de\-formed partial derivatives with
$q$-de\-formed space-time coordinates. Concretely, we have the following
$q$-de\-formed Leibniz rules
\cite{CarowWatamura:1990zp,Wess:1990vh,Wachter:2020A}:%
\begin{align}
\partial_{B}X^{A} &  =\delta_{B}^{A}+q^{4}\hat{R}{^{AC}}_{BD}\,X^{D}%
\partial_{C},\nonumber\\
\partial_{A}X^{0} &  =X^{0}\hspace{0.01in}\partial_{A},\nonumber\\
\partial_{0}\hspace{0.01in}X^{A} &  =X^{A}\hspace{0.01in}\partial
_{0},\nonumber\\
\partial_{0}\hspace{0.01in}X^{0} &  =1+X^{0}\hspace{0.01in}\partial
_{0}.\label{DifKalExtEukQuaDreUnk}%
\end{align}
$\hat{R}{^{AC}}_{BD}$ denotes the vector representation of the R-matrix for
the three-di\-men\-sion\-al $q$-de\-formed Euclidean space. Introducing
$\hat{\partial}_{A}=q^{6}\partial_{A}$ and $\hat{\partial}_{0}=\partial_{0}$,
we can write the Leibniz rules of the second differential calculus in the
following form:%
\begin{align}
\hat{\partial}_{B}\hspace{0.01in}X^{A} &  =\delta_{B}^{A}+q^{-4}(\hat{R}%
^{-1}){^{AC}}_{BD}\,X^{D}\hat{\partial}_{C},\nonumber\\
\hat{\partial}_{A}\hspace{0.01in}X^{0} &  =X^{0}\hspace{0.01in}\hat{\partial
}_{A},\nonumber\\
\hat{\partial}_{0}\hspace{0.01in}X^{A} &  =X^{A}\hspace{0.01in}\hat{\partial
}_{0},\nonumber\\
\hat{\partial}_{0}\hspace{0.01in}X^{0} &  =1+X^{0}\hspace{0.01in}\hat
{\partial}_{0}.\label{DifKalExtEukQuaDreKon}%
\end{align}

Using the Leibniz rules in Eq.$~$(\ref{DifKalExtEukQuaDreUnk}) or
Eq.$~$(\ref{DifKalExtEukQuaDreKon}), we can calculate how partial derivatives
act on nor\-mal-or\-dered monomials of noncommutative coordinates. We can
carry over these actions to commutative coordinate monomials with the help of
the Mo\-yal-Weyl mapping:%
\begin{equation}
\partial^{i}\triangleright(x^{+})^{n_{+}}(x^{3})^{n_{3}}(x^{-})^{n_{-}%
}t^{\hspace{0.01in}n_{0}}=\mathcal{W}^{\hspace{0.01in}-1}\big (\partial
^{i}\triangleright(X^{+})^{n_{+}}(X^{3})^{n_{3}}(X^{-})^{n_{-}}(X^{0})^{n_{0}%
}\big ).
\end{equation}
Since the Mo\-yal-Weyl mapping is linear, we can apply the action above to a
power series in commutative space-time coordinates:%
\begin{equation}
\partial^{i}\triangleright f(\mathbf{x},t)=\mathcal{W}^{\hspace{0.01in}%
-1}\big (\partial^{i}\triangleright\mathcal{W}(f(\mathbf{x},t))\big ).
\end{equation}

If we use the normal-ordered monomials in Eq.~(\ref{StePro0}) of the previous
chapter, the Leibniz rules in Eq.~(\ref{DifKalExtEukQuaDreUnk})\ lead to the
following operator representations \cite{Bauer:2003}:%
\begin{align}
\partial_{+}\triangleright f(\mathbf{x},t) &  =D_{q^{4},\hspace{0.01in}x^{+}%
}f(\mathbf{x},t),\nonumber\\
\partial_{3}\triangleright f(\mathbf{x},t) &  =D_{q^{2},\hspace{0.01in}x^{3}%
}f(q^{2}x^{+},x^{3},x^{-},t),\nonumber\\
\partial_{-}\triangleright f(\mathbf{x},t) &  =D_{q^{4},\hspace{0.01in}x^{-}%
}f(x^{+},q^{2}x^{3},x^{-},t)+\lambda\hspace{0.01in}x^{+}D_{q^{2}%
,\hspace{0.01in}x^{3}}^{2}f(\mathbf{x},t).\label{UnkOpeDarAbl}%
\end{align}
The derivative $\partial_{0}$, however, is represented on the commutative
space-time algebra by the standard time derivative:%
\begin{equation}
\partial_{0}\triangleright\hspace{-0.01in}f(\mathbf{x},t)=\frac{\partial
f(\mathbf{x},t)}{\partial t}.\label{OpeDarZeiAblExtQuaEuk}%
\end{equation}

Using the Leibniz rules in Eq.$~$(\ref{DifKalExtEukQuaDreKon}), we get
operator representations for the partial derivatives $\hat{\partial}_{i}$. The
Leibniz rules in Eq.$~$(\ref{DifKalExtEukQuaDreUnk}) and Eq.$~$%
(\ref{DifKalExtEukQuaDreKon}) are transformed into each other by the following
substitutions:%
\begin{gather}
q\rightarrow q^{-1},\quad X^{-}\rightarrow X^{+},\quad X^{+}\rightarrow
X^{-},\nonumber\\
\partial^{\hspace{0.01in}+}\rightarrow\hat{\partial}^{\hspace{0.01in}-}%
,\quad\partial^{\hspace{0.01in}-}\rightarrow\hat{\partial}^{\hspace{0.01in}%
+},\quad\partial^{\hspace{0.01in}3}\rightarrow\hat{\partial}^{\hspace
{0.01in}3},\quad\partial^{\hspace{0.01in}0}\rightarrow\hat{\partial}%
^{\hspace{0.01in}0}.\label{UebRegGedUngAblDreQua}%
\end{gather}
Thus, we obtain the operator representations of the partial derivatives
$\hat{\partial}_{A}$ from those of the partial derivatives $\partial_{A}$ [cf.
Eq.~(\ref{UnkOpeDarAbl})] if we replace $q$ by $q^{-1}$ and exchange the
indices $+$ and $-$:%
\begin{align}
\hat{\partial}_{-}\,\bar{\triangleright}\,f(\mathbf{x},t) &  =D_{q^{-4}%
,\hspace{0.01in}x^{-}}f(\mathbf{x},t),\nonumber\\
\hat{\partial}_{3}\,\bar{\triangleright}\,f(\mathbf{x},t) &  =D_{q^{-2}%
,\hspace{0.01in}x^{3}}f(q^{-2}x^{-},x^{3},x^{+},t),\nonumber\\
\hat{\partial}_{+}\,\bar{\triangleright}\,f(\mathbf{x},t) &  =D_{q^{-4}%
,\hspace{0.01in}x^{+}}f(x^{-},q^{-2}x^{3},x^{+},t)-\lambda\hspace{0.01in}%
x^{-}D_{q^{-2},\hspace{0.01in}x^{3}}^{2}f(\mathbf{x},t).\label{KonOpeDarAbl}%
\end{align}
Once again, $\hat{\partial}_{0}$ is represented on the commutative space-time
algebra by the standard time derivative:%
\begin{equation}
\hat{\partial}_{0}\,\bar{\triangleright}\,f(\mathbf{x},t)=\frac{\partial
f(\mathbf{x},t)}{\partial t}.\label{OpeDarZeiAblExtQuaEukKon}%
\end{equation}
Due to the substitutions given in\ Eq.~(\ref{UebRegGedUngAblDreQua}), the
actions in Eqs.~(\ref{KonOpeDarAbl}) and (\ref{OpeDarZeiAblExtQuaEukKon})
refer to nor\-mal-or\-dered monomials different from those in
Eq.~(\ref{StePro0}) of the previous chapter:%
\begin{equation}
\widetilde{\mathcal{W}}\left(  t^{\hspace{0.01in}n_{0}}(x^{+})^{n_{+}}%
(x^{3})^{n_{3}}(x^{-})^{n_{-}}\right)  =(X^{0})^{n_{0}}(X^{-})^{n_{-}}%
(X^{3})^{n_{3}}(X^{+})^{n_{+}}.\label{UmNor}%
\end{equation}

We can also commute $q$-de\-formed partial derivatives from the \textit{right}
side of a nor\-mal-or\-dered monomial to the left side using the Leibniz
rules. This way, we get\ the so-called \textit{right}-re\-pre\-sen\-ta\-tions
of partial derivatives, for which we write $f\,\bar{\triangleleft}%
\,\partial^{i}$ or $f\triangleleft\hat{\partial}^{i}$. Conjugation transforms
left actions of partial derivatives into right actions and vice versa
\cite{Bauer:2003}:%
\begin{align}
\overline{\partial^{i}\triangleright f} &  =-\bar{f}\,\bar{\triangleleft
}\,\partial_{i}, & \overline{f\,\bar{\triangleleft}\,\partial^{i}} &
=-\hspace{0.01in}\partial_{i}\triangleright\bar{f},\nonumber\\
\overline{\hat{\partial}^{i}\,\bar{\triangleright}\,f} &  =-\bar
{f}\triangleleft\hat{\partial}_{i}, & \overline{f\triangleleft\hat{\partial
}^{i}} &  =-\hspace{0.01in}\hat{\partial}_{i}\,\bar{\triangleright}\,\bar
{f}.\label{RegConAbl}%
\end{align}

The operator representations in Eqs.~(\ref{UnkOpeDarAbl}) and
(\ref{KonOpeDarAbl}) consist of two terms which we call $\partial
_{\operatorname*{cla}}^{A}$ and $\partial_{\operatorname*{cor}}^{A}$:%
\begin{equation}
\partial^{A}\triangleright F=\left(  \partial_{\operatorname*{cla}}%
^{A}+\partial_{\operatorname*{cor}}^{A}\right)  \triangleright F.
\end{equation}
In the undeformed limit $q\rightarrow1$, $\partial_{\operatorname*{cla}}^{A}$
becomes a standard partial derivative, and $\partial_{\operatorname*{cor}%
}^{A}$ disappears. We get a solution to the difference equation $\partial
^{A}\triangleright F=f$ by using the following formula
\cite{Wachter:2004A}:%
\begin{align}
F &  =(\partial^{A})^{-1}\triangleright f=\left(  \partial
_{\operatorname*{cla}}^{A}+\partial_{\operatorname*{cor}}^{A}\right)
^{-1}\triangleright f\nonumber\\
&  =\sum_{k\hspace{0.01in}=\hspace{0.01in}0}^{\infty}\left[  -(\partial
_{\operatorname*{cla}}^{A})^{-1}\partial_{\operatorname*{cor}}^{A}\right]
^{k}(\partial_{\operatorname*{cla}}^{A})^{-1}\triangleright f.
\end{align}
Applying the above formula to the operator representations in
Eq.~(\ref{UnkOpeDarAbl}), we get%
\begin{align}
(\partial_{+})^{-1}\triangleright f(\mathbf{x},t) &  =D_{q^{4},\hspace
{0.01in}x^{+}}^{-1}f(\mathbf{x},t),\nonumber\\
(\partial_{3})^{-1}\triangleright f(\mathbf{x},t) &  =D_{q^{2},\hspace
{0.01in}x^{3}}^{-1}f(q^{-2}x^{+},x^{3},x^{-},t),\label{InvParAbl1}%
\end{align}
and%
\begin{gather}
(\partial_{-})^{-1}\triangleright f(\mathbf{x},t)=\nonumber\\
=\sum_{k\hspace{0.01in}=\hspace{0.01in}0}^{\infty}q^{2k\left(  k\hspace
{0.01in}+1\right)  }\left(  -\lambda\,x^{+}D_{q^{4},\hspace{0.01in}x^{-}}%
^{-1}D_{q^{2},\hspace{0.01in}x^{3}}^{2}\right)  ^{k}D_{q^{4},\hspace
{0.01in}x^{-}}^{-1}f(x^{+},q^{-2\left(  k\hspace{0.01in}+1\right)  }%
x^{3},x^{-},t).\label{InvParAbl2}%
\end{gather}
Note that $D_{q,\hspace{0.01in}x}^{-1}$ stands for a Jackson integral, with
$x$ being the variable of integration \cite{Jackson:1908}. The explicit form
of this Jackson integral depends on its limits of integration and the value
for the deformation parameter $q$. If $x>0$ and $q>1$, for example, the
following applies:%
\begin{equation}
\int_{0}^{\hspace{0.01in}x}\text{d}_{q}z\hspace{0.01in}f(z)=(q-1)\hspace
{0.01in}x\sum_{j=1}^{\infty}q^{-j}f(q^{-j}x).
\end{equation}
%
The integral for the time coordinate has the same form as in the undeformed
case [cf. Eq.~(\ref{OpeDarZeiAblExtQuaEuk})]:%
\begin{equation}
(\partial_{0})^{-1}\triangleright f(\mathbf{x},t)\hspace{0.01in}=\int
\text{d}t\,f(\mathbf{x},t).
\end{equation}

The considerations above also apply to the partial derivatives with a hat.
However, we can obtain the representations of $\hat{\partial}_{i}$ from those
of $\partial_{i}$ if we replace $q$ with $q^{-1}$ and exchange the indices $+$
and $-$. Applying these substitutions to the expressions in
Eqs.~(\ref{InvParAbl1}) and (\ref{InvParAbl2}), we immediately get the
corresponding results for the partial derivatives $\hat{\partial}_{i}$.

By successively applying the integral operators given in
Eqs.~(\ref{InvParAbl1}) and (\ref{InvParAbl2}), we can define an integration
over all space \cite{Wachter:2004A,Wachter:2007A}:%
\begin{equation}
\int_{-\infty}^{+\infty}\text{d}_{q}^{3}x\,f(x^{+},x^{3},x^{-})=(\partial
_{-})^{-1}\big |_{-\infty}^{+\infty}\,(\partial_{3})^{-1}\big |_{-\infty
}^{+\infty}\,(\partial_{+})^{-1}\big |_{-\infty}^{+\infty}\triangleright
f.\label{DefIntSpa}%
\end{equation}
On the right-hand side of the above relation, the different integral operators
can be simplified to Jackson integrals \cite{Wachter:2004A,Jambor:2004ph}:%
\begin{equation}
\int\text{d}_{q}^{3}x\,f=\int_{-\infty}^{+\infty}\text{d}_{q}^{3}%
x\,f(\mathbf{x})=D_{q^{2},\hspace{0.01in}x^{-}}^{-1}\big |_{-\infty}^{+\infty
}\,D_{q,x^{3}}^{-1}\big |_{-\infty}^{+\infty}\,D_{q^{2},\hspace{0.01in}x^{+}%
}^{-1}\big |_{-\infty}^{+\infty}\,f(\mathbf{x}).
\end{equation}
In the above formula, the Jackson integrals refer to a smaller $q$-lat\-tice.
Using such a smaller $q$-lat\-tice ensures that the integral over all space is
a scalar with trivial braiding properties \cite{Kempf:1994yd}.

There are $q$-de\-formed versions of \textit{Stokes' theorem} for the
$q$-in\-te\-gral over all space \cite{Wachter:2007A,Jambor:2004ph}:%
\begin{align}
\int_{-\infty}^{+\infty}\text{d}_{q}^{3}x\,\partial^{A}\triangleright f &
=\int_{-\infty}^{+\infty}\text{d}_{q}^{3}x\,f\,\bar{\triangleleft}%
\,\partial^{A}=0,\nonumber\\
\int_{-\infty}^{+\infty}\text{d}_{q}^{3}x\,\hat{\partial}^{A}\,\bar
{\triangleright}\,f &  =\int_{-\infty}^{+\infty}\text{d}_{q}^{3}%
x\,f\triangleleft\hat{\partial}^{A}=0.\label{StoThe}%
\end{align}
The $q$-de\-formed Stokes' theorem implies rules for integration by parts:%
\begin{align}
\int_{-\infty}^{+\infty}\text{d}_{q}^{3}x\,f\circledast(\partial
^{A}\triangleright g) &  =\int_{-\infty}^{+\infty}\text{d}_{q}^{3}%
x\,(f\triangleleft\partial^{A})\circledast g,\nonumber\\
\int_{-\infty}^{+\infty}\text{d}_{q}^{3}x\,f\circledast(\hat{\partial}%
^{A}\,\bar{\triangleright}\,g) &  =\int_{-\infty}^{+\infty}\text{d}_{q}%
^{3}x\,(f\,\bar{\triangleleft}\,\hat{\partial}^{A})\circledast
g.\label{PatIntUneRaumInt}%
\end{align}
Finally, the $q$-in\-te\-gral over all space behaves as follows under quantum
space conjugation:%
\begin{equation}
\overline{\int_{-\infty}^{+\infty}\text{d}_{q}^{3}x\,f}=\int_{-\infty
}^{+\infty}\text{d}_{q}^{3}x\,\bar{f}.\label{KonEigVolInt}%
\end{equation}

\section{Exponentials and Translations\label{KapExp}}

\begin{figure}
[ptb]
\begin{center}
\centerline{\psfig{figure=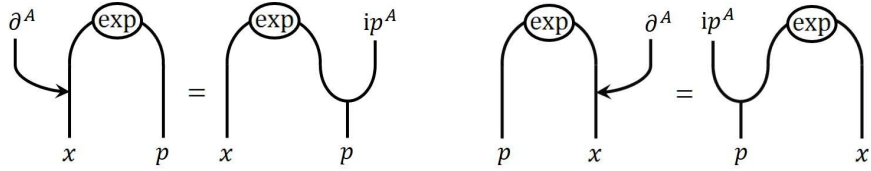,width=4.555in}}%
\caption{Eigenvalue equations of $q$-exponentials.}%
\label{Fig1}%
\end{center}
\end{figure}

An exponential of a $q$-de\-formed quantum space is an eigenfunction of each
$q$-de\-formed partial derivative
\cite{Majid:1993ud,Schirrmacher:1995,Wachter:2004ExpA}. In the following, we
consider $q$-de\-formed exponentials that are eigenfunctions for left actions
or right actions of partial derivatives:%
\begin{align}
\text{i}^{-1}\partial^{A}\triangleright\exp_{q}(\mathbf{x}|\text{i}\mathbf{p})
&  =\exp_{q}(\mathbf{x}|\text{i}\mathbf{p})\circledast p^{A},\nonumber\\
\exp_{q}(\text{i}^{-1}\mathbf{p}|\hspace{0.01in}\mathbf{x})\,\bar
{\triangleleft}\,\partial^{A}\text{i}^{-1} &  =p^{A}\circledast\exp
_{q}(\text{i}^{-1}\mathbf{p}|\hspace{0.01in}\mathbf{x}).\label{EigGl1N}%
\end{align}
The above eigenvalue equations are shown graphically in Fig.~\ref{Fig1}. The
$q$-ex\-po\-nen\-tials are defined by their eigenvalue equations and the
following normalization conditions:%
\begin{align}
\exp_{q}(\mathbf{x}|\text{i}\mathbf{p})|_{x\hspace{0.01in}=\hspace{0.01in}0}
&  =\exp_{q}(\mathbf{x}|\text{i}\mathbf{p})|_{p\hspace{0.01in}=\hspace
{0.01in}0}=1,\nonumber\\
\exp_{q}(\text{i}^{-1}\mathbf{p}|\hspace{0.01in}\mathbf{x})|_{x\hspace
{0.01in}=\hspace{0.01in}0} &  =\exp_{q}(\text{i}^{-1}\mathbf{p}|\hspace
{0.01in}\mathbf{x})|_{p\hspace{0.01in}=\hspace{0.01in}0}=1.\label{NorBedExp}%
\end{align}%

Using the operator representation in Eq.~(\ref{UnkOpeDarAbl}) of the last
chapter, we have found the following expressions for the $q$-ex\-ponen\-tials
of three-di\-men\-sion\-al Euclidean quantum space \cite{Wachter:2004ExpA}:%
\begin{align}
\exp_{q}(\text{i}^{-1}\mathbf{p}|\mathbf{x}) &  =\sum_{\underline{n}%
\,=\,0}^{\infty}\frac{(\text{i}^{-1}p^{+})^{n_{+}}(\text{i}^{-1}p^{3})^{n_{3}%
}(\text{i}^{-1}p^{-})^{n_{-}}(x_{-})^{n_{-}}(x_{3})^{n_{3}}(x_{+})^{n_{+}}%
}{[[\hspace{0.01in}n_{+}]]_{q^{4}}!\,[[\hspace{0.01in}n_{3}]]_{q^{2}%
}!\,[[\hspace{0.01in}n_{-}]]_{q^{4}}},\nonumber\\
\exp_{q}(\mathbf{x}|\text{i}\mathbf{p}) &  =\sum_{\underline{n}\,=\,0}%
^{\infty}\frac{(x^{+})^{n_{+}}(x^{3})^{n_{3}}(x^{-})^{n_{-}}(\text{i}%
p_{-})^{n_{-}}(\text{i}p_{3})^{n_{3}}(\text{i}p_{+})^{n_{+}}}{[[\hspace
{0.01in}n_{+}]]_{q^{4}}!\,[[\hspace{0.01in}n_{3}]]_{q^{2}}!\,[[\hspace
{0.01in}n_{-}]]_{q^{4}}!}.\label{ExpEukExp}%
\end{align}
If we substitute $q$ with $q^{-1}$ in both expressions of Eq.~(\ref{ExpEukExp}%
), we get two more $q$-ex\-ponen\-tials which we designate $\overline{\exp
}_{q}(x|$i$\mathbf{p})$ and $\overline{\exp}_{q}($i$^{-1}\mathbf{p}|x)$. We
obtain the eigenvalue equations and normalization conditions of these two
$q$-ex\-ponen\-tials by applying the following substitutions to
Eqs.~(\ref{EigGl1N}) and (\ref{NorBedExp}):%
\begin{equation}
\exp_{q}\rightarrow\hspace{0.01in}\overline{\exp}_{q},\qquad\triangleright
\,\rightarrow\,\bar{\triangleright},\qquad\bar{\triangleleft}\,\rightarrow
\,\triangleleft,\qquad\partial^{A}\rightarrow\hat{\partial}^{A}%
.\label{ErsRegQExp}%
\end{equation}

We can use $q$-ex\-ponen\-tials to calculate $q$-trans\-la\-tions
\cite{Chryssomalakos:1993zm}. If we replace the momentum coordinates in the
expressions for $q$-ex\-ponen\-tials with derivatives, it applies
\cite{Carnovale:1999,Majid:1993ud,Wachter:2007A}%
\begin{align}
\exp_{q}(x|\partial_{y})\triangleright g(\hspace{0.01in}y) &  =g(x\,\bar
{\oplus}\,y),\nonumber\\
\overline{\exp}_{q}(x|\hat{\partial}_{y})\,\bar{\triangleright}\,g(\hspace
{0.01in}y) &  =g(x\oplus y),\label{q-TayN}%
\end{align}
and%
\begin{align}
g(\hspace{0.01in}y)\,\bar{\triangleleft}\,\exp_{q}(-\hspace{0.01in}%
\partial_{y}|\hspace{0.01in}x) &  =g(\hspace{0.01in}y\,\bar{\oplus
}\,x),\nonumber\\
g(\hspace{0.01in}y)\triangleleft\hspace{0.01in}\overline{\exp}_{q}%
(-\hspace{0.01in}\hat{\partial}_{y}|\hspace{0.01in}x) &  =g(\hspace
{0.01in}y\oplus x).\label{q-TayRecN}%
\end{align}
In the case of the three-di\-men\-sion\-al $q$-de\-formed Euclidean space, for
example, we can get the following formula for calculating $q$-trans\-la\-tions
\cite{Wachter:2004phengl}:%
\begin{align}
f(\mathbf{x}\oplus\mathbf{y})= &  \sum_{i_{+}=\hspace{0.01in}0}^{\infty}%
\sum_{i_{3}=\hspace{0.01in}0}^{\infty}\sum_{i_{-}=\hspace{0.01in}0}^{\infty
}\sum_{k\hspace{0.01in}=\hspace{0.01in}0}^{i_{3}}\frac{(-q^{-1}\lambda
\lambda_{+})^{k}}{[[2k]]_{q^{-2}}!!}\frac{(x^{-})^{i_{-}}(x^{3})^{i_{3}%
-\hspace{0.01in}k}(x^{+})^{i_{+}+\hspace{0.01in}k}\,(\hspace{0.01in}y^{-}%
)^{k}}{[[i_{-}]]_{q^{-4}}!\,[[i_{3}-k]]_{q^{-2}}!\,[[i_{+}]]_{q^{-4}}%
!}\nonumber\\
&  \qquad\times\big (D_{q^{-4},\hspace{0.01in}y^{-}}^{i_{-}}D_{q^{-2}%
,\hspace{0.01in}y^{3}}^{i_{3}+\hspace{0.01in}k}\hspace{0.01in}D_{q^{-4}%
,\hspace{0.01in}y^{+}}^{i_{+}}f\big )(q^{2(k\hspace{0.01in}-\hspace
{0.01in}i_{3})}y^{-},q^{-2i_{+}}y^{3}).
\end{align}

\begin{figure}
[ptb]
\begin{center}
\includegraphics[width=0.40\textwidth]{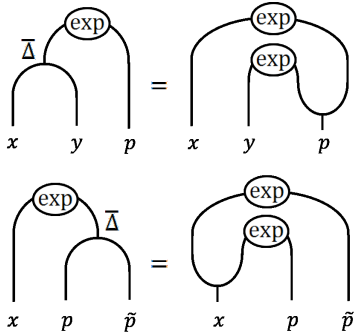}
\caption{Laws of addition for $q$-exponentials}%
\label{Fig2}%
\end{center}
\end{figure}

In analogy to the undeformed case, $q$-ex\-ponen\-tials satisfy the addition
theorems \cite{Majid:1993ud,Schirrmacher:1995,Wachter:2007A}%
\begin{align}
\exp_{q}(\mathbf{x}\,\bar{\oplus}\,\mathbf{y}|\text{i}\mathbf{p}) &  =\exp
_{q}(\mathbf{x}|\exp_{q}(\hspace{0.01in}\mathbf{y}|\text{i}\mathbf{p}%
)\circledast\text{i}\mathbf{p}),\nonumber\\
\exp_{q}(\text{i}\mathbf{x}|\mathbf{p}\,\bar{\oplus}\,\mathbf{p}^{\prime}) &
=\exp_{q}(\mathbf{x}\circledast\exp_{q}(\mathbf{x}|\hspace{0.01in}%
\text{i}\mathbf{p})|\hspace{0.01in}\text{i}\mathbf{p}^{\prime}%
),\label{AddTheExp}%
\end{align}
and%
\begin{align}
\overline{\exp}_{q}(\mathbf{x}\oplus\mathbf{y}|\text{i}\mathbf{p}) &
=\overline{\exp}_{q}(\mathbf{x}|\overline{\exp}_{q}(\hspace{0.01in}%
\mathbf{y}|\text{i}\mathbf{p})\circledast\text{i}\mathbf{p}),\nonumber\\
\overline{\exp}_{q}(\text{i}\mathbf{x}|\mathbf{p}\oplus\mathbf{p}^{\prime}) &
=\overline{\exp}_{q}(\mathbf{x}\circledast\overline{\exp}_{q}(\mathbf{x}%
|\text{i}\mathbf{p})|\hspace{0.01in}\text{i}\mathbf{p}^{\prime}).
\end{align}
We can obtain further addition theorems from the above identities by
substituting position coordinates with momentum coordinates and vice versa. In
Fig.~\ref{Fig2}, we have given graphic representations of the two addition
theorems in Eq.~(\ref{AddTheExp}).%

The $q$-de\-formed quantum spaces considered so far are so-called braided Hopf
algebras \cite{Majid:1996kd}. From this point of view, the two versions of
$q$-trans\-lations are nothing else than the representations of two braided
co-pro\-ducts $\underline{\Delta}$ and $\underline{\bar{\Delta}}$ on the
corresponding commutative coordinate algebras \cite{Wachter:2007A}:%
\begin{align}
f(\mathbf{x}\oplus\mathbf{y}) &  =((\mathcal{W}^{\hspace{0.01in}-1}%
\otimes\mathcal{W}^{\hspace{0.01in}-1})\circ\underline{\Delta})(\mathcal{W}%
(f)),\nonumber\\[0.08in]
f(\mathbf{x}\,\bar{\oplus}\,\mathbf{y}) &  =((\mathcal{W}^{\hspace{0.01in}%
-1}\otimes\mathcal{W}^{-1})\circ\underline{\bar{\Delta}})(\mathcal{W}%
(f)).\label{KonReaBraCop}%
\end{align}
The braided Hopf algebras have braided antipodes $\underline{S}$ and
$\underline{\bar{S}}$ as well. We can represent these antipodes on the
corresponding commutative algebras, too:%
\begin{align}
f(\ominus\,\mathbf{x}) &  =(\mathcal{W}^{\hspace{0.01in}-1}\circ\underline
{S}\hspace{0.01in})(\mathcal{W}(f)),\nonumber\\[0.08in]
f(\bar{\ominus}\,\mathbf{x}) &  =(\mathcal{W}^{\hspace{0.01in}-1}%
\circ\underline{\bar{S}}\hspace{0.01in})(\mathcal{W}(f)).\label{qInvDef}%
\end{align}
In the following, we refer to the operations in Eq.~(\ref{qInvDef})\ as
$q$\textit{-in\-ver\-sions}. In the case of the $q$-de\-formed Euclidean
space, for example, we have found the following operator representation for
$q$-in\-ver\-sions \cite{Wachter:2004phengl}:%
\begin{gather}
\hat{U}^{-1}f(\ominus\,\mathbf{x})=\nonumber\\
=\sum_{i=0}^{\infty}(-\hspace{0.01in}q\lambda\lambda_{+})^{i}\,\frac
{(x^{+}x^{-})^{i}}{[[2i]]_{q^{-2}}!!}\,q^{-2\hat{n}_{+}(\hat{n}_{+}%
+\hspace{0.01in}\hat{n}_{3})-2\hat{n}_{-}(\hat{n}_{-}+\hspace{0.01in}\hat
{n}_{3})-\hat{n}_{3}\hat{n}_{3}}\nonumber\\
\times D_{q^{-2},\hspace{0.01in}x^{3}}^{2i}\,f(-\hspace{0.01in}q^{2-4i}%
x^{-},-\hspace{0.01in}q^{1-2i}x^{3},-\hspace{0.01in}q^{2-4i}x^{+}).
\end{gather}
The operators $\hat{U}$ and $\hat{U}^{-1}$ act on a commutative function
$f(x^{+},x^{3},x^{-})$ as follows [cf. Eq.~(\ref{NOpeDef}) in
App.~\ref{KapQuaZeiEle}]:%
\begin{align}
\hat{U}f &  =\sum_{k\hspace{0.01in}=\hspace{0.01in}0}^{\infty}\left(
-\lambda\right)  ^{k}\frac{(x^{3})^{2k}}{[[k]]_{q^{-4}}!}\,q^{-2\hat{n}%
_{3}(\hat{n}_{+}+\hspace{0.01in}\hat{n}_{-}+\hspace{0.01in}k)}D_{q^{-4}%
,\hspace{0.01in}x^{+}}^{k}D_{q^{-4},\hspace{0.01in}x^{-}}^{k}f,\nonumber\\
\hat{U}^{-1}f &  =\sum_{k\hspace{0.01in}=\hspace{0.01in}0}^{\infty}\lambda
^{k}\hspace{0.01in}\frac{(x^{3})^{2k}}{[[k]]_{q^{4}}!}\,q^{2\hat{n}_{3}%
(\hat{n}_{+}+\hspace{0.01in}\hat{n}_{-}+\hspace{0.01in}k)}D_{q^{4}%
,\hspace{0.01in}x^{+}}^{k}D_{q^{4},\hspace{0.01in}x^{-}}^{k}f.
\end{align}

The braided co-prod\-ucts and braided antipodes satisfy the axioms (also see
Ref.~\cite{Majid:1996kd})%
\begin{align}
m\circ(\underline{S}\otimes\operatorname*{id})\circ\underline{\Delta} &
=m\circ(\operatorname*{id}\otimes\,\underline{S}\hspace{0.01in})\circ
\underline{\Delta}=\underline{\varepsilon},\nonumber\\
m\circ(\underline{\bar{S}}\otimes\operatorname*{id})\circ\underline
{\bar{\Delta}} &  =m\circ(\operatorname*{id}\otimes\,\underline{\bar{S}%
}\hspace{0.01in})\circ\underline{\bar{\Delta}}=\underline{\bar{\varepsilon}%
},\label{HopfVerAnfN}%
\end{align}
and%
\begin{align}
(\operatorname*{id}\otimes\,\underline{\varepsilon})\circ\underline{\Delta} &
=\operatorname*{id}=(\underline{\varepsilon}\otimes\operatorname*{id}%
)\circ\underline{\Delta},\nonumber\\
(\operatorname*{id}\otimes\,\underline{\bar{\varepsilon}})\circ\underline
{\bar{\Delta}} &  =\operatorname*{id}=(\underline{\bar{\varepsilon}}%
\otimes\operatorname*{id})\circ\underline{\bar{\Delta}}.\label{HopfAxi2}%
\end{align}
In the identities above, we denote multiplication on the braided Hopf algebra
by $m$. Both co-units $\underline{\varepsilon},\underline{\bar{\varepsilon}}$
of the two braided Hopf structures are linear mappings vanishing on the
coordinate generators:%
\begin{equation}
\varepsilon(X^{i})=\underline{\bar{\varepsilon}}(X^{i})=0.
\end{equation}
For this reason, we can represent the co-units $\underline{\varepsilon}$ and
$\underline{\bar{\varepsilon}}$ on a commutative coordinate algebra as
follows:%
\begin{equation}
\underline{\varepsilon}(\mathcal{W}(f))=\underline{\bar{\varepsilon}%
}(\mathcal{W}(f))=\left.  f(\mathbf{x})\right\vert _{x\hspace{0.01in}%
=\hspace{0.01in}0}=f(0).\label{ReaVerZopNeuEleKomAlg}%
\end{equation}
Next, we translate the Hopf algebra axioms in Eqs.~(\ref{HopfVerAnfN}) and
(\ref{HopfAxi2}) into corresponding rules for $q$-trans\-la\-tions and
$q$-in\-ver\-sions \cite{Wachter:2007A}, i.~e.%
\begin{align}
f((\ominus\,\mathbf{x})\oplus\mathbf{x}) &  =f(\mathbf{x}\oplus(\ominus
\,\mathbf{x}))=f(0),\nonumber\\
f((\bar{\ominus}\,\mathbf{x})\,\bar{\oplus}\,\mathbf{x}) &  =f(\mathbf{x}%
\,\bar{\oplus}\,(\bar{\ominus}\,\mathbf{x}))=f(0),\label{qAddN}%
\end{align}
and%
\begin{align}
f(\mathbf{x}\oplus\mathbf{y})|_{y\hspace{0.01in}=\hspace{0.01in}0} &
=f(\mathbf{x})=f(\mathbf{y}\oplus\mathbf{x})|_{y\hspace{0.01in}=\hspace
{0.01in}0},\nonumber\\
f(\mathbf{x}\,\bar{\oplus}\,\mathbf{y})|_{y\hspace{0.01in}=\hspace{0.01in}0}
&  =f(\mathbf{x})=f(\mathbf{y}\,\bar{\oplus}\,\mathbf{x})|_{y\hspace
{0.01in}=\hspace{0.01in}0}.\label{qNeuEle}%
\end{align}

\begin{figure}
[ptb]
\begin{center}
\includegraphics[width=0.55\textwidth]{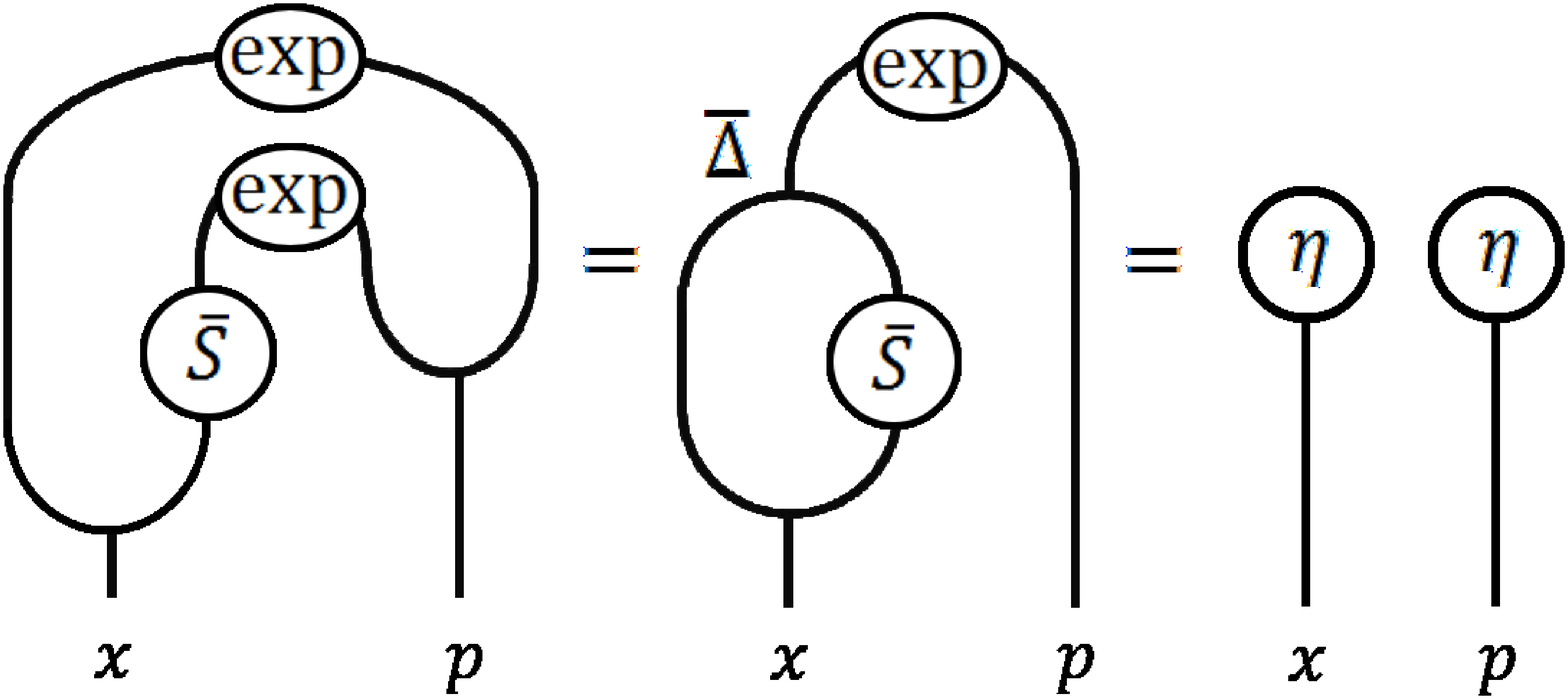}
\caption{Invertibility of $q$-exponentials}%
\label{Fig3}%
\end{center}
\end{figure}

Using $q$-in\-ver\-sions, we are also able to introduce inverse $q$%
-ex\-po\-nen\-tials:%
\begin{equation}
\exp_{q}(\bar{\ominus}\,\mathbf{x}|\text{i}\mathbf{p})=\exp_{q}(\text{i}%
\mathbf{x}|\text{{}}\bar{\ominus}\,\mathbf{p}).\label{InvExpAlgDefKom}%
\end{equation}
Due to the addition theorems and the normalization conditions of our
$q$-ex\-po\-nen\-tials, the following holds:%
\begin{equation}
\exp_{q}(\text{i}\mathbf{x}\circledast\exp_{q}(\bar{\ominus}\,\mathbf{x}%
|\hspace{0.01in}\text{i}\mathbf{p})\circledast\mathbf{p})=\exp_{q}%
(\mathbf{x}\,\bar{\oplus}\,(\bar{\ominus}\,\mathbf{x})|\hspace{0.01in}%
\text{i}\mathbf{p})=\exp_{q}(\mathbf{x}|\text{i}\mathbf{p})|_{x=0}=1.
\end{equation}
In Fig.~\ref{Fig3}, we have given graphic representations of these
identities.\footnote{You find some explanations of this sort of graphical
calculations in Ref.~\cite{Majid:2002kd}.}%
The conjugate $q$-ex\-po\-nen\-tials $\overline{\exp}_{q}$ are subject to
similar rules obtained from the above identities by using the following
substitutions:%
\begin{equation}
\exp_{q}\rightarrow\overline{\exp}_{q},\qquad\bar{\oplus}\,\rightarrow
\,\oplus,\qquad\bar{\ominus}\,\rightarrow\,\ominus.
\end{equation}

\begin{figure}
[ptbptb]
\begin{center}
\includegraphics[width=0.38\textwidth]{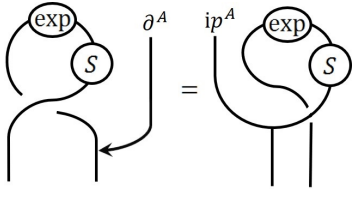}
\caption{Eigenvalue equation of twisted $q$-exponential}%
\label{Fig4}%
\end{center}
\end{figure}

Next, we describe another way of obtaining $q$-ex\-ponen\-tials. We exchange
the two tensor factors of a $q$-ex\-ponen\-tial using the inverse of the
so-called universal R-ma\-trix (also see the graphic representation in
Fig.~\ref{Fig4}):%
\begin{align}
\exp_{q}^{\ast}(\text{i}\mathbf{p}|\hspace{0.01in}\mathbf{x}) &  =\tau
\circ\lbrack(\mathcal{R}_{[2]}^{-1}\otimes\mathcal{R}_{[1]}^{-1}%
)\triangleright\exp_{q}(\text{i}\mathbf{x}|\hspace{-0.03in}\ominus
\hspace{-0.01in}\mathbf{p})],\nonumber\\
\exp_{q}^{\ast}(\mathbf{x}|\text{i}\mathbf{p}) &  =\tau\circ\lbrack
(\mathcal{R}_{[2]}^{-1}\otimes\mathcal{R}_{[1]}^{-1})\triangleright\exp
_{q}(\ominus\hspace{0.02in}\mathbf{p}|\hspace{0.01in}\text{i}\mathbf{x}%
)].\label{DuaExp2}%
\end{align}
In the expressions above, $\tau$ denotes the ordinary twist operator. One can
show that the new $q$-ex\-ponen\-tials satisfy the following eigenvalue
equations (see Fig.~\ref{Fig4}):%
\begin{align}
\exp_{q}^{\ast}(\text{i}\mathbf{p}|\hspace{0.01in}\mathbf{x})\triangleleft
\partial^{A} &  =\text{i}p^{A}\circledast\exp_{q}^{\ast}(\text{i}%
\mathbf{p}|\hspace{0.01in}\mathbf{x}),\nonumber\\
\partial^{A}\,\bar{\triangleright}\,\exp_{q}^{\ast}(\mathbf{x}|\text{i}%
^{-1}\mathbf{p}) &  =\exp_{q}^{\ast}(\mathbf{x}|\text{i}^{-1}\mathbf{p}%
)\circledast\text{i}p^{A}.\label{EigGleExpQueAbl}%
\end{align}

Similar considerations apply to the conjugate $q$-ex\-ponen\-tials. We only
need to modify Eqs.~(\ref{DuaExp2}) and (\ref{EigGleExpQueAbl}) by performing
the following substitutions:%
\begin{gather}
\exp_{q}^{\ast}\rightarrow\overline{\exp}_{q}^{\ast},\qquad\mathcal{R}%
_{[2]}^{-1}\otimes\mathcal{R}_{[1]}^{-1}\rightarrow\mathcal{R}_{[1]}%
\otimes\mathcal{R}_{[2]},\qquad\ominus\,\rightarrow\,\bar{\ominus},\nonumber\\
\bar{\triangleright}\,\rightarrow\,\triangleright,\qquad\triangleleft
\,\rightarrow\,\bar{\triangleleft},\qquad\partial^{A}\rightarrow\hat{\partial
}^{A}.
\end{gather}

The $q$-ex\-ponen\-tials in Eq.~(\ref{DuaExp2}) are related to the conjugate
$q$-ex\-ponen\-tials. To show this, we rewrite the eigenvalue equations in
(\ref{EigGleExpQueAbl}) using the identity $\hat{\partial}^{A}=q^{6}%
\partial^{A}$:%
\begin{align}
\exp_{q}^{\ast}(\text{i}\mathbf{p}|\hspace{0.01in}\mathbf{x})\triangleleft
\hat{\partial}^{A} &  =\text{i}q^{6}p^{A}\circledast\exp_{q}^{\ast}%
(\text{i}\mathbf{p}|\hspace{0.01in}\mathbf{x}),\nonumber\\
\hat{\partial}^{A}\,\bar{\triangleright}\,\exp_{q}^{\ast}(\mathbf{x}%
|\text{i}^{-1}\mathbf{p}) &  =\exp_{q}^{\ast}(\mathbf{x}|\text{i}%
^{-1}\mathbf{p})\circledast\text{i}q^{6}p^{A}.
\end{align}
These are the eigenvalue equations for $\overline{\exp}_{q}($i$^{-1}%
q^{6}\mathbf{p}|\mathbf{x})$ and $\overline{\exp}_{q}(\mathbf{x}|$%
i$q^{6}\mathbf{p})$, so the following identifications are valid:%
\begin{equation}
\exp_{q}^{\ast}(\text{i}\mathbf{p}|\mathbf{x})=\overline{\exp}_{q}%
(\text{i}^{-1}q^{6}\mathbf{p}|\mathbf{x}),\qquad\exp_{q}^{\ast}(\mathbf{x}%
|\hspace{0.01in}\text{i}^{-1}\mathbf{p})=\overline{\exp}_{q}(\mathbf{x}%
|\text{i}q^{6}\mathbf{p}).\label{IdeSteExpEpxKon1}%
\end{equation}
%


For the sake of completeness, we write down how the $q$-ex\-ponen\-tials of
the $q$-de\-formed Euclidean space behave under quantum space conjugation:%
\begin{align}
\overline{\exp_{q}(\mathbf{x}|\text{i}\mathbf{p})} &  =\exp_{q}(\text{i}%
^{-1}\mathbf{p}|\mathbf{x}), & \overline{\overline{\exp}_{q}(\mathbf{x}%
|\text{i}\mathbf{p})} &  =\overline{\exp}_{q}(\text{i}^{-1}\mathbf{p}%
|\mathbf{x}),\nonumber\\
\overline{\exp_{q}^{\ast}(\text{i}\mathbf{p}|\mathbf{x})} &  =\exp_{q}^{\ast
}(\mathbf{x}|\text{i}^{-1}\mathbf{p}), & \overline{\overline{\exp}_{q}^{\ast
}(\text{i}\mathbf{p}|\mathbf{x})} &  =\overline{\exp}_{q}^{\ast}%
(\mathbf{x}|\text{i}^{-1}\mathbf{p}).\label{KonEigExpQua}%
\end{align}

\section{Hopf structures and L-matrices\label{KapHofStr}}

The three-di\-men\-sio\-nal $q$-de\-formed Euclidean space $\mathbb{R}_{q}%
^{3}$ is a three-di\-men\-sio\-nal representation of the Drin\-feld-Jim\-bo
al\-ge\-bra $\mathcal{U}_{q}(\operatorname*{su}_{2})$. The latter is a
deformation of the universal enveloping algebra of the Lie algebra
$\operatorname*{su}_{2}$ \cite{Kulish:1983md}. Accordingly, the algebra
$\mathcal{U}_{q}(\operatorname*{su}_{2})$ has the three generators $T^{+}$,
$T^{-}$, and $T^{3}$, subject to the following relations \cite{Lorek:1993tq}:%
\begin{align}
q^{-1}\hspace{0.01in}T^{+}T^{-}-q\,T^{-}T^{+} &  =T^{3},\nonumber\\
q^{\hspace{0.01in}2}\hspace{0.01in}T^{3}T^{+}-q^{-2}\hspace{0.01in}T^{+}T^{3}
&  =(q+q^{-1})\hspace{0.01in}T^{+},\nonumber\\
q^{\hspace{0.01in}2}\hspace{0.01in}T^{-}T^{3}-q^{-2}\hspace{0.01in}T^{3}T^{-}
&  =(q+q^{-1})\hspace{0.01in}T^{-}.\label{VerRel1UqSU2}%
\end{align}
These relations are compatible with the following conjugation assignment:%
\begin{equation}
\overline{T^{+}}=q^{-2}T^{-},\qquad\overline{T^{-}}=q^{2}T^{+},\qquad
\overline{T^{3}}=T^{3}.
\end{equation}

The algebra of the $q$-de\-formed partial derivatives $\partial^{A}$,
$A\in\{+,3,-\}$, together with $\mathcal{U}_{q}(\operatorname*{su}_{2})$ form
the cross-prod\-uct algebra $\mathbb{R}_{q}^{3}\rtimes\hspace{0.01in}%
\mathcal{U}_{q}(\operatorname*{su}\nolimits_{2})$
\cite{majid-1993-34,Weixler:1993ph}. We know that the algebra $\mathbb{R}%
_{q}^{3}\rtimes\mathcal{U}_{q}(\operatorname*{su}\nolimits_{2})$ is a Hopf
algebra \cite{Klimyk:1997eb}. Accordingly, the $q$-de\-formed partial
derivatives as elements of $\mathbb{R}_{q}^{3}\rtimes\hspace{0.01in}%
\mathcal{U}_{q}(\operatorname*{su}\nolimits_{2})$ have a co-prod\-uct, an
antipode, and a co-unit.

There are two ways of choosing the Hopf structure of the $q$-de\-formed
partial derivatives. The two different co-pro\-ducts of the $q$-de\-formed
partial derivatives are related to the two versions of Leibniz rules given in
Eq.~(\ref{DifKalExtEukQuaDreUnk}) or Eq.~(\ref{DifKalExtEukQuaDreKon}). We can
generalize these Leibniz rules by introducing the L-ma\-tri\-ces
$\mathcal{L}_{\partial}$ and $\mathcal{\bar{L}}_{\partial}$ ($u\in
\mathbb{R}_{q}^{3}$):%
\begin{align}
\partial^{A}\hspace{0.01in}u &  =(\partial_{(1)}^{A}\triangleright
u)\,\partial_{(2)}^{A}=\partial^{A}\triangleright u\hspace{0.01in}%
+\big ((\mathcal{L}_{\partial}){^{A}}_{B}\triangleright u\big )\partial
^{B},\nonumber\\
\hat{\partial}^{A}\hspace{0.01in}u &  =(\hat{\partial}_{(\bar{1})}^{A}%
\,\bar{\triangleright}\,u)\,\hat{\partial}_{(\bar{2})}^{A}=\hat{\partial}%
^{A}\,\bar{\triangleright}\,u\hspace{0.01in}+\big ((\mathcal{\bar{L}%
}_{\partial}){^{A}}_{B}\triangleright u\big )\hat{\partial}^{B}%
.\label{AllVerRelParAblEle1}%
\end{align}
From the above identities, you can see that the two L-ma\-trices determine the
two co-prod\-ucts\footnote{We write the co-product in the so-called Sweedler
notation, i.~e. $\Delta(a)=a_{(1)}\otimes a_{(2)}.$} of the $q$-de\-formed
partial derivatives \cite{ogievetsky1992}:%
\begin{align}
\partial_{(1)}^{A}\otimes\partial_{(2)}^{A} &  =\partial^{A}\otimes
1+(\mathcal{L}_{\partial}){^{A}}_{B}\otimes\partial^{B},\nonumber\\
\hat{\partial}_{(1)}^{A}\otimes\hat{\partial}_{(2)}^{A} &  =\hat{\partial}%
^{A}\otimes1+(\mathcal{\bar{L}}_{\partial}){^{A}}_{B}\otimes\hat{\partial}%
^{B}.\label{KopParAllg}%
\end{align}

The entries of the two L-ma\-tri\-ces consist of generators of the Hopf
algebra $\mathcal{U}_{q}(\operatorname*{su}_{2})$ and powers of a unitary
scaling operator $\Lambda$ [also see Eq.~(\ref{SkaOpeWir})]. For this reason,
the L-ma\-tri\-ces can act on any element of $\mathbb{R}_{q}^{3}$. In this
respect, an element of $\mathbb{R}_{q}^{3}$ has trivial braiding if the
L-ma\-tri\-ces act on it as follows:%
\begin{align}
(\mathcal{L}_{\partial}){^{A}}_{B}\triangleright u &  =\delta_{B}^{A}%
\hspace{0.01in}u,\nonumber\\
(\mathcal{\bar{L}}_{\partial}){^{A}}_{B}\triangleright u &  =\delta_{B}%
^{A}\hspace{0.01in}u.\label{trivBrai}%
\end{align}
Moreover, the two L-ma\-tri\-ces transform into each other by conjugation:%
\begin{equation}
\overline{(\mathcal{L}_{\partial}){^{A}}_{B}}=g_{AC}\hspace{0.01in}%
(\mathcal{\bar{L}}_{\partial}){^{C}}_{D}\hspace{0.01in}g^{DB},\qquad
\overline{(\mathcal{\bar{L}}_{\partial}){^{A}}_{B}}=g_{AC}\hspace
{0.01in}(\mathcal{L}_{\partial}){^{C}}_{D}\hspace{0.01in}g^{DB}%
.\label{KonLMat}%
\end{equation}

In Ref.~\cite{Bauer:2003} and Ref.~\cite{Mikulovic:2006}, we have written down
the co-prod\-ucts of the partial derivatives $\partial^{A}$ or $\hat{\partial
}^{A}$, $A\in\{+,3,-\}$. By taking into account Eq.~(\ref{KopParAllg}), you
can read off the entries of the L-ma\-tri\-ces $\mathcal{L}_{\partial}$ and
$\mathcal{\bar{L}}_{\partial}$ from these co-prod\-ucts. You find, for
example:\footnote{Instead of $T^{3}$, one often uses $\tau=1-(q-q^{-1}%
)\hspace{0.01in}T^{3}$.}%
\begin{equation}
(\mathcal{L}_{\partial}){^{-}}_{-}=\Lambda^{1/2}\hspace{0.01in}\tau
^{-1/2}\text{ and }(\mathcal{\bar{L}}_{\partial}){^{+}}_{+}=\Lambda
^{-1/2}\hspace{0.01in}\tau^{-1/2}.
\end{equation}

The scaling operator $\Lambda$ acts on the spatial coordinates or the
corresponding partial derivatives as follows:%
\begin{equation}
\Lambda\triangleright X^{A}=q^{4}X^{A},\qquad\Lambda\triangleright\partial
^{A}=q^{-4}\partial^{A}.\label{SkaOpeWir}%
\end{equation}
These actions imply the commutation relations%
\begin{equation}
\Lambda\hspace{0.01in}X^{A}=q^{4}X^{A}\Lambda,\qquad\Lambda\hspace
{0.01in}\partial^{A}=q^{-4}\partial^{A}\Lambda\label{VerSkaKooQuaEukDrei}%
\end{equation}
if we take into account the Hopf structure of $\Lambda$ \cite{ogievetsky1992}:%
\begin{equation}
\Delta(\Lambda)=\Lambda\otimes\Lambda,\qquad S(\Lambda)=\Lambda^{-1}%
,\qquad\varepsilon(\Lambda)=1.\label{HopStrLam}%
\end{equation}

The Hopf structure of the partial derivatives also includes an
antipode and a co-unit. Regarding the co-unit of the partial derivatives, the
following holds \cite{ogievetsky1992}:%
\[
\varepsilon(\partial^{A})=0.
\]
We can obtain the antipodes of the partial derivatives from their
co-prod\-ucts using the following Hopf algebra axioms:%
\begin{equation}
a_{(1)}\cdot S(a_{(2)})=\varepsilon(a)=S(a_{(1)})\cdot a_{(2)}.
\end{equation}
Due to this axiom, we have:%
\begin{equation}
S(\partial^{A})=-\hspace{0.01in}S(\mathcal{L}_{\partial}){^{A}}_{B}%
\hspace{0.01in}\partial^{B},\qquad\bar{S}(\hat{\partial}^{A})=-\hspace
{0.01in}S^{-1}(\mathcal{\bar{L}}_{\partial}){^{A}}_{B}\hspace{0.01in}%
\hat{\partial}^{B}.\label{AlgForInvAntAbl}%
\end{equation}
For example, we get the following expressions for the antipodes of the partial
derivatives $\partial^{-}$ and $\hat{\partial}^{+}$ (also see
Ref.~\cite{Bauer:2003}):%
\begin{equation}
S(\partial^{-})=-\Lambda^{-1/2}\hspace{0.01in}\tau^{1/2}\hspace{0.01in}%
\partial^{-},\qquad\bar{S}(\hat{\partial}^{+})=-\Lambda^{1/2}\hspace
{0.01in}\tau^{1/2}\hspace{0.01in}\hat{\partial}^{+}.
\end{equation}

With the help of the antipodes of partial derivatives, we can write the left
actions of partial derivatives as right actions and vice versa. Concretely, we
have%
\begin{align}
\partial^{A}\triangleright f &  =f\triangleleft S(\partial^{A}%
)=-\big (f\triangleleft S(\mathcal{L}_{\partial}){^{A}}_{B}\big )\triangleleft
\partial^{B}\nonumber\\
&  =-\big ((\mathcal{L}_{\partial}){^{A}}_{B}\triangleright
f\big )\triangleleft\partial^{B},\nonumber\\[0.06in]
\hat{\partial}^{A}\,\bar{\triangleright}\,f &  =f\,\bar{\triangleleft}%
\,\bar{S}(\hat{\partial}^{A})=-\big (f\triangleleft S(\mathcal{\bar{L}%
}_{\partial}){^{A}}_{B}\big )\,\bar{\triangleleft}\,\hat{\partial}%
^{B}\nonumber\\
&  =-\big ((\mathcal{\bar{L}}_{\partial}){^{A}}_{B}\triangleright
f\big )\,\bar{\triangleleft}\,\hat{\partial}^{B},\label{LinkRechtDarN}%
\end{align}
and%
\begin{align}
f\triangleleft\hat{\partial}^{A} &  =S^{-1}(\hat{\partial}^{A})\triangleright
f=-\hspace{0.01in}\hat{\partial}^{B}\triangleright\big (S^{-1}(\mathcal{L}%
_{\partial}){^{A}}_{B}\triangleright f\big )\nonumber\\
&  =-\hspace{0.01in}\hat{\partial}^{B}\triangleright\big (f\triangleleft
(\mathcal{L}_{\partial}){^{A}}_{B}\big ),\nonumber\\[0.06in]
f\,\bar{\triangleleft}\,\partial^{A} &  =\bar{S}^{-1}(\partial^{A}%
)\,\bar{\triangleright}\,f=-\hspace{0.01in}\partial^{B}\,\bar{\triangleright
}\,\big (S^{-1}(\mathcal{\bar{L}}_{\partial}){^{A}}_{B}\triangleright
f\big )\nonumber\\
&  =-\hspace{0.01in}\partial^{B}\,\bar{\triangleright}\,\big (f\triangleleft
(\mathcal{\bar{L}}_{\partial}){^{A}}_{B}\big ).\label{RechtsLinksDarN}%
\end{align}

{\normalsize
\bibliographystyle{unsrt}
\bibliography{book,habil}
}
\end{document}